\documentclass[11pt,letter]{article}
\usepackage{epsfig}
\usepackage{tabularx}
\usepackage[titletoc,title]{appendix}
\usepackage{setspace}
\usepackage{lscape}
\usepackage{booktabs}
\usepackage{array}
\usepackage[english]{babel}
\usepackage{makeidx}
\usepackage{latexsym}
\usepackage{amssymb}
\usepackage{qsymbols}
\usepackage{graphicx}
\usepackage{amsmath, amsthm, amssymb}
\usepackage{nomencl}
\usepackage{booktabs,caption}
\usepackage{threeparttable}

\usepackage{multirow}
\usepackage{multicol}
\usepackage{xcolor,soul}
\usepackage{lscape}
\usepackage{tabularx}
\usepackage{pdflscape}
\usepackage{epstopdf}
\usepackage{color}
\usepackage{dsfont}
\usepackage{natbib}

\newcommand{\bs}{\boldsymbol}

\def\b#1{\mbox{\boldmath $#1$}}    % - \b: bold in formulas
\def\m#1{\mbox{#1}}                % - \m: text in formulas
\def\ml#1{\mbox{\scriptsize #1}} % - \ml: little text in formulas
\def\ha#1{\mbox{$\hat{\b #1}$}}

\newcommand{\nothing}[1]{}

\allowdisplaybreaks

\newlength{\defbaselineskip}
\newcommand{\setlinespacing}[1]%
           {\setlength{\baselineskip}{#1 \defbaselineskip}}

\title{\vspace*{-1.5cm} {{Semi-Parametric Empirical Best Prediction for small area estimation of unemployment indicators }} \vspace*{5mm}}
\author{
M.F. Marino\thanks{Department of Statistics, Computer Science, Applications, Universit\`a degli Studi di Firenze, Italy. {\tt mariafrancesca.marino@unifi.it}} 
\and M.G. Ranalli \thanks{Department of Political Science, Universit\`a degli Studi di Perugia, Italy. {\tt giovanna.ranalli@unipg.it}}
\and N. Salvati \thanks{Department of Economics and Management, Universit\`a di Pisa, Italy. {\tt  nicola.salvati@unipi.it}} 
\and M. Alf\`o \thanks{Department of Statistical Science, Sapienza Universit\`a di Roma, Italy. {\tt marco.alfo@uniroma1.it}} 
}

\date{}
\begin{document}

\maketitle

\begin{abstract}
The Italian National Institute for Statistics regularly provides estimates of unemployment indicators using data from the Labor Force Survey. However, direct estimates of unemployment incidence cannot be released for Local Labor Market Areas. These are unplanned domains defined as clusters of municipalities; many are out-of-sample areas and the majority is characterized by a small sample size, which render direct estimates inadequate. The Empirical Best Predictor represents an appropriate, model-based, alternative.  
However, for non-Gaussian responses, its computation and the computation of the analytic approximation to its Mean Squared Error require the solution of (possibly) multiple integrals that, generally, have not a closed form. 
To solve the issue, Monte Carlo methods and parametric bootstrap are common choices, even though the computational burden is a non trivial task.
{In this paper, we propose a Semi-Parametric Empirical Best Predictor for a (possibly) non-linear mixed effect model by leaving the distribution of the area-specific random effects unspecified and  estimating it from the observed data. 
This approach is known to lead to a discrete mixing distribution which helps avoid unverifiable parametric assumptions and heavy integral approximations.  We also derive a second-order, bias-corrected, analytic approximation to the corresponding Mean Squared Error.} Finite sample properties of the proposed approach are tested via a large scale simulation study. Furthermore, the proposal is applied to unit-level data from the $2012$ Italian Labor Force Survey to estimate unemployment incidence for $611$ Local Labor Market Areas using auxiliary information from administrative registers and the 2011 Census.  
\end{abstract}

%\myspacing

\section{Introduction}
\label{sec:intro}

The Italian National Institute for  Statistics (ISTAT) regularly provides  estimates of unemployment indicators based on data obtained through the Italian Labor Force Survey (ILFS). 
{The ILFS allows to obtain quarterly estimates of the main aggregates regarding the labor market; these are particularly important both at the local and the central government levels for the development of labor market policies. 
These estimates are planned to be reliable at a given, chosen a priori, geographical level, and may not be suitable to all needs. For example,
direct estimates of unemployment indicators cannot be disseminated for Local Labor Market Areas (LLMAs).} These are 611 unplanned domains obtained as clusters of municipalities, defined at the Census on the basis of daily working commuting flows.
In this context, direct survey estimates of unemployment incidence cannot be computed and/or published for most LLMAs. This is due to the presence of out-of-sample areas and to many LLMAs having a small sample size which leads to estimates with an unacceptable large coefficient of variation. For these reasons, ISTAT has implemented the use of indirect, model-based, small area estimators to produce official, yearly, estimates of unemployment incidence for Italian LLMAs \citep{d2012use,d2017space}.  

Small Area Estimation (SAE) has received considerable attention in the past decades in terms of theoretical developments and applications to Official Statistics. An updated appraisal of available approaches for SAE is given in \citet{rao2015small}. In this context, Generalized Liner Mixed Models  \citep[GLMMs, ][]{LairdWare1982} represent a typical tool of analysis. 
Area-specific random effects are used to account for sources of unobserved heterogeneity that are not captured by the covariates and describe correlation between units within the same small area. 
For Gaussian data, \citet{battese:88} introduced and \citet{pra:rao:1990} developed an Empirical Best Linear Unbiased Predictor (EBLUP) to estimate small area characteristics. Tailored to the purpose  of the ILFS, \citet{d2017space} developed unit-level linear mixed models with area- and time-specific random effects, which, based on data from different survey cycles, implement estimation using aggregate data to manage a large number of records. In fact, the ILFS is a continuous survey that collects, every year, information on almost $250,000$ households in $1,400$  municipalities for a total of $600,000$ individuals. However, many survey variables, such as the unemployment status, are categorical in nature and, therefore, SAE methods based on linear mixed models may not be fully appropriate.

{{\citet{Jiang2001} developed an Empirical Best Prediction (EBP) method for the area-specific random effects under a mixed logistic model providing a second-order, bias-corrected, estimator for the corresponding Mean Squared Error (MSE). \citet{jiang:2003} extended this approach to deal with GLMMs for general responses in the Exponential Family. Several functions of area-specific model effects are also investigated by the author.
More recently, \citet{Boubeta2015, Boubeta2017} derived the EBP and the corresponding (second-order) approximation to the MSE under an area-level mixed Poisson model for small area counts, while \cite{Hobza2016} specifically focused on the development of an EBP for small area proportions under the unit-level mixed logistic model according to \cite{jiang:2003} and investigated the empirical behavior of the proposal through a large-scale simulation study. An extension of this latter approach to deal with longitudinal responses was also recently proposed by \cite{Hobza2018}.}
}

In all of these approaches, the area-specific random effects are assumed to be iid draws from a Gaussian distribution. 
One of the drawbacks associated with this assumption entails the computational burden required to derive parameter estimates, compute the EBP and, in particular, provide the corresponding measure of reliability. For non-Gaussian responses, we need to deal with (possibly) multiple integrals that do not admit a closed form expression and, therefore, need to be approximated. Numerical approaches, based e.g. on (adaptive) Gaussian quadrature or Laplace approximations \citep[see e.g.][]{PinheiroBates1995}, or using Monte Carlo approximations  \citep[see e.g.][]{Mcculloch1997} are frequently used for this purpose. 
For this reason, ad-hoc alternatives, mainly based on plug-in predictors and Taylor linearizations, were proposed and are currently largely applied \citep{Saei:Chamb:2003, manteiga:2007, molina:2007, Lopez2013}. 

In this paper, we describe a further alternative and develop  a Semi-Parametric EBP  (sp-EBP) for the small area parameters of interest and a second-order, bias-corrected, approximation to the corresponding MSE. 
{In particular, we propose to leave the distribution of the area-specific random effects (the mixing distribution) unspecified and estimate it from the observed data via a NonParametric Maximum Likelihood approach \citep[NPML -][]{Simar1976, Laird1978, Lindsay1983a, Lindsay1983b}.
This estimate known to be a discrete distribution defined over a finite number of locations leading to a (semi-parametric) finite mixture model with a conditional kernel in the Exponential Family.
The proposed approach offers a number of advantages. First, it allows us to avoid unverifiable assumptions on the random effect distribution; second, since mixture parameters are directly estimated from the data and are completely free to vary over the corresponding support, extreme and/or asymmetric departures from the homogeneous model can be easily accommodated. 
Last and more important, the discrete nature of the mixing distribution allows us to avoid integral approximations and considerably reduces the computational effort. 
The gain with respect to the parametric alternatives is particularly evident when analyzing non-Gaussian responses. }

We present the proposed approach for a general small area parameter, starting from a general response with density in the Exponential Family and, later, focusing on the relevant case of binary data. We compare our proposal to the EBP \citep{jiang:2003} and to the plug-in estimator \citep[e.g.][]{Saei:Chamb:2003,manteiga:2007} in terms of prediction accuracy and computational burden in a large scale simulation study. 
Then, we prove the benefits from using the proposed sp-EBP approach on data from the ILFS to estimate unemployment incidence for the 611 LLMAs using auxiliary information from administrative registers and the 2011 Census. We compare the proposed approach with direct estimates, and with the two aforementioned approaches based on parametric mixed logistic models. 

The paper is organized as follows. Section \ref{sec:data} presents the Italian Labor Force Survey, the estimation problem, and the auxiliary information available. Section \ref{sec:notation} introduces the notation and a brief review of the EBP and its MSE approximation. 
In Section \ref{sec:proposal}, we describe the proposed approach: section \ref{sec:NPML} entails maximum likelihood estimation, while the proposed sp-EBP and its MSE approximation are detailed in Section \ref{sec:EBP}. Section \ref{sec:binary} focuses on the  case of binary responses. Section \ref{sec:sim} reports the results of the simulation study, while Section \ref{sec:application} entails the application of the proposed approach to the ILFS data. Last, Section \ref{sec:concl} summarizes our findings and provides guidelines for future research.

\section{The ILFS data} 
\label{sec:data}
The ILFS is the most important statistical source of information on the Italian labor market. The target population includes the members of all Italian households who regularly live within the national borders, have Italian or foreign citizenship, and are regularly enrolled in the municipal lists. Households registered as resident in Italy who habitually live abroad and permanent members of collective facilities (hospices, children's homes, religious institutions, barracks, etc.) are excluded. A two-stage, municipality-household, sampling design is used to collect data. Primary sampling units are stratified by province (LAU1) and population size. Secondary sampling units are selected with equal probabilities. All individuals with usual residence in the dwelling are interviewed.

The ILFS provides quarterly estimates of the main aggregates for the labor market, such as employment status, type of work and work experience, by gender, age, and region (NUTS2).
Here, we focus on data from the first quarter of 2012 which consist of measurements taken on $93,217$ units aged 15-65 and distributed in $453$  LLMAs. 
{LLMAs refer to 611 unplanned domains obtained as clusters of municipalities where the bulk of the labour force lives and works, and where establishments can find the largest amount of the labour force necessary to occupy the offered jobs. They respond to the need for meaningfully, comparable, sub-regional labour market areas for the reporting and the analysis of statistics. LLMAs are defined on a functional basis, the key criterion being the proportion of commuters who cross the LLMA boundary on their way to work. In $2011$, with the last Census, LLMAs were re-defined by the analysis of daily working commuting flows using a new allocation process, an evolution of the previous algorithm. Nearly half of the LLMAs stands in the size class from $10,000$ up to $50,000$ inhabitants, whereas the highest proportion of the population (35.0$\%$) lives in LLMAs with a dimension between $100,000$ and $500,000$ inhabitants. In 332 LLMAs (over 70$\%$ of the national population), more than three quarter of the labour force lives and
works in the same LLMA, that is self-containment is more than 75\%}. 

\begin{figure}
\caption{Distribution of LLMAs by sample size (a) and  percent coefficient of variation of direct estimates of unemployment incidence (b). First quarter, 2012.}
\centering
\label{fig:samplesizes}
\includegraphics[width=0.45\textwidth]{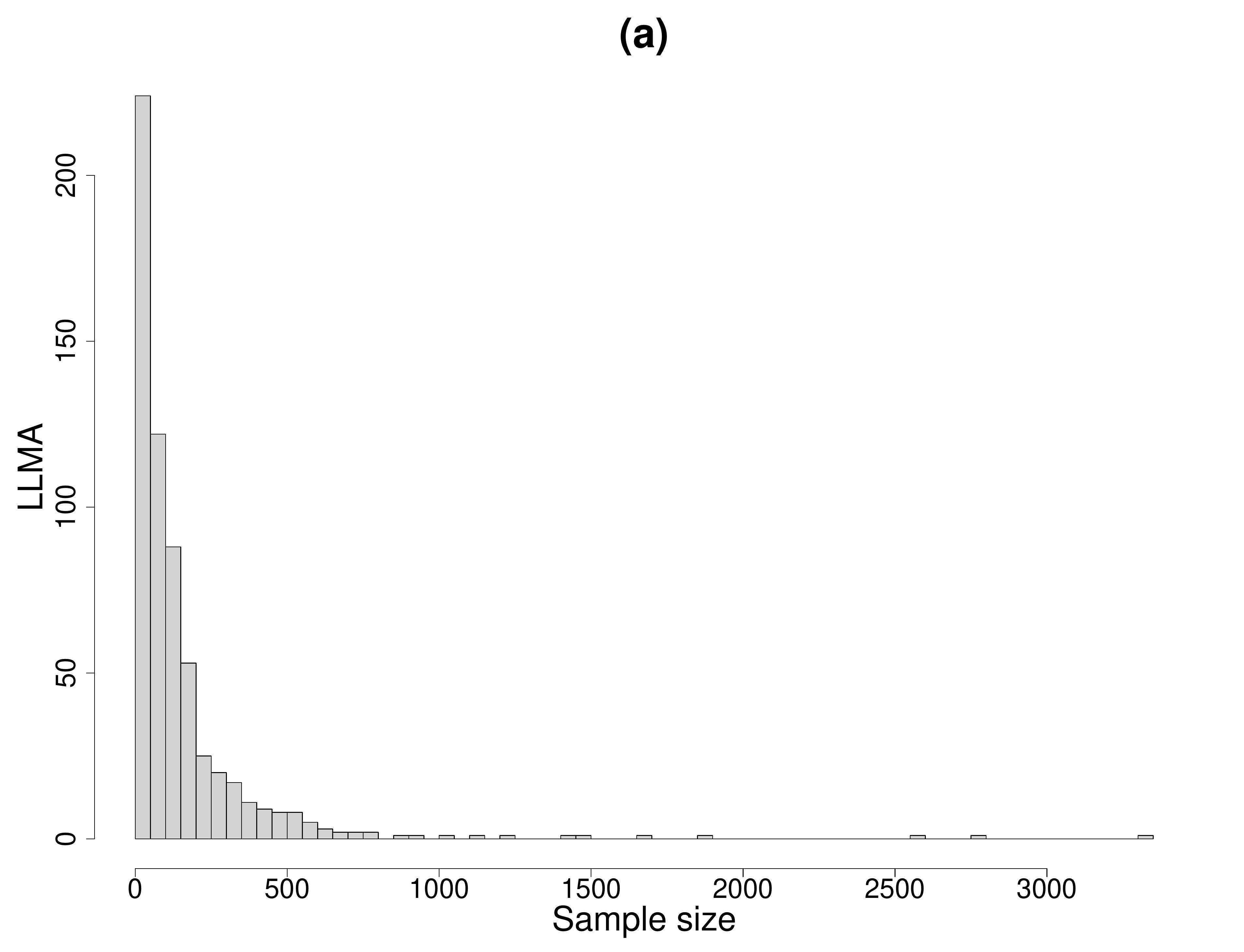} \includegraphics[width=0.45\textwidth]{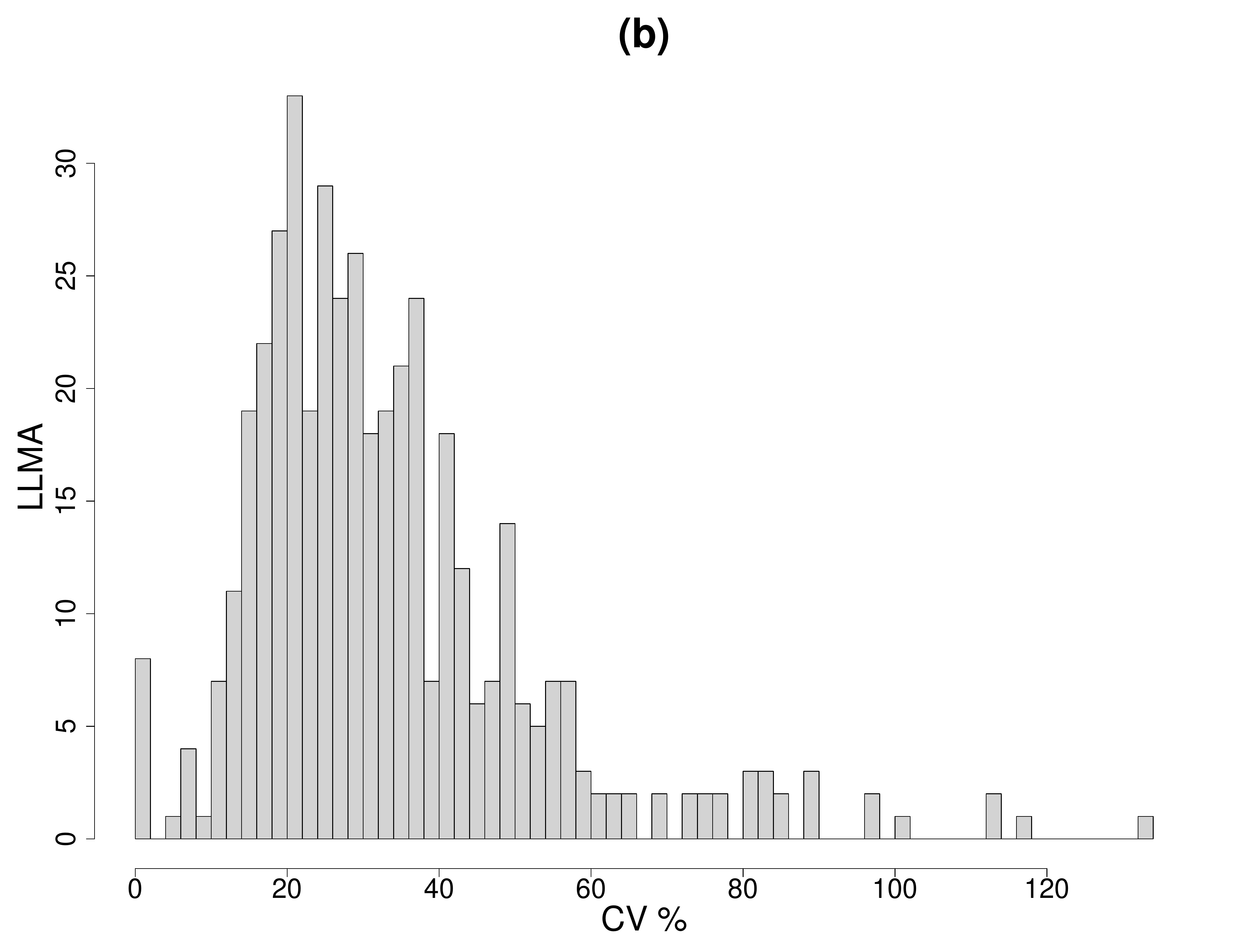} 

\end{figure}

Figure \ref{fig:samplesizes}a shows the distribution of the LLMAs by  sample size. This plot does not include the 158 areas with zero sample size. Among the observed LLMAs, the sample size ranges between $13$ (Acqui Terme, Piedmont Region) and $3,301$ (Milan, Lombardy Region). The mean value is equal to $205.8$, while quartiles are $61\, (25\%)$, $122 \,(50\%)$, and $223 \, (75\%)$, respectively. That is, several LLMAs are characterized by a very small sample size that hinders reliability of direct estimates. Figure \ref{fig:samplesizes}b reports the distribution of the (percent) coefficient of variation (CV) for the direct estimates of unemployment incidence. {The vast majority of estimates have a CV that is larger than $33\%$ that is usually considered as a threshold for reliability. }

Our main interest is on the \textit{Employment Status} variable which can take one out of three different categories: employed ($53.6\%$), unemployed ($6.6\%$), and inactive ($39.8\%$). 
Together with information on employment status for sampled individuals, the following explanatory variables are also available. 
\textit{Sex-Age}: a categorical variable with six categories corresponding to female or male (F/M) and three age groups (15-24, 25-34, and 35-65);
\textit{Educational Level}: a categorical variable with four categories corresponding to no education or primary school diploma, secondary school diploma, high school diploma, and university degree or beyond;
\textit{U-count}: a discrete variable measuring the number of unemployed in a given sex-age group for each LLMA according to the $2011$ Census.

To have a first insight on the data, we report in Tables \ref{tab_status_clsolari}-\ref{tab_status_Edu} the sample distribution of the \textit{Employment Status} by \textit{Sex-Age} and \textit{Educational Level}, respectively. From these tables, we may observe that unemployment incidence is generally higher for people aged 25-34 in the sample, regardless of gender; for the other age groups, unemployment is less frequent among females. By looking at the last column of the table, we notice that females are more frequently {\em inactive} when compared to males, regardless of the age group. This is likely due to their engagement in housekeeping and explains why unemployment incidence is lower in this group. Similarly, by looking at Table \ref{tab_status_Edu}, we may observe that the percentage of unemployment is relatively higher for individuals with higher education. Also in this case, by looking at the last column, it is evident that such a finding is mainly related to job hunting. For instance, $70\%$ of individuals with a primary school diploma or less are out of the job market, as they are not actively looking for a job, and this can be explained by a relatively older age. 

\begin{table}
\centering
\caption{Sample percentage distribution {with standard errors (S.E.)} of \textit{Unemployed status} by \textit{Sex-Age}}
\scalebox{0.8}{
\begin{tabular}{lrrrrrrr}
\toprule
 & Unemployed & S.E. & Employed & S.E. & Inactive & S.E. \\ 
\midrule
  M: 15-24 & 11.0 & 0.4 & 21.5 & 0.5 & 67.6 & 0.6 \\ 
  M: 25-34 & 11.7 & 0.4 & 71.6 & 0.5 & 16.7 & 0.4 \\ 
  M: 35-65 & 5.3 & 0.1 & 70.8 & 0.3 & 23.8 & 0.2 \\ 
  F: 15-24 & 8.8 & 0.3 & 14.1 & 0.4 & 77.1 & 0.5 \\ 
  F: 25-34 & 11.7 & 0.4 & 53.9 & 0.6 & 34.5 & 0.5 \\ 
  F: 35-65 & 4.2 & 0.1 & 48.3 & 0.3 & 47.5 & 0.3 \\

\bottomrule
\end{tabular}
}
\label{tab_status_clsolari}
\vspace{4mm}
%\end{table}
%\begin{table}
\centering
\caption{Sample percentage distribution {with standard errors (S.E.)} of \textit{Unemployed status} by \textit{Educationial level}}
\scalebox{0.8}{
\begin{tabular}{lrrrrrrr}
\toprule
 & Unemployed & S.E.& Employed & S.E.& Inactive & S.E.\\ 
\midrule

Primary school or less & 4.6 & 0.2 & 24.9 & 0.4 & 70.5 & 0.5 \\ 
Middle school & 6.9 & 0.1 & 44.5 & 0.3 & 48.6 & 0.3 \\ 
High school & 7.2 & 0.1 & 62.8 & 0.3 & 30.0 & 0.2 \\ 
University degree or beyond & 5.6 & 0.2 & 75.7 & 0.4 & 18.8 & 0.4 \\

\bottomrule
  \end{tabular}
  }
  \label{tab_status_Edu}
\end{table}

{As highlighted before, the prediction of unemployment incidence for the Italian LLMAs cannot be based on direct survey estimation as direct estimates cannot be computed and/or published for most of the LLMAs.} For these reason, unit-level SAE methods may provide a viable tool to obtain such estimates. In the following section, we introduce the EBP approach by \cite{jiang:2003} for the estimation of small area parameters, together with the approach to approximate the corresponding MSE. As stated in Section \ref{sec:intro}, one of the main drawbacks of such a method is the computational complexity we have face with non-Gaussian data and a large number of observations/small areas, as for the ILFS data. 
In Section \ref{sec:proposal}, we develop a computationally efficient alternative based on a semi-parametric approach.

\section{The Empirical Best Prediction}
\label{sec:notation}
Let $U$ denote a finite population of size $N$, which can be partitioned into $m$ non-overlapping small areas/domains, with $U_i$ denoting the $i$-th small area  with size $N_i, i = 1, \dots, m$.
For a given small area $i$, data consist of $N_i$ measurements of a response variable $Y_{ij}$ and a $p$-dimensional vector of covariates $\b x_{ij} = (x_{ij1}, \dots, x_{ijp})^\prime$, with $j= 1, \dots, N_i$.
{Also, let $\b {\alpha}_{1}, \dots, \b \alpha_m$ be iid, $q$-dimensional, vectors of area-specific random effects ($q \leq p$) with density $f_\alpha(\cdot)$, $E_\alpha(\b \alpha_i) = 0$, and $E_\alpha(\b \alpha_i \b \alpha_i^\prime)= \b \Sigma$ for all $i = 1, \dots, m$. Last, let $\b w_{ij}$ denote a $q$-dimensional subset of $\b x_{ij}$ associated to $\b \alpha_i$.}
We assume that a sample of size $n$ is drawn from the above population and denote by $s_i$ the set containing the $n_i$ population indexes of sample units belonging to small area $i$, with  $n = \sum_{i = 1}^{m} n_i$. On the other hand, the set $r_i\subseteq U_i$ contains the $N_i-n_i$ indexes for non-sampled units in small area $i$. For ease of notation, we assume that all areas are sampled, even though the presence of  out of sample areas can be easily accommodated. We further assume that values of $Y_{ij}$ are known only for the sample ($i = 1, \dots, m, j  \in s_i)$, while the values of $\b x_{ij}$ and $\b w_{ij}$, are known for all units in the population ($i = 1, \dots, m, j = 1, \dots, N_i$). This assumption can be quite restrictive  in some real-world applications, since it implies the availability of individual population information. However, when the auxiliary variables are categorical and/or take a finite number of values, the assumption can be relaxed. We will discuss this issue in more details in the application to the  ILFS data. Last, we assume that sampling is non-informative for the small area distribution of $Y_{ij} \mid \b x_{ij}$, allowing us to use population level models with sample data.

\subsection{The model}
According to a local independence assumption, we assume that, conditional on the area-specific random effects $ \b{\alpha}_{i}$, responses $Y_{ij}$ from the same small area $i$ are independent with density in the Exponential Family 
$$
f_{y\mid\alpha} (y_{ij} \mid  \b\alpha_{i}; \b x_{ij}) = \exp\left\{\frac{y_{ij} \theta_{ij} - b (\theta_{ij})}{a(\phi)} + c(y_{ij}, \phi)\right\},
$$
for $i = 1, \dots, m$ and $j = 1, \dots, N_i$.
In the previous expression, $\phi$ is a dispersion parameter, $a(\cdot), b(\cdot)$ and $c(\cdot)$ are known functions and $\theta_{ij}$ is the canonical parameter for the chosen member of the family. Let $\b \beta$ denote a $p$-dimensional vector of fixed regression coefficients and let us assume that $\theta_{ij}$ is modeled via the following regression model 
$$
\theta_{ij} = \eta_{ij} = \b x_{ij}^\prime \bs \beta +  \b w_{ij}^\prime \b {\alpha}_{i}.
$$
The joint distribution of $\b y_i = (y_{i1}, \dots, y_{iN_i})^\prime$ for the $i$-th small area, conditional on the {vector of area-specific random effects $\b {\alpha}_{i}$}, is obtained by exploiting conditional independence
\[
f_{y\mid\alpha} (\b y_{i} \mid \b \alpha_{i}; \b X_i) = \prod_{j=1}^{N_i} f_{y\mid\alpha} (y_{ij} \mid  \b\alpha_{i}; \b x_{ij}) = 
\exp\left\{\sum_{j=1}^{N_i} \frac{y_{ij}\,\theta_{ij} - b (\theta_{ij})}{a(\phi)} + c(y_{ij}, \phi)\right\},
\]
where $\b X_i$ denotes the matrix of covariates associated to units in the $i$-th area.
The marginal distribution of the area-specific sequence $\b y_i$ is obtained by integrating out $\b \alpha_i$:
\begin{align*}
f_y (\b y_i; \b X_i) &= \int_{\mathbb{R}^q} f_{y\mid\alpha} (\b y_{i} \mid \b \alpha_{i};  \b X_i) f_\alpha( {\b \alpha}_{i}) {d} \b {\alpha}_{i}. 
\end{align*}
Typically, a parametric specification for $f_\alpha(\b \alpha_i)$ is adopted, with a common choice being the zero mean, multivariate, Gaussian distribution. 
It is worth noticing that an implicit exogeneity assumption of observed covariates $\b  x_{ij}$ is taken, that is $f_{\alpha} (\b \alpha_i \mid \b X_i) = f_{\alpha} (\b \alpha_i)$ or $E(\b \alpha_i\mid  \b X_i)  = E(\b \alpha_i) = \b 0$. When this assumption is not fulfilled, the auxiliary regression approach by \cite{Mundlak1978} can be adopted. This slightly modifies the linear predictor above and produces area-specific random effects that are (linearly) free of $\b X_i$, see \cite{NeuhausMcCulloch2006}. In the following, we will assume that, if needed, such an approach is applied and that $f_\alpha(\b \alpha_i \mid \b X_i) = f_\alpha(\b \alpha_i)$.

\subsection{EBP and MSE approximation}
We are interested in using sample data on responses $Y_{ij}$ ($i = 1, \dots, m, j \in s_i$) and population data on covariates $\b x_{ij}$ ($i=1, \dots, m, j = 1, \dots, N_i$) to predict a (possibly) non-linear function of fixed and random effects, say $\zeta(\b\beta, \b\alpha, \b \Sigma)$, with ${\b \alpha = (\b \alpha_1, \dots, \b \alpha_m)}$.
According to \cite{jiang:2003}, the Best Predictor (BP) of $\zeta$ in terms of minimum MSE is given by 
\begin{align}
\tilde{\zeta}^{BP} & = 
E_{\alpha \mid y} \left[\zeta(\b\beta, \b\alpha, \b\Sigma) \mid \b y\right] = 
\int_{\mathbb R^{m \times q}}\zeta(\b\beta, \b \alpha, \b \Sigma) 
f_{\alpha\mid y}(\b \alpha \mid \b y) d\b \alpha,
\label{eq:BPestimator_Jiang}
\end{align}
where 
\[
f_{\alpha\mid y}(\b \alpha \mid \b y) = 
\frac{\prod_{i=1}^m f_{y\mid \alpha}(\b y_{i} \mid \b \alpha_i; \b X_i) f_{\alpha} (\b \alpha_i)}
{ \prod_{i=1}^m \int_{\mathbb R^q} f_{y\mid \alpha}(\b y_{i} \mid \b \alpha_i; \b X_i) f_{\alpha} (\b \alpha_i) d \b \alpha_i},
\]
and $f_{y\mid \alpha}(\b y_{i} \mid \b \alpha_i; \b X_i)  = \prod_{j \in s_i} f_{y\mid \alpha}( y_{ij} \mid \b \alpha_i; \b x_{ij})$.
Since model parameters $\bs \Phi = (\bs \beta, \phi, \b \Sigma)$ are unknown, they need to be estimated. Estimation can be accomplished by maximizing the observed data likelihood function:
\begin{align}
\label{eq_lik}
L(\bs \Phi) = \prod_{i = 1}^m f_y (\b y_{i};\b X_i) = \prod_{i = 1}^m  \int_{\mathbb{R}^q} f_{y\mid\alpha} (\b y_{i} \mid  \b \alpha_{i}; \b X_i)f_\alpha(\b {\alpha}_{i}) d  \b {\alpha}_{i},
\end{align}
where, as before, $f_{y\mid \alpha}(\b y_{i} \mid \b \alpha_i; \b X_i)$ refers to sample data only. To maximize equation \eqref{eq_lik}, we need to evaluate an integral defined over the support of the area-specific random effects and this can be directly done in few cases, for instance when $f_{y\mid\alpha}(\cdot\mid \cdot)$ and $f_{\alpha}(\cdot)$ are conjugate. In all other cases, numerical approximations (e.g. Gaussian quadrature techniques) or simulation based methods (e.g. Monte Carlo integration) need to be used, often leading to a  non trivial computational complexity. To overcome the issue, \cite{Jiang1998} suggested to derive estimates by exploiting the method of moments. A Penalized Quasi Likelihood (PQL) approach \citep[e.g.,][]{BreslowClayton1993} represents a further alternative which is less computationally demanding, even though it may provide inconsistent model parameter estimates \citep[see e.g.][]{RodriguesGoldman1995}.

Once parameters are estimated, we may compute the EBP of $\zeta$, that is $\hat{\zeta}^{EBP}=\tilde \zeta^{BP}(\ha\beta, \ha\alpha, \hat{\b \Sigma})$. 
{To evaluate the quality of such predictions, the second-order MSE estimator detailed by \cite{jiang:2003} can be considered. Under mild regularity conditions, the following decomposition holds: 
\begin{align}
\text{MSE}(\hat{\zeta}^{EBP}) &= E[(\hat{\zeta}^{EBP} -{\zeta})^2]  = \frac{1}{m}e(\bs \Phi) + d(\bs \Phi)+o_p \left(1/m\right),
\label{eq:MSE_Jiang}
\end{align}
where 
\begin{align}
e(\b \Phi) & = E_y \left[ \left( \frac{\partial \tilde \zeta^{BP}}{\partial \bs \Phi} \right)^\prime m V ({\hat{\bs \Phi}}) \left( \frac{\partial \tilde \zeta^{BP}}{\partial \bs \Phi} \right)\right], \label{eq:ePhi}
\\[3mm]
d(\b\Phi) & = E_\alpha [({\zeta})^2] -  E_y [(\tilde{\zeta}^{BP})^2 ] \nonumber \\
& = \int_{\mathbb R^{m\times q}} \zeta(\b \beta, \b \alpha, \b \Sigma)^2 \, f_\alpha(\b \alpha) \, d \b \alpha - E_y \left[ \left( 
\int_{\mathbb R^{m \times q}}\zeta(\b\beta, \b \alpha, \b \Sigma) f_{\alpha \mid y}(\b \alpha \mid \b y) \, d\b \alpha
\right)^2
\right],\label{eq:dPhi}
\end{align}
and $f_\alpha(\b \alpha) = \prod_{i=1}^m f_\alpha(\b \alpha_i)$ denotes the joint density of the random effects $\b \alpha_i, i = 1, \dots, m.$
An estimator of $\text{MSE}(\hat \zeta^{EBP})$ can be obtained by replacing $\bs \Phi$ in equation \eqref{eq:MSE_Jiang} by a consistent estimator, that is
\[
\widehat{\text{MSE}}(\hat{\zeta}^{EBP}) = \frac{1}{m}e(\hat{\bs \Phi})+d(\hat {\bs \Phi}) .
\]
However, as outlined by \cite{jiang:2003}, while
we get an error of order $o_p(m^{-1})$ when we replace ${\b \Phi}$  by $\hat{\b \Phi}$ into $e(\bs \Phi)$, a bias correction is needed to obtain an unbiased estimator for $d({\bs \Phi})$. We discuss this issue in more detail in the following.

As it is clear, computing the MSE requires the solution of (multiple) integrals that may not admit a closed form expression. As stated before, Monte Carlo approximations or numerical integration techniques are required and this makes the computation extremely time-consuming. Bootstrap may represent a further alternative, particularly when dealing with a limited number of small areas. However, when $m$ is large, as in the case of the ILFS data, neither the analytic MSE approximation nor the bootstrap represent viable strategies due to computational issues. 
{\cite{manteiga:2007} proposed a non-optimal Prasad-Rao-type MSE estimator derived from a Taylor series approximation. This estimator fails when sample sizes are too small, while its behavior is proved to be reliable in the case of large sample sizes.}

\section{The Semi-Parametric Empirical Best Prediction}\label{sec:proposal}
As highlighted before, deriving the  EBP of small area parameters and the corresponding MSE approximation as detailed by \cite{jiang:2003} is a non trivial task. In this 
section, we develop a computationally convenient alternative that allows us to avoid unverifiable parametric assumption on the random effect distribution. In Section \ref{sec:NPML}, we present the proposed approach to derive model parameter estimates within a maximum likelihood framework. In Section \ref{sec:EBP}, we detail the proposed Semi-Parametric Empirical Best Predictor (sp-EBP) and the corresponding second-order, bias-corrected, MSE estimator. 

\subsection{Model parameter estimation}
\label{sec:NPML}
When dealing with non-Gaussian responses and GLMMs with Gaussian random effects, maximum likelihood (ML) estimators, although optimal, can be time consuming as we need to approximate (possibly multi-dimensional) integrals that do not admit a closed form expression. 
{An alternative may be based on leaving the distribution of $\b \alpha_i$ completely unspecified and follow the approach detailed by \cite{Aitkin1996, Aitkin1999}. The area-specific random effects are treated as nuisance parameters and a NonParametric Maximum Likelihood (NPML) estimate of their distribution is derived. 
Different contributions to the theory of NPML can be found in the literature \citep{Simar1976, Laird1978, Bohning1982, Lindsay1983a, Lindsay1983b}. Results by 
\citet{Lindsay1983a, Lindsay1983b} show that, as long as the (log-) likelihood function is bounded, it is maximized by a discrete distribution defined on, at most, as many support points as the number of distinct area profiles in the sample. 
In particular, the mixing distribution estimate is a discrete distribution which puts masses $\pi_g > 0 $ on locations $\b \xi_g = (\xi_{g1}, \dots, \xi_{gq})^\prime$, $g = 1, \dots, G$, where the constraint $\sum_{g = 1} ^ G \pi_g = 1$ holds. In a regression context, the number of locations $G$ is bounded from above by the number of different profiles $(\b{y}_{i}, {\b X}_{i})$ in the sample. That is, in the presence of categorical covariates, the number of locations does not necessarily grow with $m$.}

Let $\b \Phi$ denote the global {vector} of model parameters, $\bs \Phi = (\bs \beta, \phi,  \b \xi_1, \dots, \b \xi_G, \pi_1, \dots, \pi_G)^\prime$; the observed data likelihood is approximated by 
\begin{equation} \label{eq:NPML_lik}
L (\bs \Phi ) =  
\prod_{i = 1}^m \int_{\mathbb R^{q}} f_{y\mid \alpha} (\b y_{i} \mid  \b \alpha_i; \b X_i) \, d \b \alpha_i \simeq 
\prod_{i = 1}^m \sum_{g = 1}^G f_{y\mid \alpha} (\b y_{i} \mid  \b \xi_g; \b X_i) \pi_g,
\end{equation} 
where $f_{y\mid \alpha} (\b y_{i} \mid  \b \xi_g; \b X_i) = \prod_{j \in s_i} f_{y\mid\alpha}(y_{ij} \mid  \b \alpha_i = \b \xi_g; \b x_{ij})$ denotes the product of densities in the Exponential Family with canonical parameter $\theta_{ijg}$ defined by the following (mixed) model: 
\begin{equation*}
\theta_{ijg} =  \eta_{ijg} = \b x_{ij}^\prime \bs \beta + \b w_{ij}^\prime  \b {\xi}_{g}.
\end{equation*}
{As it is clear, expression \eqref{eq:NPML_lik} resembles the likelihood of a finite mixture of distributions, with weights $\pi_g=\text{Pr}(\b \alpha_i = \b \xi_g)$. That is, $\b \alpha_i\sim\sum_{g=1}^G\pi_g\delta({\b \xi_g})$, where $\delta(a)$ is a one-point distribution putting a unit mass at $a$.  }
{It is worth noticing that, while the discrete nature of the estimate for $f_\alpha(\cdot)$ may seem unappealing, most approximation techniques (e.g. based on Gaussian quadrature or Monte Carlo approaches) applied when a parametric specification is considered, are exactly of the type in equation \eqref{eq:NPML_lik}. The only substantial difference is that locations $\b \xi_g$ and masses $\pi_g$ in the present proposal are estimated to best fit observed data.}

To maximize the likelihood in \eqref{eq:NPML_lik}, the EM algorithm \citep{Dempster1977} can be employed. A drawback of such an algorithm is that it does not directly provide estimates for the covariance matrix of model parameters. A frequent solution to this issue is based on the use of the Oakes' formula \citep{Oakes1999}, as detailed in Sections 1 and 2 of the on-line Supplementary Material, where the EM algorithm is described. A crucial point in the proposed approach is the choice of the number of mixture components in \eqref{eq:NPML_lik}. {A simple and frequently used solution is as follows: parameter estimates are computed for varying values of $G$ and the model with the best fit, typically measured by penalized likelihood criteria (such as AIC or BIC), is retained.} 
{Typically, the optimal $G$ increases either (\textit{i}) when the variability of the random effect distribution increases or (\textit{ii}) when  the number of small areas increases as, in this case, this may lead to a higher number of distinct area profiles in the sample. }
As long as convergence is entailed, the order for the mixing distribution estimate is {$O_{p}(m^{-1/4})$}, as compared to $O_{p}(m^{-1/2})$ for ML parameter estimates in regular models \citep[see][]{Chen:95}.
However, according to \citet{Lindsay:95}, some smooth functionals, such as the empirical Bayes estimates, can be estimated at the usual $O_{p}(m^{-1/2})$ rate. Furthermore, as shown by \citet{Redner:84}, when the order of the mixture is finite and known, that is when  $\b \alpha_i \sim \sum_{g = 1}^G \pi_g \delta(\b \xi_g)$ is the true mixing, with $G$ known, the usual ML asymptotics apply.

\subsection{Semi-Parametric EBP and MSE approximation }\label{sec:EBP}

Let us now turn to the main problem of interest, where we have a finite population of size $N$ which can be partitioned into $m$ non-overlapping domains or small areas. Furthermore, let $\b\xi=(\b \xi_1,\ldots,\b \xi_G)^\prime$ and $\b\pi=(\pi_1,\ldots,\pi_G)^\prime$ denote the vectors of locations and masses of the finite mixture, respectively. 
We aim at predicting a (possibly) non-linear function of fixed and random effects, $\zeta(\b\beta, \b\xi, \b\pi)$ by exploiting sample data on responses $Y_{ij}$ and populations data on covariates $\b x_{ij}$. 
Under the proposed approach, the Semi-Parametric Best Predictor (sp-BP) of $\zeta$ is defined according to the following expression: 
\begin{align}
\tilde{\zeta}^{\ml{sp-BP}} & = 
E_{\alpha \mid y} \left[\zeta(\b\beta, \b\xi, \b\pi)\mid \b y\right]= 
\sum_{g_1 \dots g_m}\zeta(\b\beta, \b \xi_{g_1, \dots, g_m}, \b \pi)
\prod_{i=1}^m \tau_{ig_i},
\label{eq:npBPestimator}
\end{align}
where $\sum_{g_1 \dots g_m}$ is a shorthand for $\sum_{g_1=1}^G \cdots \sum_{g_m=1}^G$, $\b \xi_{g_1, \dots, g_m} = (\b \xi_{g_1}, \dots, \b \xi_{g_m})^\prime$, and $\tau_{ig}$ denotes the posterior probability for the $i$-th small area to belong to the $g$-th component of the finite mixture. In particular, denoting by $z_{ig}, i = 1, \dots, m, g = 1, \dots, G,$ the component membership indicator for the $i$-th small area, $\tau_{ig}$ is defined by
\begin{equation}
\tau _{ig} = \Pr\left(z_{ig} =1 \mid \b y_i \right) =
\frac{\pi_g \, f_{y\mid \alpha} (\b y_{i} \mid  \b \xi_g; \b X_i) }
{\sum_{l =1}^G \pi_l \, f_{y\mid \alpha} (\b y_{i} \mid  \b \xi_l; \b X_i) },
\label{eq:tau_ig}
\end{equation}
where, as before, $f_{y\mid \alpha} (\b y_{i} \mid  \b \xi_g; \b X_i)$ refers to sample data only.
As it is clear, expression \eqref{eq:npBPestimator} denotes the expected value of $\zeta(\b\beta, \b \xi, \b\pi)$, with respect to the posterior distribution of the random effects $\b \alpha$. Since this is a discrete distribution, the integral approximation which is required in equation \eqref{eq:BPestimator_Jiang} directly translates into simpler summations. 

An estimate of $\tilde \zeta^{\ml{sp-BP}}$ can be obtained by replacing model parameters $\b \beta, \b \xi$, and $\b \pi$ by consistent estimates. Here, we consider the estimates derived by the EM algorithm described in Section 1 of the {on-line Supplementary Material}. In the following, we will refer to such a quantity as the Semi-Parametric Empirical Best Predictor (sp-EBP) of $\zeta$, denoted by $\hat\zeta^{\ml{sp-EBP}}=\tilde \zeta^{\ml{sp-BP}}(\ha\beta, \ha\xi, \ha\pi)$.

To evaluate the quality of predictions, we develop an analytic approximation to the MSE of $\hat\zeta^{\ml{sp-EBP}}$ based on the approach by \cite{jiang:2003}, but considering a maximum likelihood estimator. Starting from equation \eqref{eq:MSE_Jiang}, the MSE of the sp-EBP is given by: 
\begin{align}\label{eq:MSE}
\text{MSE}(\hat{\zeta}^{\ml{sp-EBP}}) &=  \frac{1}{m} e^{\ml{sp}}(\b \Phi) + d^{\ml{sp}}(\b \Phi)+ o_p\left(1/m\right),
\end{align}
where the former term, $e^{\ml{sp}}(\b \Phi)$, is defined according to expression \eqref{eq:ePhi} and can be derived by computing model derivatives with respect to $ \b \beta$,  $\b \alpha$, and $\b \pi$, together with the covariance matrix of model parameter estimates, $ V(\hat{\b \Phi})$. See Section 2 in the on-line Supplementary Material for computational details.
On the other hand, $d^{\ml{sp}}(\b \Phi)$ can be derived as follows
\begin{align*}
d^{\ml{sp}}(\b \Phi) &= E_\alpha [({\zeta})^2] -  E_y [(\tilde{\zeta}^{\ml{sp-BP}})^2 ]  \\
& = \sum_{g_1 \cdots g_m} \zeta(\b \beta, \b \xi_{g_1, \dots, g_m}, \b \pi)^2 \, 	\prod_{i=1}^m \pi_{g_i}
%\Pr(\b \alpha_i = \b \xi_{g_i})
- 
E_y \left[ \left( \sum_{g_1 \cdots g_m}  \zeta(\b \beta, \b \xi_{g_1, \dots, g_m}, \b \pi) \prod_{i=1}^m \tau_{ig_i}\right)^2
\right]. \\
%&= \sum_{g_1 \cdots g_m}  \zeta(\b \beta, \b \xi_{g_1, \dots, g_m}, \b \pi)^2 \, \pi_g^m - 
%E_y \left[ \left( \sum_{g = 1}^G \zeta(\b \beta, \b \xi_g, \pi_g) \prod_{i=1}^m \tau_{ig}\right)^2
%\right]
\end{align*}
The computational burden to obtain the above quantities is substantially lower than that required for the approach by \cite{jiang:2003}. Intractable integrals appearing in equations \eqref{eq:ePhi} and \eqref{eq:dPhi} all translate into simple summations which can be solved analytically.

An estimator of $\text{MSE}(\hat \zeta^{\ml{sp-EBP}})$ is obtained by replacing $\bs \Phi$ in \eqref{eq:MSE} by a consistent estimator such as that obtained by maximizing the observed data likelihood in equation \eqref{eq:NPML_lik}. That is,
\begin{equation}\label{eq:MSE_NP_noBias}
\widehat{\text{MSE}}(\hat{\zeta}^{\ml{sp-EBP}}) =   \frac{1}{m}e^{\ml{sp}}(\hat{\bs \Phi})+d^{\ml{sp}}(\hat {\bs \Phi}) .
\end{equation}
However, as we remarked before, this approach does not directly lead to an unbiased estimator of $\text{MSE}(\hat \zeta^{\ml{sp-EBP}})$. 
When replacing $\hat{\b \Phi}$ in $d^{\ml {sp}}(\b \Phi)$, we get an error of order $O_p(m^{-1/2})$ and a bias correction term needs to be considered. \cite{jiang:2003} provided an explicit expression for such a term when model parameters are estimated by the method of moments. Clearly, under the current approach, these results do directly not hold but, rather, need to be adapted. 

Let $\b \Phi_0$ denote the ``true'' vector of model parameters and let us consider a second-order Taylor expansion of $d^{\ml{sp}}({\b \Phi})$ around $\b \Phi_0$ evaluated at $\hat{\b \Phi}$:
\begin{equation}\label{eq:d_sp}
d^{\ml{sp}}(\hat{\b \Phi}) = d^{\ml{sp}}(\b \Phi_0) + \left(\frac{\partial d^{\ml{sp}}}{\partial \bs \Phi} \right)^\prime\Bigg\rvert_{\b \Phi_0} (\hat{\bs \Phi} - \bs \Phi_0) + \frac{1}{2} (\hat{\bs \Phi} - \bs \Phi_0)^\prime \left(\frac{\partial^2 d^{\ml{sp}}}{\partial \bs \Phi \bs \Phi^\prime} \right)(\hat{\bs \Phi} - \bs \Phi_0) + o_p(m^{-1}),
\end{equation}
where $ d^{\ml{sp}}$ is a shorthand for $ d^{\ml{sp}}(\b \Phi)$. From expression \eqref{eq:d_sp}, it is easy to see that
$$
E[d^{\ml{sp}}(\hat{\b \Phi})] = d^{\ml{sp}}({\b \Phi}) 
+ \frac{1}{m} b^{\ml{sp}}(\b \Phi) +  o_p(m^{-1}),
$$
where $b^{\ml{sp}}(\b \Phi)$ denotes a bias correction defined as 
\begin{align}\label{eq:biasNP}
b^{\ml{sp}}(\bs \Phi) 
& = \left(\frac{\partial d^{\ml{sp}}}{\partial \bs \Phi} \right)^\prime\Bigg\rvert_{\b \Phi_0} m E(\hat{\bs \Phi} - \bs \Phi_0) + \frac{m}{2}E\left[(\hat{\bs \Phi} - \bs \Phi_0)^\prime \left(\frac{\partial^2 d^{\ml{sp}}}{\partial \bs \Phi \bs \Phi^\prime} \right)\Bigg\rvert_{\b \Phi_0}(\hat{\bs \Phi} - \bs \Phi_0)\right] \nonumber\\
& = b_1^{\ml{sp}}(\hat{\b \Phi}) + b_2^{\ml{sp}}(\hat{\b \Phi}).
\end{align}
As it is shown in Section 3 of the on-line Supplementary Material, the former term on the right hand side of equation \eqref{eq:biasNP} is given by
\begin{align*}
b_1^{\ml{sp}}(\hat{\b \Phi}) &= \left(\frac{\partial d^{\ml{sp}}}{\partial \bs \Phi} \right)^\prime\Bigg \rvert_{\b \Phi_0}  \frac{m}{2} I_e({\bs \Phi_0})^{-1} E\left\{ tr\left[
I_e(\b \Phi_0)^{-1}   
\left[\frac{\partial^2 S^k(\b \Phi)}{\partial \b \Phi \partial \b \Phi^\prime}\right]\Bigg\rvert_{\b \Phi_0} I_e(\b \Phi_0)^{-1} I_e(\b \Phi_0)  \right]_{1\leq k \leq K}
\right\}
\\
& -\left(\frac{\partial d^{\ml{sp}}}{\partial \bs \Phi} \right)^\prime\Bigg \rvert_{\b \Phi_0}  \frac{m}{2} I_e({\bs \Phi_0})^{-1} tr\left\{ I_e(\b \Phi_0)^{-1}   
\left[
\frac{  \partial I_{e}^{k}(\b \Phi_0)}{ \partial \b \Phi^\prime}\right]_{1\leq k \leq K}
\right\}.
\end{align*}
Here, $I_{e}(\b \Phi_0)$ denotes the expected information matrix, while $S^k(\b \Phi)$ and $I_{e}^{k}(\b \Phi_0)$ denote the $k$-th element of the score function $S(\b \Phi)$ and the $k$-th row of the expected information matrix $I_e(\b \Phi_0)$, respectively.

On the other hand, it can be shown that the latter term in equation \eqref{eq:biasNP} , $b_2^{\ml{sp}}(\hat{\b \Phi})$, can be computed as
\begin{equation*}
b_2^{\ml{sp}}(\hat{\b \Phi}) = \frac{{m}}{2}tr \left\{\left(\frac{\partial^2 d^{\ml{sp}}}{\partial \bs \Phi \partial \bs \Phi^\prime} \right)\Bigg\rvert_{\b \Phi_0} V(\hat{\b \Phi}) \right\}.
\end{equation*}
%where $V(\hat{\b \Phi})$ is the covariance matrix of model parameter estimates. 
A second order, bias corrected, estimator of $\text{MSE}(\hat \zeta^{\ml{sp-EBP}})$ is then given by
\begin{equation}\label{eq:bias_mseNP}
\widehat{\text{MSE}}^*(\hat{\zeta}^{\ml{sp-EBP}}) =  d^{\ml{sp}}(\hat {\bs \Phi}) + \frac{1}{m}\left[e^{\ml{sp}}(\hat{\bs \Phi})-b^{\ml{sp}}(\hat{ \b {\Phi}})\right].
\end{equation}
We report the computational details required to derive $b_1^{\ml{sp}}(\hat{ \b {\Phi}})$ and $b_2^{\ml{sp}}(\hat{ \b {\Phi}})$ in Section 3  of the on-line Supplementary Material. 
 
\section{A special case: binary data}
\label{sec:binary}
In this section, we focus on the relevant case of binary responses modeled via a mixed logistic model with random intercepts. Let $Y_{ij}$ denote the binary response associated to unit $j$ in the $i$-th small area ($i = 1, \dots, m, j = 1, \dots, N_i$), and let $\alpha_i$ denote an area-specific random effect. Again, let $\b x_{ij}$ denote a $p$-dimensional vector of covariates, and  $\b X_{i}$ the matrix of covariates for the $i$-th small area. We assume that, conditional on $\alpha_i$, responses for units in the $i$-th small area are independent Bernoulli random variables with success probability $p_{ij}$, described by the following mixed logistic model: 
\begin{equation}
\label{eq_glmm_pij}
\theta_{ij} = \log\frac{p_{ij}}{1-p_{ij}} =  \eta_{ij} =  \alpha_{i} + \b x_{ij} ^\prime \bs \beta. 
\end{equation}
In the equation above, $\bs \beta$ is a $p$-dimensional vector of fixed model parameters that describes the effect of observed covariates on the logit transform of $p_{ij}$. 
{We consider the practical problem of predicting small area proportions 
$$\bar{Y}_i = \frac{1}{N_i} \sum_{j = 1}^{N_i} Y_{ij},$$
using the GLMM in equation \eqref{eq_glmm_pij}. To this end, we will use the EBP for the quantity 
\begin{equation}
\label{eq_glmm_pi}
p_i = \frac{1}{N_i} \sum_{j = 1}^{N_i} p_{ij}.
\end{equation}
In fact, since $N_i$ is usually very large in most applications, as it is the case in the one at hand, the EBP for $p_i$ can also be used to predict the indicator $\bar{Y}_i$. 
} Let us assume that responses $Y_{ij}$ are observed for sampled units only ($i = 1, \dots, m, j \in s_i$), while covariates $\b x_{ij}$ are available at the population level ($i = 1, \dots, m, j = 1, \dots, N_i$).
Following the approach detailed in the previous sections, we leave the distribution of the area-specific random effects in equation \eqref{eq_glmm_pij} unspecified and approximate it via a discrete distribution that puts masses  $\pi_g>0$ on locations $\xi_1, \dots, \xi_G$, with $\sum_{g = 1}^G \pi_g = 1$. By adopting a canonical link function, the logistic transform of the success probability for a generic area $i$ in the $g$-th component of the finite mixture is given by 
\begin{equation*}
\label{eq_glmm_pijg}
\theta_{ijg} = \log\frac{p_{ijg}}{1-p_{ijg}} =  \eta_{ijg} =  \xi_{g} + \b x_{ij} ^\prime \bs \beta.
\end{equation*}
Using the standard notation for the Exponential Family, the joint conditional density for the observed responses in the $i$-th small area and the $g$-th component is 
\begin{equation*}
f_{ig} = f_{y \mid \alpha} (\b y_i \mid \xi_g; \b X_i) = \exp \left\{ \sum_{j\in s_i} \left[y_{ij} \theta_{ijg} - \log \left(1 + e^{\theta_{ijg}}\right)\right]\right\}.
\end{equation*}
Turning back to the problem of estimating $p_i$ in equation \eqref{eq_glmm_pi}, the corresponding sp-BP is given by
\begin{align*}
\tilde{p}_{i}^{\ml{sp-BP}} & = 
 \sum_{g = 1}^G p_{ig} \, \frac{\exp \left[ \sum_{j\in s_i} y_{ij} \eta_{ijg} - \sum_{j\in s_i}\log \left(1 + e^{\eta_{ijg}}\right)
\right]\pi_g}
{\sum_{l = 1}^G \exp \left[ \sum_{j\in s_i} y_{ij} \eta_{ijl} -\sum_{j\in s_i} \log \left(1 + e^{\eta_{ijl}}\right)\right]
\pi_l} \nonumber \\
& = \sum_{g = 1}^G   p_{ig} \,  \frac{\exp \left[ \alpha_g y_{i\cdot} - \sum_{j\in s_i} \log \left(1 + e^{\eta_{ijg}}\right)
\right]\pi_g}
{\sum_{l = 1}^G \exp \left[ \alpha_l y_{i\cdot} - \sum_{j\in s_i} \log \left(1 + e^{\eta_{ijl}})\right)
\right]\pi_l}  
%\nonumber \\
%& =  \sum_{g = 1}^G p_{ig} \, \tau_{igy_{i\cdot}}, 
\end{align*}
where $y_{i\cdot} = \sum_{j \in s_i} y_{ij}$ and $p_{ig} = N_i^{-1}\sum_{j=1}^{N_i} p_{ijg}$.
{
By letting 
\[
\tau_{igy_{i\cdot}} = \frac{\exp \left[ \alpha_g y_{i\cdot} - \sum_{j\in s_i} \log \left(1 + e^{\eta_{ijg}}\right)
\right]\pi_g}
{\sum_{l = 1}^G \exp \left[ \alpha_l y_{i\cdot} - \sum_{j\in s_i} \log \left(1 + e^{\eta_{ijl}})\right)
\right]\pi_l}, 
\]
the sp-BP of $p_i$ is given by
\begin{align}
\label{binary_BP}
\tilde{p}_{i}^{\ml{sp-BP}} = \sum_{g = 1}^G p_{ig} \, \tau_{ig(y_{i\cdot})}.
\end{align}
}
The corresponding sp-EBP, denoted by $\hat p_i^{\ml{sp-EBP}}$, is obtained by substituting ML estimates of model parameters into expression \eqref{binary_BP}:
\begin{equation}
\label{binary_EBP}
\hat{p}_{i}^{\ml{sp-EBP}}= \sum_{g = 1}^G \hat{p}_{ig} \, {\hat{\tau}_{ig(y_{i\cdot})}},
\end{equation}
while the quality of predictions obtained via $\hat p_i^{\ml{sp-EBP}}$ can be evaluated through the following MSE expression:
\begin{equation}\label{MSE_binary}
\text{MSE}(\hat p_i^{\ml{sp-EBP}}) = E_\alpha [(p_i)^2] - E_y[( \tilde p_i^{\ml{sp-BP}} )^2]  + E_\alpha [(\hat p_i^{\ml{sp-EBP}} - \tilde p_i^{\ml{sp-BP}})^2],
\end{equation}
where
\begin{align*}
E_\alpha [(p_i)^2] &=   \sum_{g = 1}^G p_{ig}^2 \pi_g 
\end{align*}
and
\begin{align*}
E_y [(\tilde p_i^{\ml{sp-BP}})^2] &=  \sum_{h = 0}^{n_i} \left(\tilde p_{i(h)}^{\ml{sp-BP}}\right)^2  \Pr\left( Y_{i\cdot} = h ; \b X_i\right). 
\end{align*}
Here, $\tilde p_{i(h)}^{\ml{sp-BP}}$ denotes the sp-BP of $p_{i}$ conditional on $y_{i.} = h$, that is 
\begin{align*}
\tilde p_{i(h)}^{\ml{sp-BP}} &= \sum_{g=1}^G 
p_{ig} 
\, \left[ \frac{\exp[\xi_g h -\sum_{j\in s_i} \log(1+e^{\theta_{ijg}})]\pi_g}{\sum_{l=1}^G\exp[\xi_l h -\sum_{j\in s_i} \log(1+e^{\theta_{ijl}})]\pi_l}\right]=
\sum_{g=1}^G 
 p_{ig}  \tau_{ig(h)}.
\end{align*}
The term $\Pr\left( Y_{i\cdot} = h; \b X_i \right)$ is obtained as
\[
\Pr\left( Y_{i\cdot} = h ; \b X_i\right) = \sum_{g=1}^G  \Pr \left(Y_{i\cdot} =h\mid \xi_g ; \b X_i\right) \pi_g, 
\]
where $\Pr \left(Y_{i\cdot} =h \mid \xi_g ; \b X_i\right)$ represents the probability of observing $h$ successes in $n_i$ independent, but non identically distributed, Bernoulli trials. This quantity can be obtained using the probability mass function of a Poisson-Binomial random variable \citep[see][]{Chen1997} with parameter $(p_{i1g}, \dots, p_{in_{i}g})$. 
The last term in equation \eqref{MSE_binary} is obtained as
\begin{equation}
\label{eq:MSE2}
E_\alpha [(\hat p_i^{\ml{sp-EBP}} - \tilde p_i^{\ml{sp-BP}})^2] = \sum_{h=0}^{n_i} \left[\left(\frac{\partial \tilde p_{i(h)}^{\ml{sp-BP}}}{\partial \bs \Phi}\right)^\prime m V (\hat{\bs \Phi}) \left(\frac{\partial \tilde p_{i(h)}^{\ml{sp-BP}}}{\partial \bs \Phi}\right) \right] \Pr \left(Y_{i\cdot} = h ; \b X_i\right),
\end{equation}
where $ V(\hat{\b \Phi})$ is the covariance matrix of model parameter estimates and ${\partial \tilde p_{i(h)}^{\ml{sp-BP}}}/{\partial \bs \Phi}$ is the vector of model derivatives conditional on $y_{i \cdot} = h$. Explicit formulas for these latter quantities are provided in Section 4 of the on-line Supplementary Material. 

The  second-order, bias-corrected, estimator of $\text{MSE}(\hat p_i^{\ml{sp-EBP}})$, that is $\widehat{\mbox{MSE}}^*(\hat p_i^{\ml{sp-EBP}})$, is obtained according to expression \eqref{eq:bias_mseNP}, after adapting the bias correction term to the binary case.

\section{Model-based simulation study}\label{sec:sim}
In this section, we evaluate the empirical properties of the proposed approach via a large scale (model-based) simulation study. This consists of $T = 1,000$ samples, where binary population data are generated under some model assumptions and sample data are selected from the simulated population. 
In particular, population data are generated considering $m = 100,200,500$ small areas; then, samples are selected by simple random sampling without replacement within each area. The population and the sample sizes are constant across areas and are fixed to $N_i =100$ and $n_i =10$, respectively. According to the simulation study discussed by \cite{manteiga:2007}, for each unit $j$ in small area $i$, we generate the target variable $Y_{ij}, i = 1, \dots, m, j = 1, \dots, N_i$, from a Bernoulli distribution with success probability defined by
\begin{equation}\label{eq:simulatedModel}
p_{ij} = \frac{\exp(\alpha_i + x_{ij}\beta)}{1+\exp(\alpha_i + x_{ij}\beta)},
\end{equation}
with $\beta=1$, $x_{ij} \sim \m{Unif}(-1, b_i)$, and $b_i= i/8,~i/16,~i/48$ for $m=100,200$ and $500$, respectively. 
To evaluate the impact of parametric assumptions on the distribution of the area-specific random effects, we considered two different scenarios. The first one ({Scenario 1}) uses area-specific random effects from a zero mean, Gaussian, distribution with standard deviation equal to $\sigma_{1}=0.5$. The second scenario ({Scenario 2}) involves area-specific random effects generated from a mixture of Gaussian distributions, $\alpha_i\sim \nu N(\mu_1, \sigma_2)+(1-\nu)N(\mu_2,\sigma_2)$, where $\nu$ represents a random draw from a Bernoulli distribution $\Pr(\nu = 1)= 0.7$, $\mu_1 = 0$, $\mu_2 = 3$, and $\sigma_2 = 0.05$. Based on this latter quantity, it is evident that, under this scenario, the random effect distribution closely resembles that of a discrete distribution putting masses $\nu$ and $1-\nu$ on locations $\mu_1$ and $\mu_2$. {In this framework, the population is made by two separate sets of small areas having different \textit{baseline} levels for the success probabilities. This may be reasonable e.g. for properly representing non-homogeneous unemployment rates typically observed in the North/South of Italy, as we will see in Section \ref{sec:application}.}
Clearly, the chosen scenarios represent two extreme situations; we expect that, in real data applications, the random effect distribution lies in between them.

{In this simulation study, our aim is that of evaluating the empirical behavior of the proposed approach. For each simulated sample, we estimated model parameters for a varying number of mixture components ($G = 2, \dots, 5$) and selected the optimal $G$ according to the AIC index. 
We report in Table \ref{tab:optimalG} the distribution of the optimal number of mixture components $G$ across simulations. As it can be observed, in most of the cases the AIC index leads to selecting a model with $G = 2$ components only. This reflects the reduced variability of the random effect distribution considered under both simulation scenarios. However, it is worth to highlight that, for higher sample sizes, the chance of selecting a higher $G$ slightly increases, especially when $\alpha_i$ is a random draw from a Gaussian density. This result is clearly related to the requirement of a higher number of components to properly approximate the ``true'', continuous, distribution of the area-specific effects.}
\begin{table}[htbp]
  \centering
  \caption{Distribution of the optimal number of mixture components across simulations.}
  \label{tab:optimalG}
    \begin{tabular}{l|rrrrrrrrr}
    \toprule
    & \multicolumn{4}{c}{Scenario 1} &       & \multicolumn{4}{c}{Scenario 2} \\
          \midrule
 m / k    & \multicolumn{1}{c}{2}     & \multicolumn{1}{c}{3}     & \multicolumn{1}{c}{4}     & \multicolumn{1}{c}{5}     &       & \multicolumn{1}{c}{2}     & \multicolumn{1}{c}{3} & \multicolumn{1}{c}{4}& \multicolumn{1}{c}{5} \\
      \midrule
100 & 0.988 & 0.012 & 0.000 & 0.000 &       & 0.968 & 0.032 & 0.000 & 0.000 \\
200 & 0.962 & 0.038 & 0.000 & 0.000 &       & 0.950 & 0.049 & 0.001 & 0.000 \\
500 & 0.858 & 0.132 & 0.001 & 0.000 &       & 0.943 & 0.056 & 0.001 & 0.000 \\
    \bottomrule
    \end{tabular}%
  \label{tab:addlabel}%
\end{table}%

{Starting from parameter estimates derived from the proposed approach, the sp-EBP for small area proportions was derived according to equation \eqref{binary_EBP}. The proposed predictor was then compared with the parametric EBP by \citet{Jiang2001} and the Naive predictor considered in \cite{manteiga:2007}, both based on the assumption of Gaussian random effects.}
For the EBP, parameter estimates were derived via the ML approach based on a Laplace approximation available in the \texttt{glmer} function from the R \texttt{lme4} package \citep{Bates2015}. Given the estimates, small area proportions and corresponding MSEs were derived by adopting the formulas detailed in Section \ref{sec:notation}. To evaluate the intractable integrals, we followed the approach suggested by \cite{Boubeta2015}. That is, we started by generating $B = 2,500$ replicates of the area-specific random effects $\alpha_i^{(b)}$ from a Gaussian density with zero mean and variance equal to the corresponding ML estimate. Then, we considered their antithetic transform $\alpha_i^{(B+b)} = -\alpha_i^{(b)}$ to obtain $2B$ random effect values. Finally, integrals were approximated by the corresponding empirical means. 
In the following, we will denote EBP estimates of small area proportions by $\hat p_{i}^{\ml{EBP}}$. Although this approach is optimal, the computational complexity greatly limits its applicability. Via the current simulation study, we aim at understanding whether the sp-EBP approach we propose could represent an effective alternative, which is optimal in terms of minimum MSE and simpler from a computational point of view. 

For completeness, we also included in the simulation study results from the Naive approach. In this case, parameter estimates were obtained using the PQL approach via the \texttt{glmmPQL} function from the {R} \texttt{MASS} package  \citep{Venables2002}. To get predictions, parameter estimates were directly plugged into the expression for the area-specific proportions: 
\begin{equation}\label{eq:naive}
\hat p_{i}^{\ml{Naive}}=\frac{1}{N_i}\sum_{j=1}^{N_i}\frac{\exp(\hat\alpha_i + x_{ij}\hat\beta)}{1+\exp(\hat\alpha_i + x_{ij}\hat\beta)}.
\end{equation}

{The performance of the small area estimators were evaluated by computing, for each area $i = 1, \dots, m,$ the bias and the Root Mean Squared Error (RMSE), defined as follows:
$$
\mbox{BIAS}_i=T^{-1}\sum_{t=1}^{T}(\hat{p}_{it}^{\ml{Model}}-p_{it}), \quad i = 1, \dots, m,
$$
$$
\mbox{RMSE}_i= \sqrt{T^{-1}\sum_{t=1}^{T}(\hat{p}_{it}^{\ml{Model}}-p_{it})^2},  \quad i = 1, \dots, m,$$
where $\hat p_{it}^{\ml{Model}}$ denotes the model-based proportion estimate for the $i$-th small area in the $t$-th simulated sample obtained via either the EBP ($\hat{p}_i^{\ml{EBP}}$), the sp-EBP ($\hat p_i^{\ml{sp-EBP}}$), or the Naive ($\hat{p}_i^{\ml{Naive}}$) approach.
%These two measures (BIAS and RMSE) are frequently used to evaluate the quality of predictions in the small area literature. However, readers should be aware that the absolute bias may not be a consistent tool to evaluate predictions obtained by posterior means \citep{Gneiting2011}. We decided to report it to be coherent with the simulation studies developed elsewhere \citep[see e.g.][]{Boubeta2015} in the small area context.
For completeness, we also report in Section 5 of the on-line Supplementary Material the distribution of the Mean Absoute Error (MAE) across small areas for the EBP, the sp-EBP, and the Naive predictor under different experimental scenarios. Together with the bias, MAE is frequently used to evaluate the quality of predictions in the small area literature, even though it may not be a consistent tool to evaluate predictions obtained by posterior means \citep{Gneiting2011}. 
}
Figures \ref{fig1} and \ref{fig2} show the BIAS and the RMSE distribution across small areas for the three estimators under investigation for Scenario 1 and $m = 100,200,500$, respectively; {the red line denotes the corresponding mean values.} As expected, when looking at the first two panels (i.e. $m = 100,~200$), the sp-EBP performs better than the Naive estimator and slightly worse than the EBP, with a gap that reduces as $m$ increases, both in terms of BIAS and RMSE.
When $m=500$, performance values of EBP are not showed due to the computational burden required to get the estimates: for one replication, we needed 161.612 minutes on an Intel(R) I5-3330 architecture -- 3.0 GHz, and, therefore, we couldn't obtain results for $T=1,000$ replications in a reasonable amount of time.

\begin{figure}
\caption{Scenario 1: Distribution of the BIAS over areas for $\hat p_i^{\ml{sp-EBP}}$, $\hat p_i^{\ml{EBP}}$, and $\hat p_i^{\ml{Naive}}$, for $m=100$ (left panel), $m = 200$ (central panel), and $m = 500$ (right panel). \label{fig1} }
\vspace{2mm}
\centering
\frame{\includegraphics[scale=0.12]{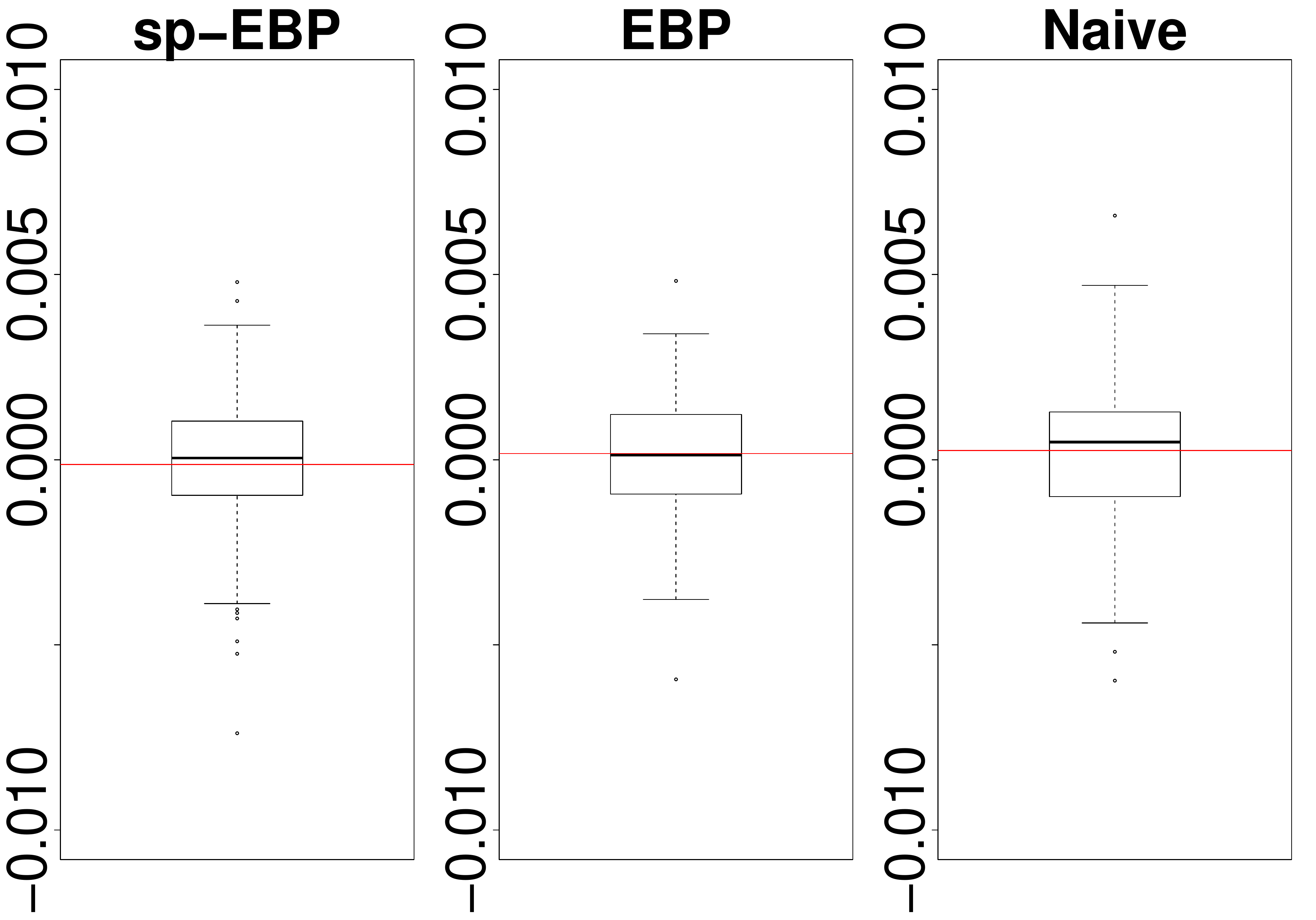} }
\frame{\includegraphics[scale=0.12]{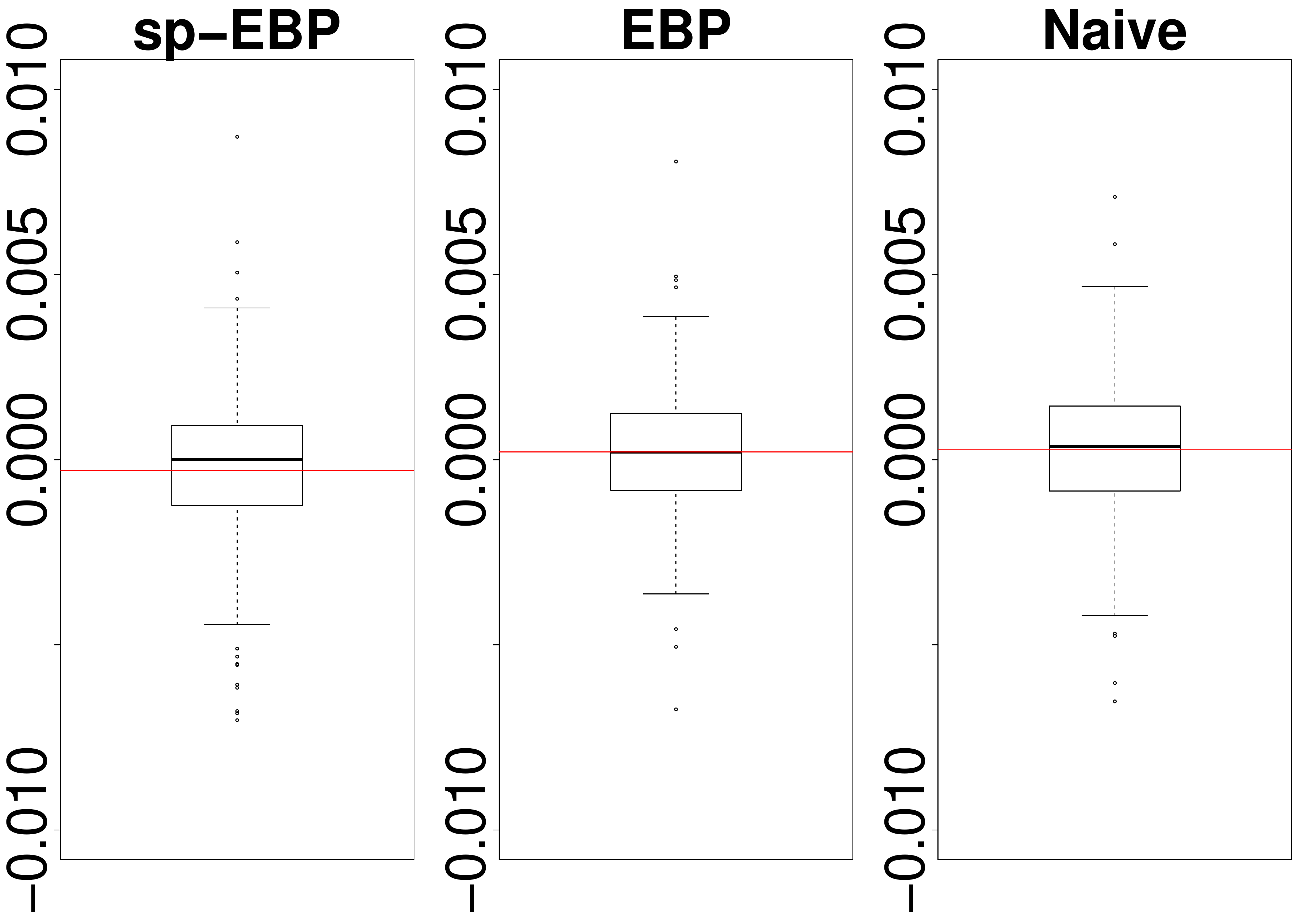}}
\frame{\includegraphics[scale=0.12]{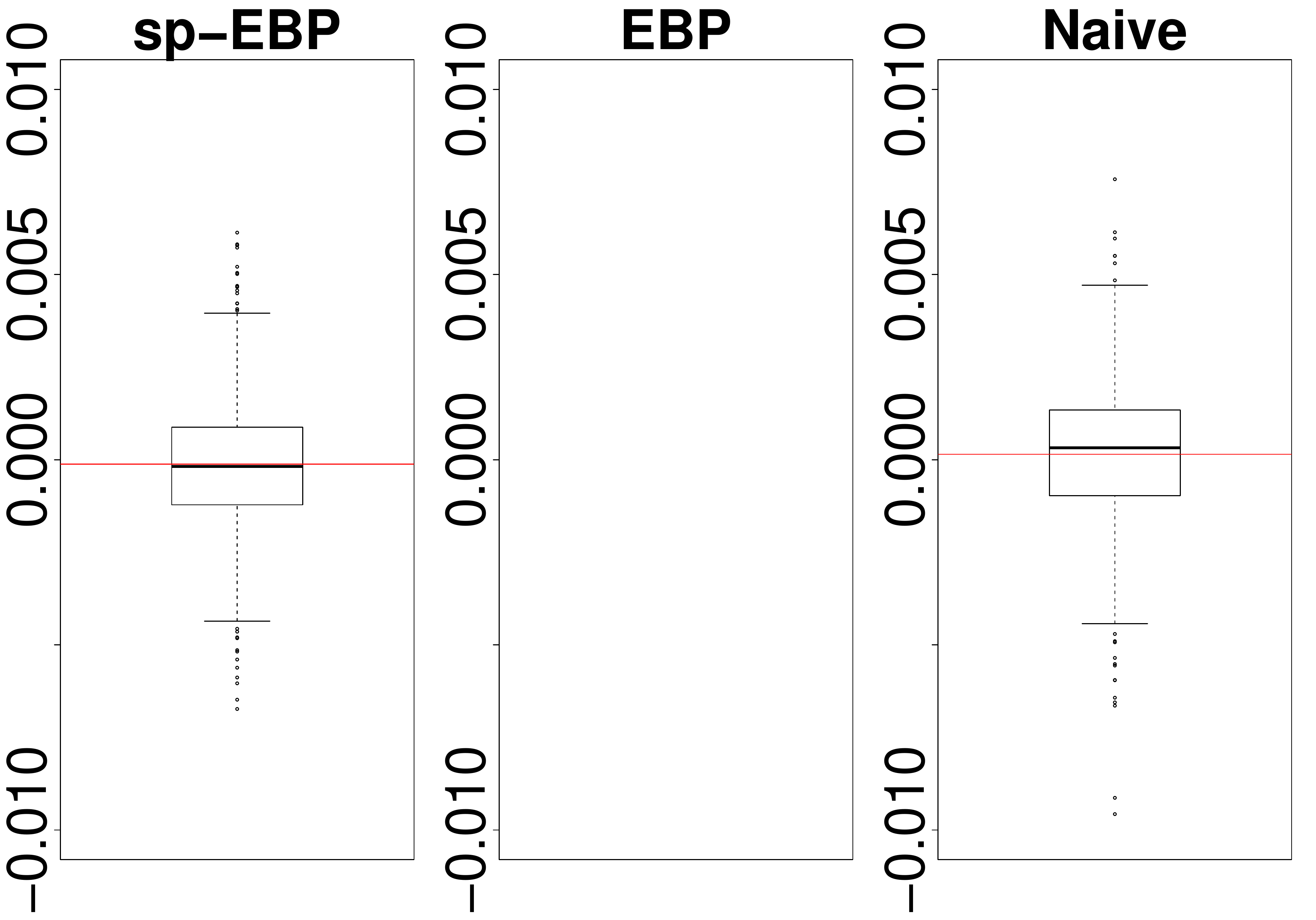} }
\end{figure}

\begin{figure}
\caption{Scenario 1:  Distribution of the RMSE over areas for $\hat p_i^{\ml{sp-EBP}}$, $\hat p_i^{\ml{EBP}}$, and $\hat p_i^{\ml{Naive}}$, for $m=100$ (left panel), $m = 200$ (central panel), and $m = 500$ (right panel). \label{fig2}}
\vspace{2mm}
\centering

\frame{\includegraphics[scale=0.12]{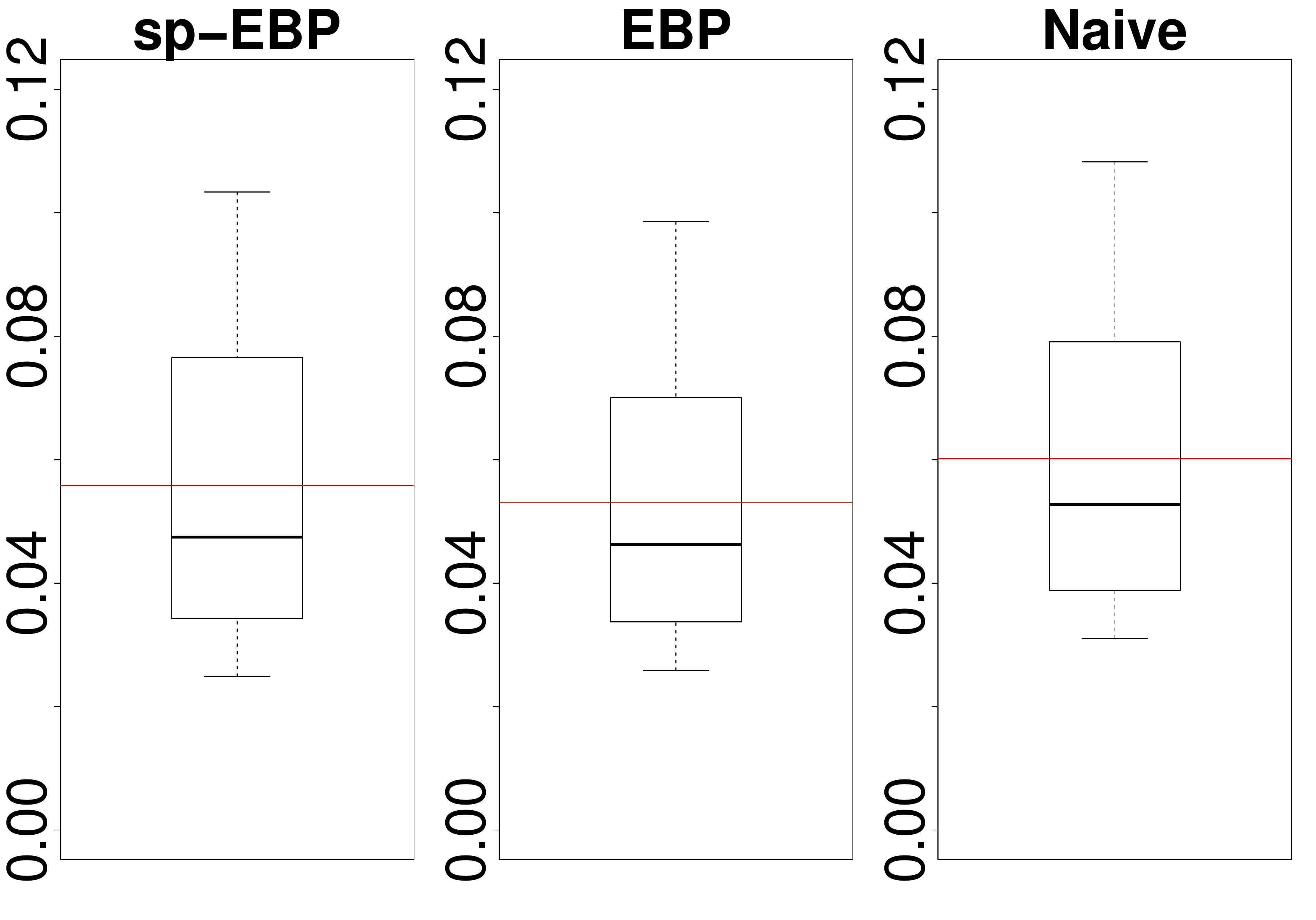} }
\frame{\includegraphics[scale=0.12]{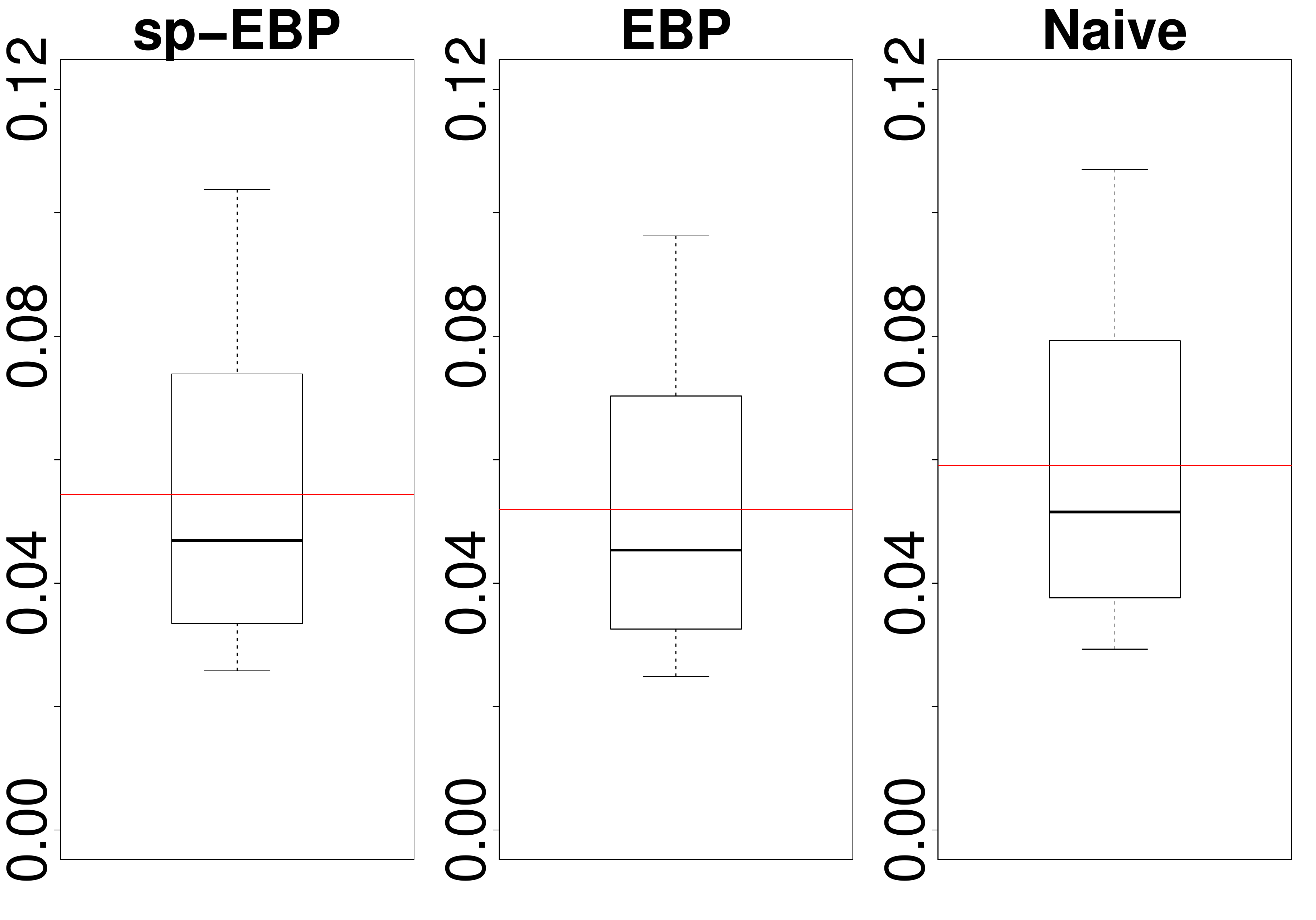}}
\frame{\includegraphics[scale=0.12]{rmse_pred_m100G.pdf}}
\end{figure}

Figures \ref{fig3} and \ref{fig4} show the performance of the estimators under Scenario 2. As before, results for the EBP approach for $m= 500$ are not showed due to computational issues. As it is clear by looking at these plots, when  the assumption of Gaussian random effects does not hold, parametric approaches seem to produce predictions with a reduced quality than those obtained via the semi-parametric alternative we propose. In particular, we may notice that  $\hat p_i^{\ml{sp-EBP}}$ clearly outperforms the two competitors in terms of both bias and RMSE. Also, results for $\hat p_i^{\ml{Naive}}$ and $\hat p_i^{\ml{EBP}}$ seem to slightly worsen as $m$ increases. This may be possibly due to the higher information available and the stronger impact of the random effect distribution on the overall response variability when the number of small areas increases.

\begin{figure}
\caption{Scenario 2:  Distribution of the BIAS over areas for $\hat p_i^{\ml{Naive}}$, $\hat p_i^{\ml{EBP}}$, and $\hat p_i^{\ml{sp-EBP}}$ for $m=100$ (left panel), $m = 200$ (central panel), and $m = 500$ (right panel). \label{fig3}}
\vspace{2mm}
\centering

\frame{\includegraphics[scale=0.12]{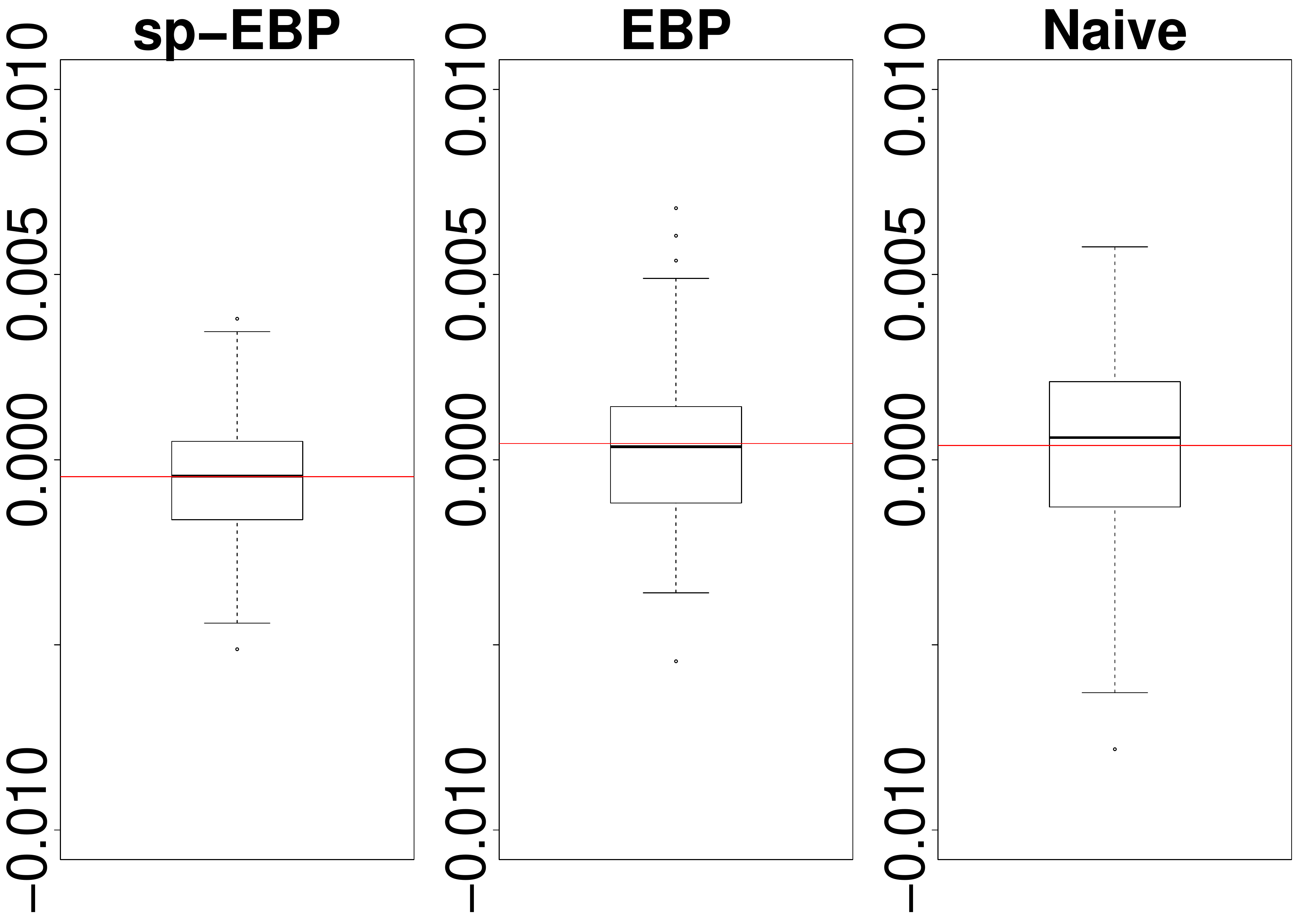} }
\frame{\includegraphics[scale=0.12]{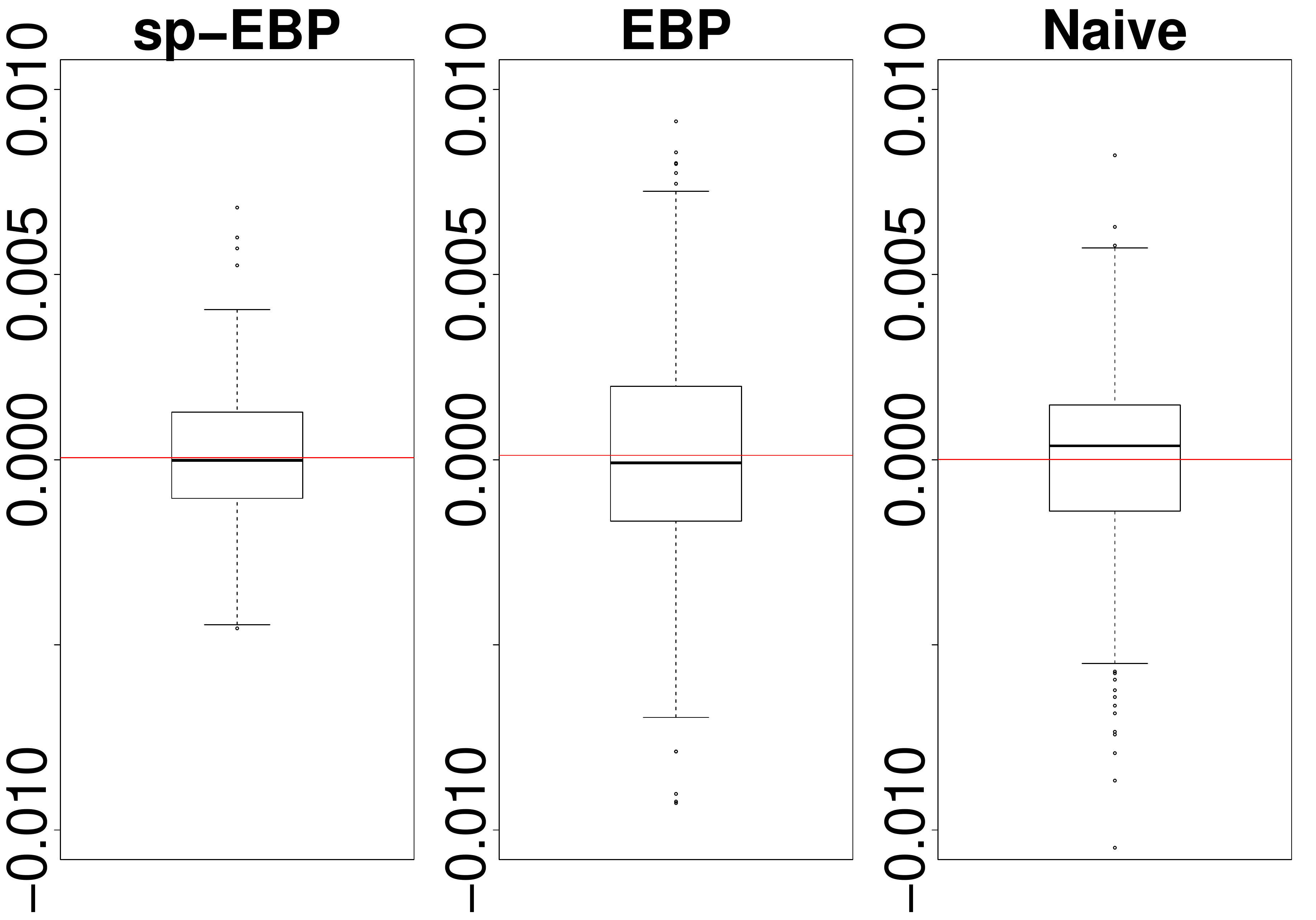}}
\frame{\includegraphics[scale=0.12]{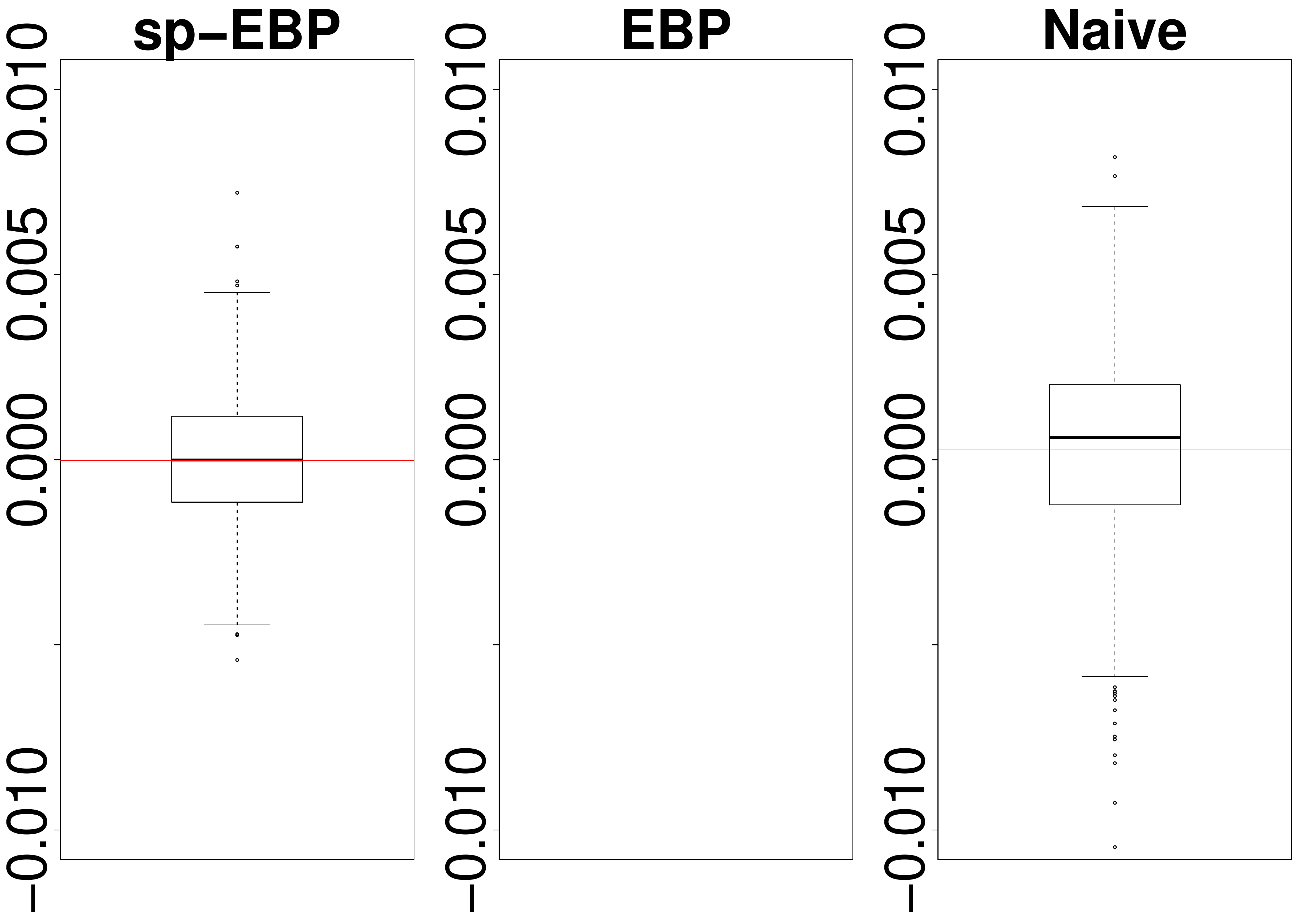} }
\end{figure}
\begin{figure}
\caption{Scenario 2:  Distribution of the RMSE over areas for $\hat p_i^{\ml{Naive}}$, $\hat p_i^{\ml{EBP}}$, and $\hat p_i^{\ml{sp-EBP}}$ for $m=100$ (left panel), $m = 200$ (central panel), and $m = 500$ (right panel). \label{fig4}}
\vspace{2mm}
\centering

\frame{\includegraphics[scale=0.12]{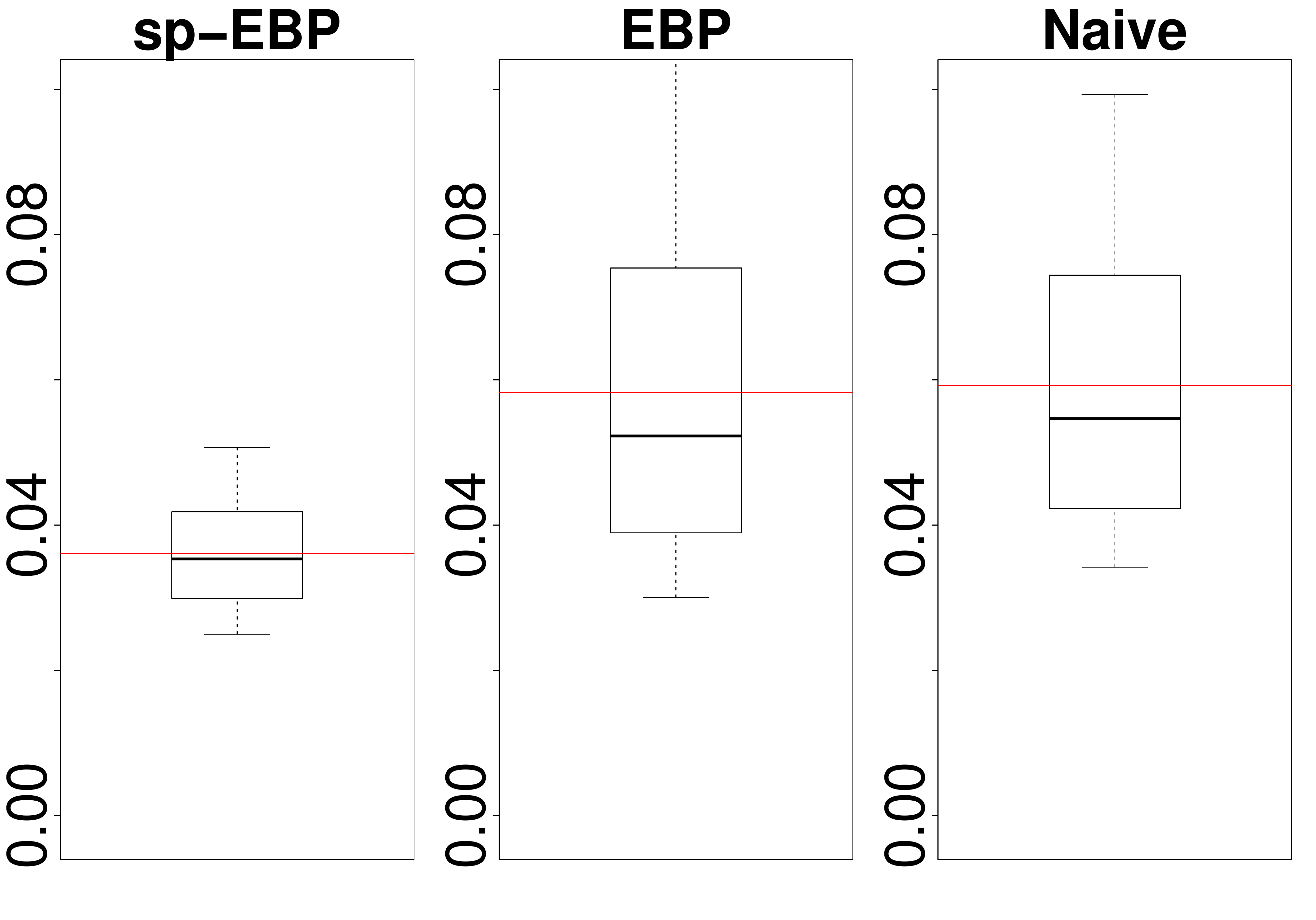} }
\frame{\includegraphics[scale=0.12]{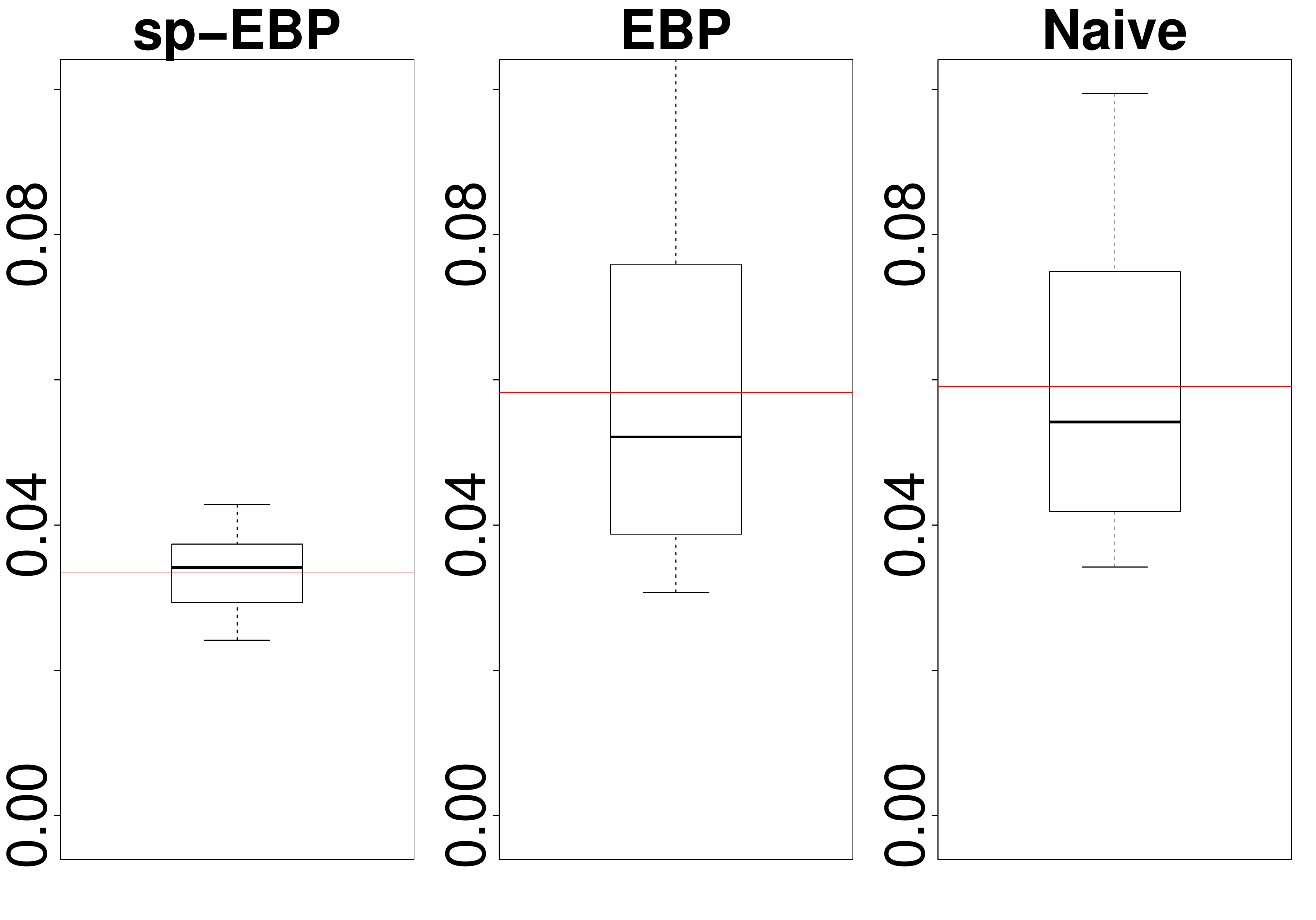}}
\frame{\includegraphics[scale=0.12]{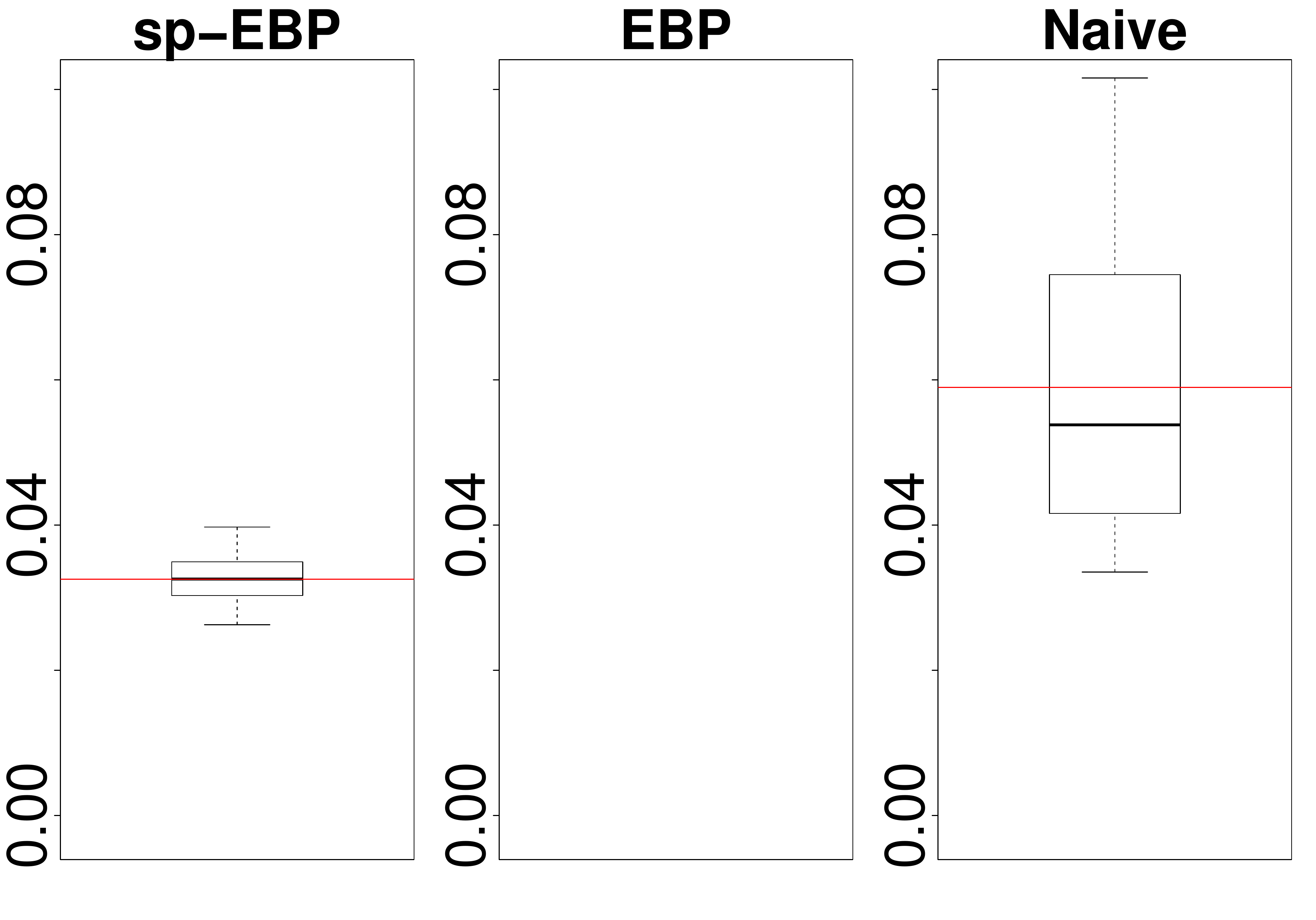} }

\end{figure}
A further purpose of this simulation study is to investigate  the performance of the MSE estimators to evaluate the accuracy of the predictions we discussed so far. In particular, for $\hat p_i^{\ml{sp-EBP}}$, we considered the proposed MSE estimator reported in equations \eqref{eq:MSE_NP_noBias} and \eqref{eq:bias_mseNP}; we will refer to the square root of these quantities as $\widehat {\mbox{RMSE}}(\hat p_i^{\ml{sp-EBP}})$ and $\widehat {\mbox{RMSE}^*}(\hat p_i^{\ml{sp-EBP}})$, respectively. For the estimator $\hat p_i^{\ml{EBP}}$, we used the approximate MSE estimator proposed by {\cite{Hobza2016}}; the corresponding square root will be denoted by $\widehat {\mbox{RMSE}}(\hat p_i^{\ml{EBP}})$. 
Last, for the Naive predictor $\hat p_i^{\ml{Naive}}$, we considered the approach suggested by \cite{manteiga:2007}, based on linearizing the GLMM in equation \eqref{eq:simulatedModel} and, then, applying the Prasad-Rao MSE approximation for the corresponding linear mixed model; the square root of such an estimator will be denoted by $\widehat {\mbox{RMSE}}(\hat p_i^{\ml{Naive}})$.

The performance of the RMSE estimators were evaluated by considering the ratio (R) between the estimated RMSE for the model-based estimates and the corresponding actual RMSE for each small area prediction, that is:
\[
\mbox{R}_i = \frac{\sum_{t=1}^{T} \widehat{\mbox{RMSE}}(\hat p_{it}^{\ml{Model}})}{ \sqrt{\sum_{t=1}^{T} (\hat p_{it}^{\ml{Model}} - p_{it})^2}}, \quad i = 1, \dots, m.
\]
The distribution over areas for such a  ratio for varying $m$ and varying random effect distributions is reported in Figure \ref{fig_Ri}. 
Under Scenario 1, $\widehat {\mbox{RMSE}}(\hat p_i^{\ml{sp-EBP}})$ and $\widehat {\mbox{RMSE}^*}(\hat p_i^{\ml{sp-EBP}})$ seem to perform generally better than the  alternatives. In particular, simulation results suggest that the former estimator is more appropriate when a reduced number of small areas is available ($m = 100, 200$), while its precision decreases in case of larger $m$. On the other hand, $\widehat {\mbox{RMSE}^*}(\hat p_i^{\ml{sp-EBP}})$ shows slight overestimation of the actual Monte Carlo RMSE for $m=100$ and $m = 200$, but it has to be preferred in the presence of a large number of small areas since the $R_i$ index is strongly concentrated around $1$. 

The estimator $\widehat{\mbox{RMSE}}( \hat{ p}_i^{\ml{EBP}})$ underestimate the actual Monte Carlo RMSE, with a ratio which is always lower than $1$ for $m = 100$. The quality of the results improves with $m$, even though it is always lower than that provided by the proposed approach. Such a finding may be possibly due to the estimation of the covariance matrix for parameter estimates which is not as accurate as expected with $B^* = 250$ bootstrap resamples. In fact, it is worth noticing that \cite{Boubeta2015} highlighted the need of a very accurate estimate of the covariance matrix of parameter estimates to ensure high quality of the results. For this reason, in their simulation study, the authors suggested to estimate $V(\hat{\b \Phi})$ by running a Monte Carlo experiment based on $10^{4}$ iterations in advance. In practice, when dealing with large sample sizes, such an approach is computationally very expensive and this is the reason why we considered a bootstrap approach based on $B^* = 250$ iterations only.  Last, the estimator suggested by \cite{manteiga:2007} overestimates the actual RMSE in all the scenarios we considered in this simulation study. 

\begin{figure}
\caption{Distribution of the RMSE ratio over areas for the sp-EBP (without bias correction), the $\mbox{sp-EBP}^*$ (with bias correction), the EBP, and the Naive approach, for $m = 100$ (left panel), $m = 200$ (central panel), and $m = 500$ (right panel), under Scenario 1 (upper panel) and Scenario 2 (lower panel). \label{fig_Ri}}
\vspace{2mm}
\centering
\frame{\includegraphics[scale=0.13]{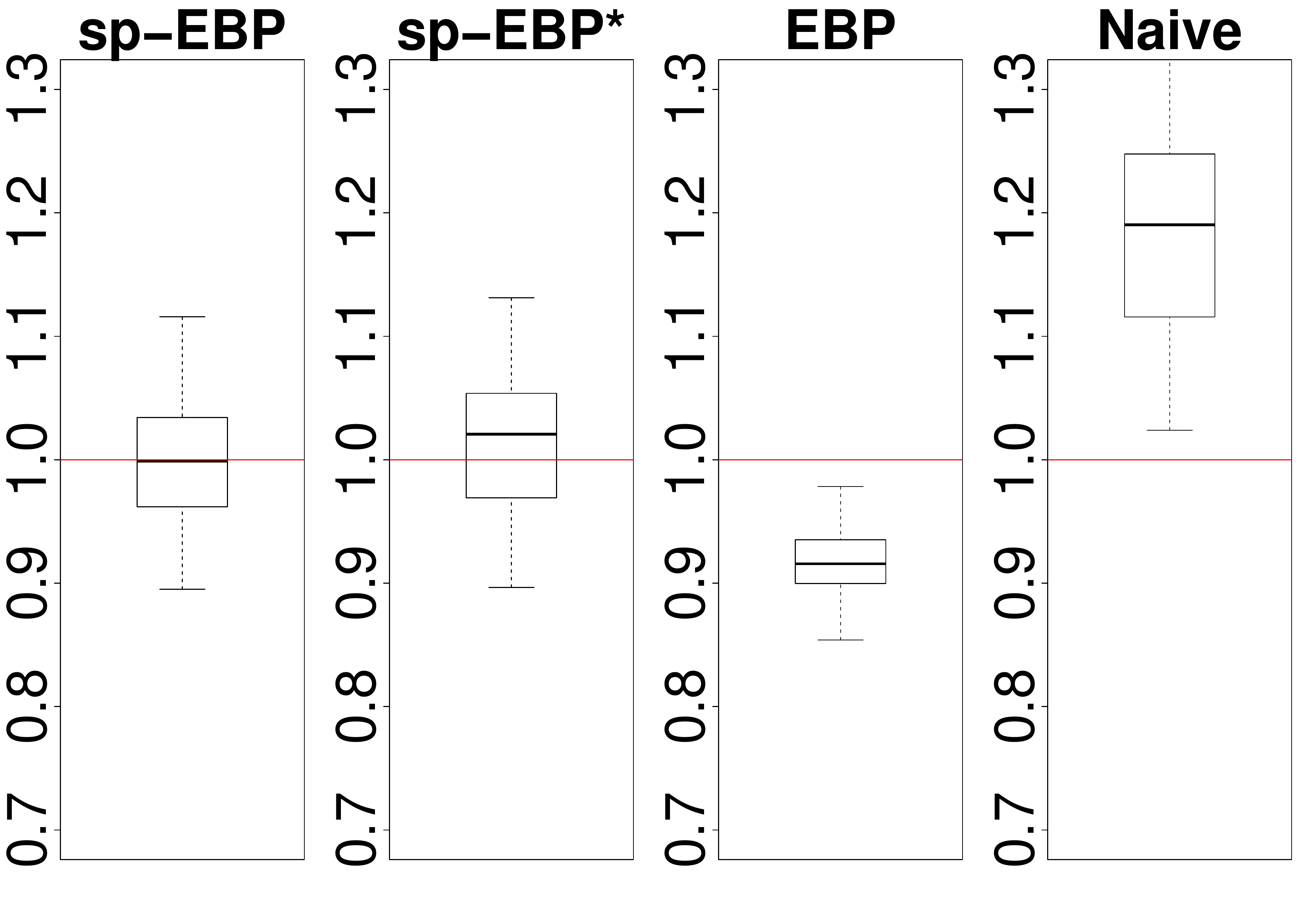}}
\frame{\includegraphics[scale=0.13]{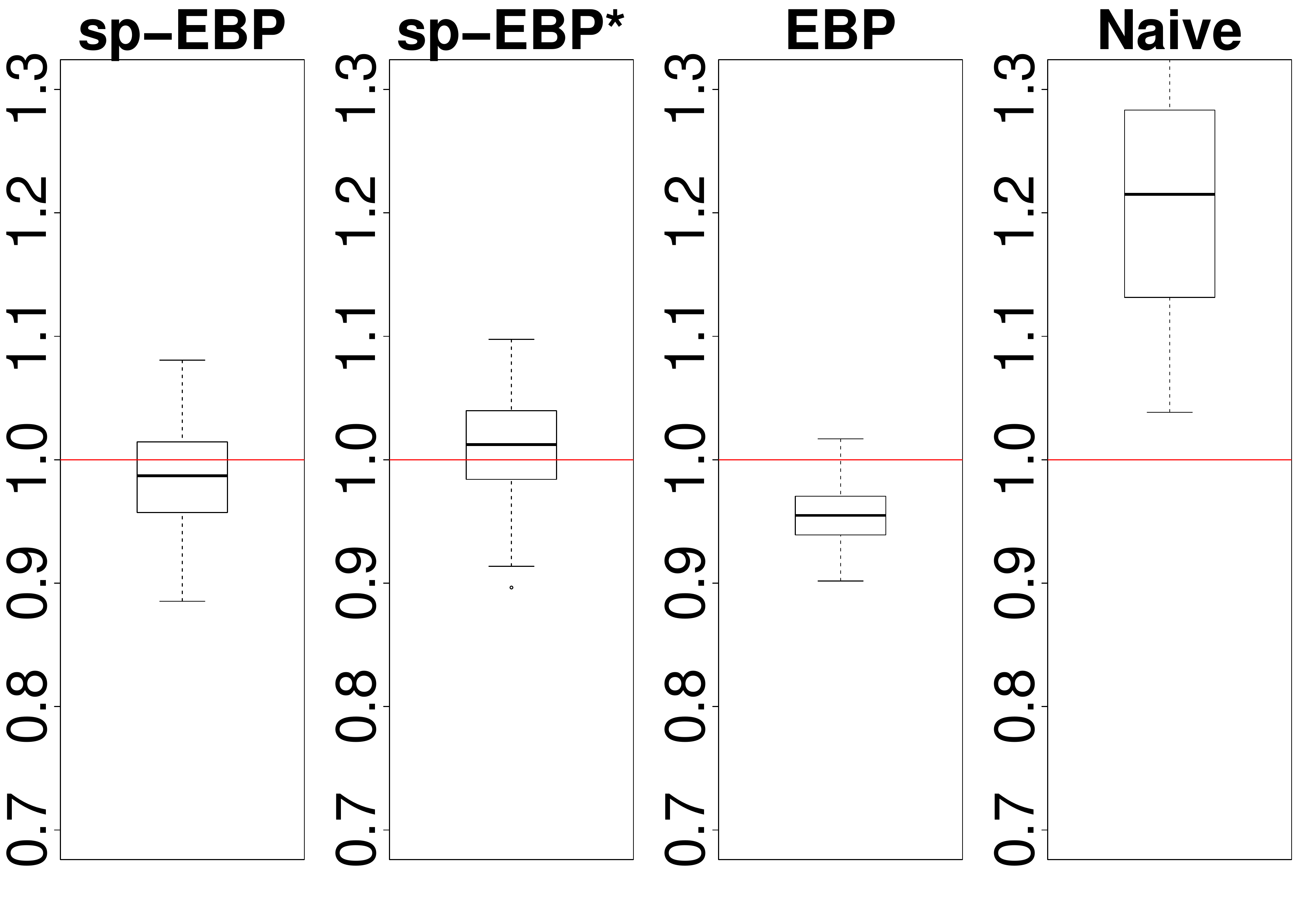}}
\frame{\includegraphics[scale=0.13]{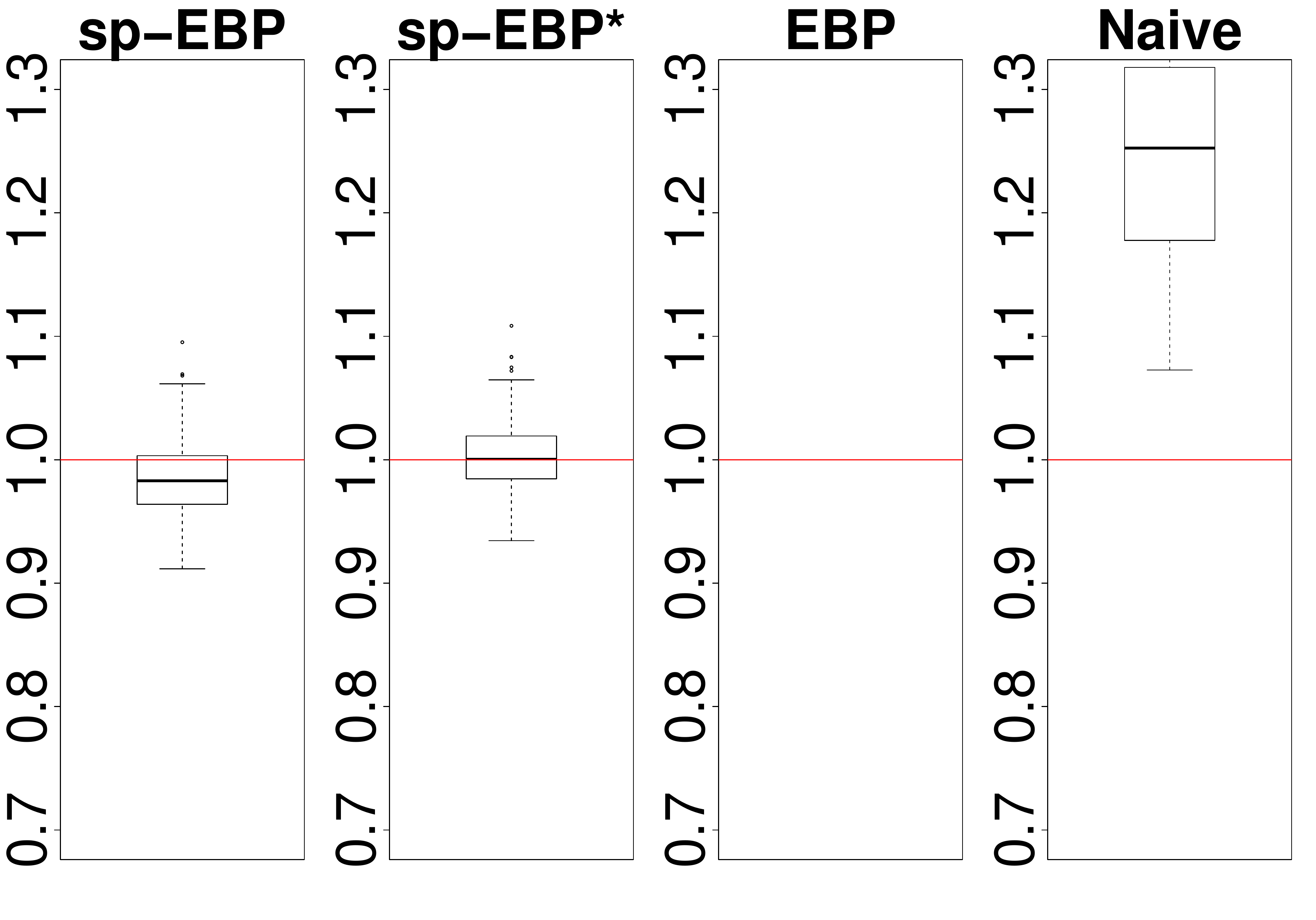}}
\\[5mm]

\frame{\includegraphics[scale=0.13]{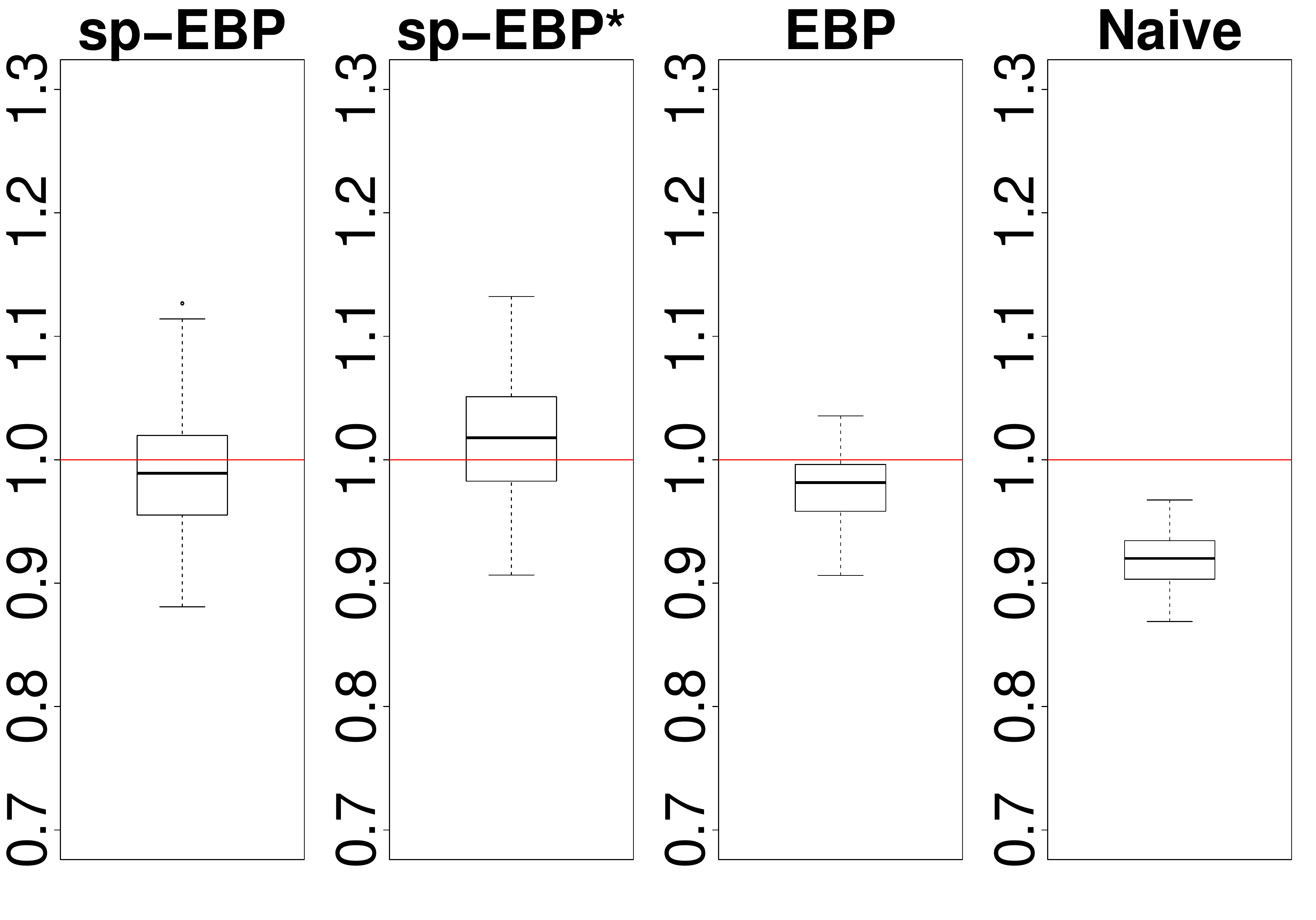}}
\frame{\includegraphics[scale=0.13]{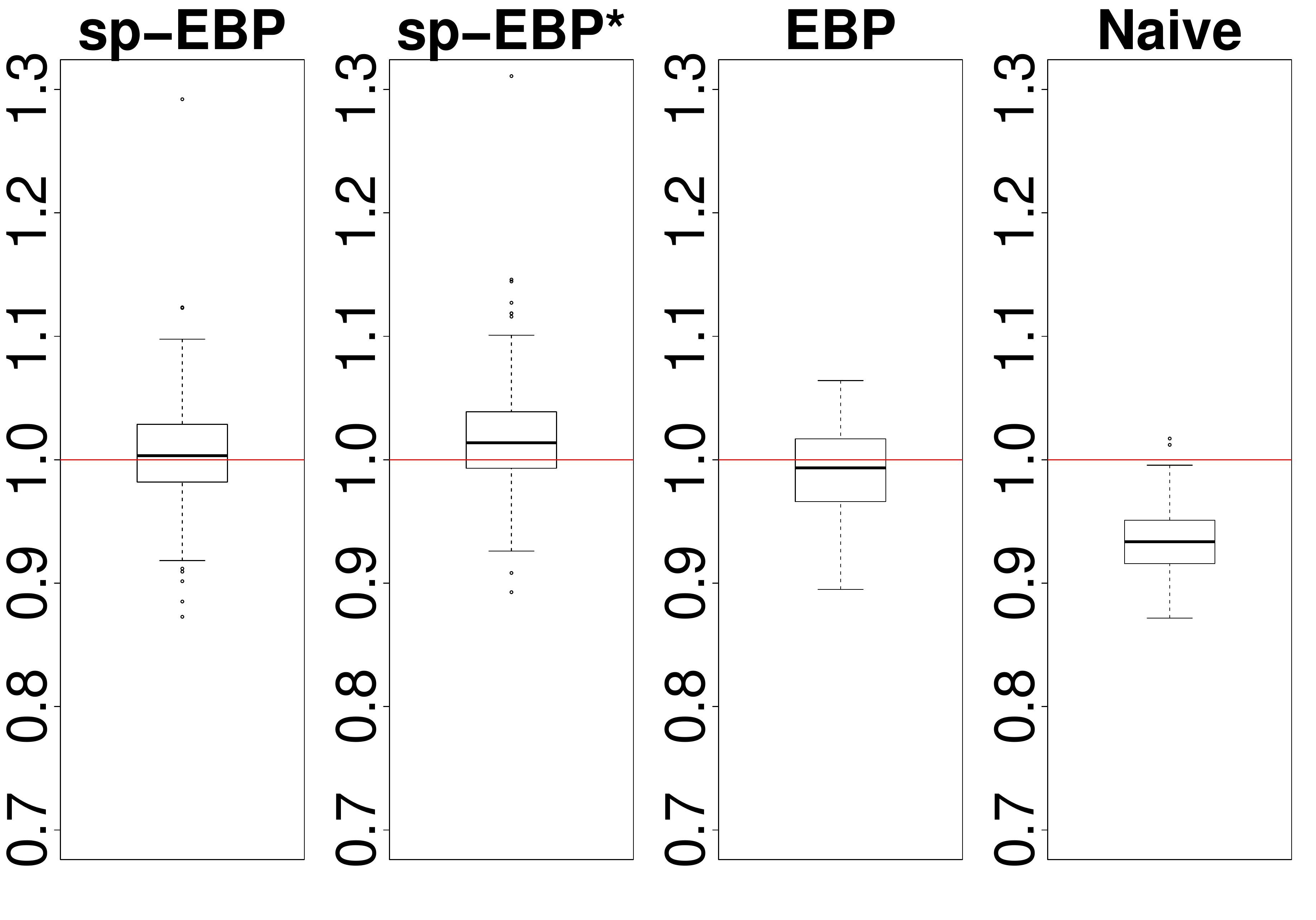}}
\frame{\includegraphics[scale=0.13]{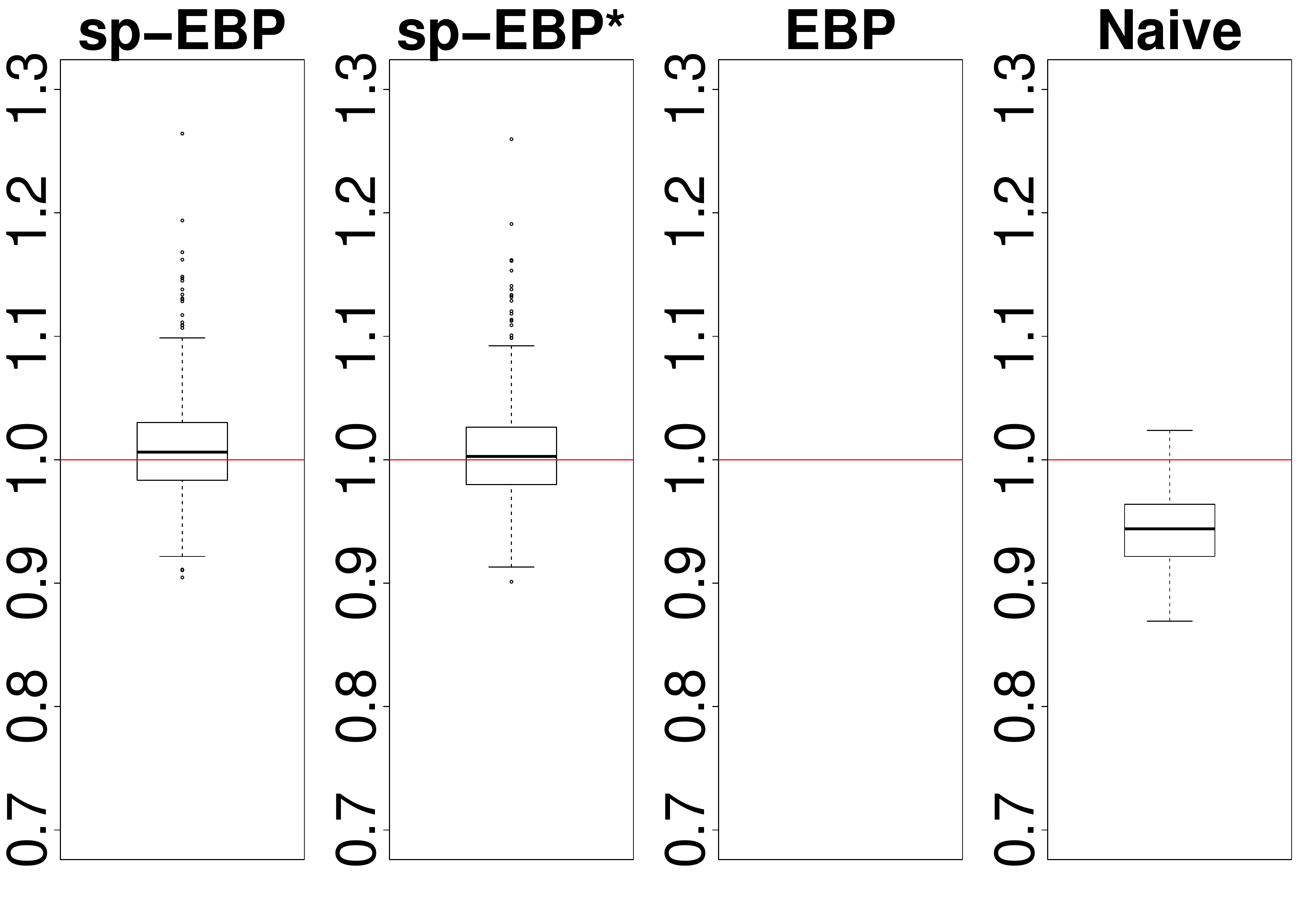}}
\end{figure}

By looking at the bottom panel in Figure \ref{fig_Ri}, we observe that, when dealing with non-Gaussian random effects, the MSE estimator of the sp-EBP has again good performances with an average ratio close to $1$ for all values of $m$. 
The effect of the bias correction term is less evident than before. When a reduced number of small areas is available, $\widehat {\mbox{RMSE}}(\hat p_i^{\ml{sp-EBP}})$ allows to estimate the actual RMSE with a higher precision than the corresponding bias-corrected version $\widehat {\mbox{RMSE}^*}(\hat p_i^{\ml{sp-EBP}})$. However, when $m = 500$, the two estimators seem to perform similarly. 
{The above results are not that surprising from our perspective. The bias correction term strongly relies on asymptotic results from ML theory. As a consequence, the quality of the approximation and, in turn, of the results, improves only when dealing with large sample sizes that render asymptotics more reliable.}
Considering that, in real applications, we expect the random effect distribution to lie in between the two ``extreme'' settings we considered in this simulation study and, also, that we generally need to deal with a large number of small areas, using $\widehat {\mbox{RMSE}^*}(\hat p_i^{\ml{sp-EBP}})$ seems to be generally more appropriate.
From Figure \ref{fig_Ri}, we may also notice that, under Scenario 2, the MSE estimator of the EBP works quite well (apart from being computationally prohibitive from large $m$), whereas that for the Naive estimator consistently underestimates the actual RMSE. 

Furthermore, Table \ref{tab_sim} shows the mean  coverage rate (CR) for nominal $95\%$ Wald-type confidence intervals over simulations, that is 
\[
\mbox{CR}_i=T^{-1}\sum_{t=1}^{T} {\mathds{1}}\left(|\hat{p}_{it}-p_{it}| \leqslant 1.96\times {\widehat{\mbox{RMSE}}(p_{it}^{\ml{Model}})}\right),\quad i = 1, \dots, m.
\]
As it is clear from the table, the proposed estimators show a good performance, with an average empirical coverage of approximately $92-94\%$ in all cases, except for $m = 100$ under Scenario 1. 
On the other hand, both the EBP and, particularly, the Naive approach show a more erratic behavior. The former approach leads to under coverage for Scenario 1 and to over coverage for Scenario 2. This behavior is reversed for the Naive estimator.

\begin{table}[h]
\caption{Average coverage rate over areas and computational time (in minutes) of $\widehat{\mbox{RMSE}}(\hat p_i^{\ml{Naive}})$, $\widehat {\mbox{RMSE}}(\hat p_{i}^{\ml{EBP}})$, $\widehat {\mbox{RMSE}}(\hat p_i^{\ml{sp-EBP}})$, and $\widehat {\mbox{RMSE}^*}(\hat p_i^{\ml{sp-EBP}})$,  for $m = 100, 200, 500$. }
\centering
\scalebox{0.8}{
\begin{tabular}{lrrrrrrrrrrrrr}
\toprule
&\multicolumn{3}{c}{Coverage} & & &  \multicolumn{3}{c}{Computational Time}\\
\midrule
m&\multicolumn{1}{c}{$100$} & \multicolumn{1}{c}{$200$} &\multicolumn{1}{c}{$500$} &&& \multicolumn{1}{c}{$100$} & \multicolumn{1}{c}{$200$} & \multicolumn{1}{c}{$500$}\\

\midrule
\texttt{Scenario} 1\\

    sp-EBP & 0.888 & 0.920 & 0.940 &   &    & 0.059 & 0.122 & 0.332 \\
    sp-EBP* & 0.890 & 0.923 & 0.944 &   &    & 0.528 & 1.042 & 2.986 \\
    EBP   & 0.864 & 0.912 &       &   &    & 16.907 & 41.385 & 161.609* \\
    Naive & 0.962 & 0.967 & 0.976 &   &    & 0.005 & 0.018 & 0.206 \\
\midrule
\texttt{Scenario} 2\\

    sp-EBP & 0.928 & 0.931 & 0.927 &       &       & 0.053 & 0.110  & 0.303 \\
    sp-EBP* & 0.933 & 0.933 & 0.928 &       &       & 0.481 & 0.959 & 2.437 \\
    EBP   & 0.966 & 0.974 &       &       &       & 17.229 & 42.112 & 162.528* \\
    Naive & 0.903 & 0.906 & 0.910  &       &       & 0.004 & 0.018 & 0.219 \\
\bottomrule
\multicolumn{8}{l}{$^*$ Results refer to a single Monte Carlo draw}
\end{tabular}	
}
\label{tab_sim}
\end{table}

To conclude, we also compared MSE estimators in terms of computational complexity. The last column of Tables \ref{tab_sim} reports the computational time (averaged over simulations) required to get the estimates on an Intel(R) I5-3330 architecture (3.0 GHz) under each  simulation setting.  
As it can be seen, the proposed MSE estimators show good performance also in this respect. When compared to the Naive approach, they clearly require a higher effort, which is, however, always under control. When compared to the EBP approach, the computational burden is considerably reduced. 
It is important to notice that, due to computational issues, results reported for the EBP approach when $m = 500$ refer to a single Monte Carlo draw in place of being the average of $T = 1,000$ draws as for the other methods. In this respect, it is clear that this approach does not represent an option for empirical applications with large $m$, as the one we discuss here. 

When comparing the two MSE estimators we propose (with and without bias correction), we may observe that deriving $\widehat {\mbox{RMSE}^*}(\hat p_i^{\ml{sp-EBP}})$ requires a higher computational effort than that required for $\widehat {\mbox{RMSE}}(\hat p_i^{\ml{sp-EBP}})$: this is clearly due to the computation of model derivatives in equation \eqref{eq:biasNP} which does not represent an easy task. However, such an effort is rewarded by the quality improvements we discussed so far, at least for large $m$.

\section{Estimating unemployment incidence for LLMAs in Italy}\label{sec:application}

In this section, we use ILFS data to estimate unemployment incidence for $611$ LLMAs in Italy. According to the simulation results in Section \ref{sec:sim}, the sp-EBP is a potentially useful approach as $(i)$ it performs better than the Naive predictor in terms of bias and efficiency;  $(ii)$ it dramatically decreases the computational complexity of the MSE estimator for the parametric EBP which becomes unfeasible for a large number of small areas and/or large sample sizes. {The use of the proposed approach is made easy by the availability of a (computationally efficient) algorithm for estimation and inference developed in \texttt{R} language from the authours. This is part of the on-line Supplementary Material at the publisher's web-site, together with an example data set similar to the real one.}

\subsection{The model}
To predict unemployment incidence in Italy, we considered a response variable $Y_{ij}$ taking value $1$ if unit $j$ in small area $i$ is unemployed and $0$ otherwise. We followed an approach similar to that used by \citet{Molina2014} and considered the variables introduced in Section \ref{sec:data} and their transformations in the linear predictor, that is \textit{Sex-Age} (reference = 15-24), \textit{Educational Level} (reference = no education or primary school diploma), and the logarithmic transform of \textit{U-count}. 
We ran {the EM algorithm described in the Supplementary Material (Section 1) for different model specifications and a varying number of components ($G = 2,\dots,6$) for the random effect distribution.} The optimal solution, corresponding to the smallest AIC value, is based on $G = 3$ components and includes in the linear predictor a random intercept and main covariate effects only.
We report in Table \ref{table1:app} model parameter estimates, standard errors, and resulting p-values, together with the corresponding log-likelihood and AIC index. {For comparison, we also report such quantities for the corresponding parametric model based on Gaussian random effects. Looking at this table, we may first observe that the AIC index suggests a better fit of the model based on an unspecified random effect distribution with respect to  its parametric counterpart, even though differences in terms of parameter estimates are rather negligible. }
In particular, looking at the estimates for \textit{Sex-Age}, we may notice that, when controlling for the effect of other explanatory variables in the model and for the effect of unobserved heterogeneity, the odds of being unemployed for younger people is higher than that for the older ones. For instance, the odds of being unemployed for a male in the 25-34 group are $e^{0.118} = 1.125$ times those of males aged 15-24. On the other hand, for a male who is aged 35-65 years, the odds are  $e^{-0.787}=0.455$, that is $54.5\%$ lower than those for the baseline category. Such differences are even stronger for females.
Turning to the \textit{Educational Level}, the odds of being unemployed for subjects with middle or high school diploma is higher than that of low educated subjects (parameter estimates for middle and high school diplomas are all positive). On the other hand, having a University degree or higher education has not a significant effect. These findings are in line with the results reported in the preliminary analysis: low educated females and relatively younger individuals (the reference category) are more frequent in the inactive category.  Last, as expected, results reported in Table \ref{table1:app} suggest that the probability of being unemployed increases as the total number of unemployed registered at the $2011$ census increases.

\begin{table}
\caption{Parameter estimates, standard errors and corresponding p-values for the mixed logistic model fitted to the ILFS data based on an unspecific (left) and a Gaussian (right) random effect distribution. \label{table1:app}}
\centering
\scalebox{0.8}{

    \begin{tabular}{lrrrrrrrr}
    \toprule
        & \multicolumn{3}{c}{Unspecific }& \multicolumn{3}{c}{Gaussian}\\
    & \multicolumn{1}{c}{Estimate} & \multicolumn{1}{c}{SE} & \multicolumn{1}{c}{p-value}     && \multicolumn{1}{c}{Estimate} & \multicolumn{1}{c}{SE} & \multicolumn{1}{c}{p-value} \\ 
    \midrule
   Intercept  & -3.052 & 0.176 &  $<$0.001  &       & -3.002 & 0.150 &  $<$0.001  \\
      M:25-34  & 0.118 & 0.057 & \multicolumn{1}{r}{0.038} &       & 0.122 & 0.054 & \multicolumn{1}{r}{0.024} \\
      M:35-65  & -0.787 & 0.054 &  $<$0.001  &       & -0.778 & 0.048 &  $<$0.001  \\
      F:15-24  & -0.222 & 0.062 &  $<$0.001  &       & -0.221 & 0.058 &  $<$0.001  \\
      F:25-34  & 0.113 & 0.05  & \multicolumn{1}{r}{0.023} &       & 0.118 & 0.054 & \multicolumn{1}{r}{0.029} \\
      F:35-65  & -1.039 & 0.052 &  $<$0.001  &       & -1.027 & 0.050 &  $<$0.001  \\
      Middle School  & 0.206 & 0.059 &  $<$0.001  &       & 0.211 & 0.054 &  $<$0.001  \\
      High School  & 0.235 & 0.062 &  $<$0.001  &       & 0.239 & 0.053 &  $<$0.001  \\
      University Degree or Beyond  & -0.032 & 0.083 & \multicolumn{1}{r}{0.682} &       & -0.029 & 0.064 & \multicolumn{1}{r}{0.653} \\
$\log($\textit{U-count})   & 0.113 & 0.012 &  $<$0.001  &       & 0.104 & 0.022 &  $<$0.001  \\
\midrule
$\ell$ & \multicolumn{3}{r}{-21833.599}&& \multicolumn{3}{r}{-21837.357}\\
AIC & \multicolumn{3}{r}{43695.199}&& \multicolumn{3}{r}{43696.714}\\

      \bottomrule
    \end{tabular}%
%
%\begin{tabular}{lrrr}
%\toprule
% & \multicolumn{1}{c}{Estimate} & \multicolumn{1}{c}{SE} & \multicolumn{1}{c}{p-value} \\ 
%\midrule
%
%Intercept & -3.052 & 0.176 & $<$0.001 \\ 
%  M:25-34 & 0.118 & 0.057 & 0.038 \\ 
%  M:35-65 & -0.787 & 0.054 & $<$0.001 \\ 
%  F:15-24 & -0.222 & 0.062 & $<$0.001 \\ 
%  F:25-34 & 0.113 & 0.050 & 0.023 \\ 
%  F:35-65 & -1.039 & 0.052 & $<$0.001 \\ 
%  Middle School & 0.206 & 0.059 & $<$0.001 \\ 
%  High School & 0.235 & 0.062 & $<$0.001 \\ 
%  University Degree or Beyond & -0.032 & 0.083 & 0.682 \\ 
%  $\log($\textit{U-count})  & 0.113 & 0.012 & $<$0.001 \\ 
%
%\bottomrule
%\end{tabular}
}
\end{table}

{Figure \ref{fig_REdistrib} shows the estimated prior (Figure 1a) and posterior distribution (Figure 1b) estimates for the random effects obtained using the proposed (semi-parametric) approach, together with the estimated posterior distribution deriving from the parametric approach (Figure 1c).} In particular, in Figure 1b, we report the posterior mean of the area-specific random intercept calculated as
\[
\hat \alpha_i = \sum_{g =1}^G (\hat \xi_g - \hat{ \bar \xi}) \hat \tau_{ig}, 
\]
where $\hat{\bar \xi} = \sum_g \hat \xi_g \hat \pi_g$ is the overall intercept estimatereported in Table \ref{table1:app}. 
{By focusing the attention on Figure 1a, we may clearly observe that observed data lead to the estimation of a random effect with a clear degree of skewness. If the standard Gaussian assumption had been reasonable, the NPML estimate of the random effect distribution 
would have been a symmetric distribution centered around zero. As a consequence, we may conclude that such an assumption may not be that adequate for the current application. Furthermore, by comparing Figures 1b and 1c, we may observe that the parametric assumption also affects the posterior mean of the area-specific intercepts, leading to a less skewed distribution than that obtained under the proposed approach. 
}

\begin{figure}[ht]
\caption{Semi-parametric approach: estimated prior (a) and posterior (b) distribution for $\alpha_i$'s;  parametric approach: posterior distribution for $\alpha_i$'s  (c).} \label{fig_REdistrib}
\vspace{2mm}
\centering
\includegraphics[scale=0.6]{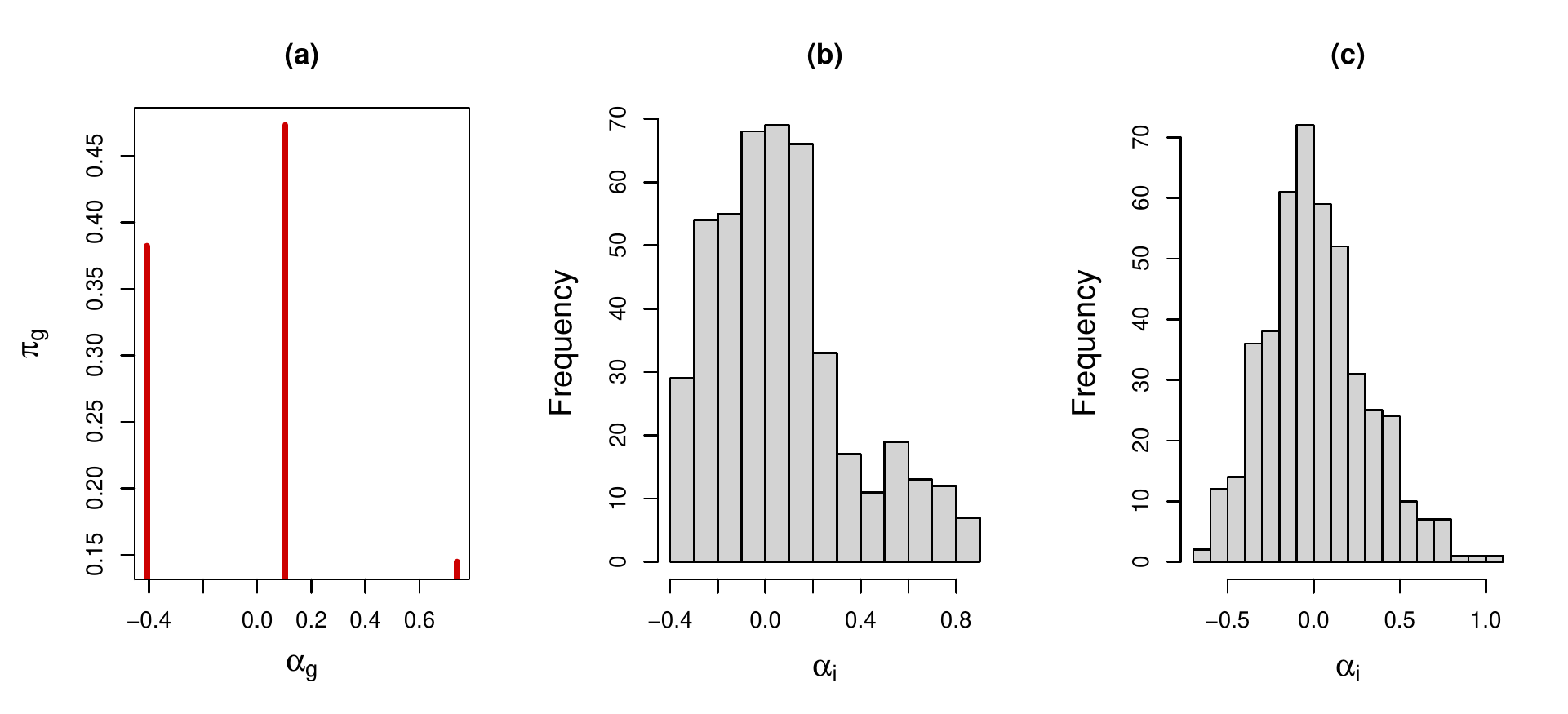} 
\end{figure}

\subsection{Small area predictions}
As highlighted in Section \ref{sec:notation}, we need the covariate values, $\b x_{ij}$, to be known for all units in the population to predict the target variable. This would require access e.g. to census micro-data. However, in the important and special case where the components of $\b x_{ij}$ are all categorical, or take a finite number of values, the method described in this paper only requires the corresponding area level cross-tabulations to be available. This is the case of the ILFS data, where information on the covariates in the model are available at an aggregate level for the whole population. Figure \ref{fig1:app} shows the map of unemployment incidence prediction for the $611$ LLMAs obtained using direct estimation, the proposed sp-EBP approach, the  parametric EBP, and the Naive approach. {Direct estimates are computed using H\'ajek-type estimators with adjusted weights that account for nonresponse and calibrate to population level information of demographic variables. } The patterns of unemployment produced by the proposed approach are consistent with those obtained by all the other methods. As expected, model-based maps are smoother when compared to direct estimates; relatively larger values for unemployment incidences are mainly located in the South of Italy and in the Islands. 

\begin{figure}
\caption{Maps of the estimated unemployment incidences for LLMAs in Italy in 2012: direct estimates, sp-EBP, EBP, and Naive estimates. \label{fig1:app}}
\vspace{2mm}
\centering
{\includegraphics[scale=0.3]{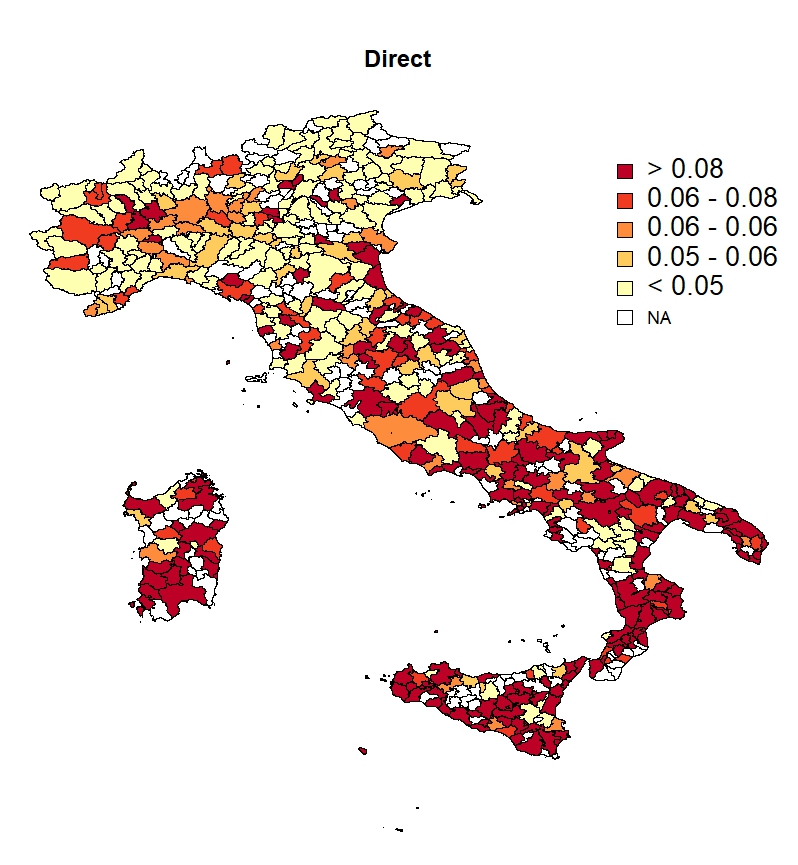}}
{\includegraphics[scale=0.4]{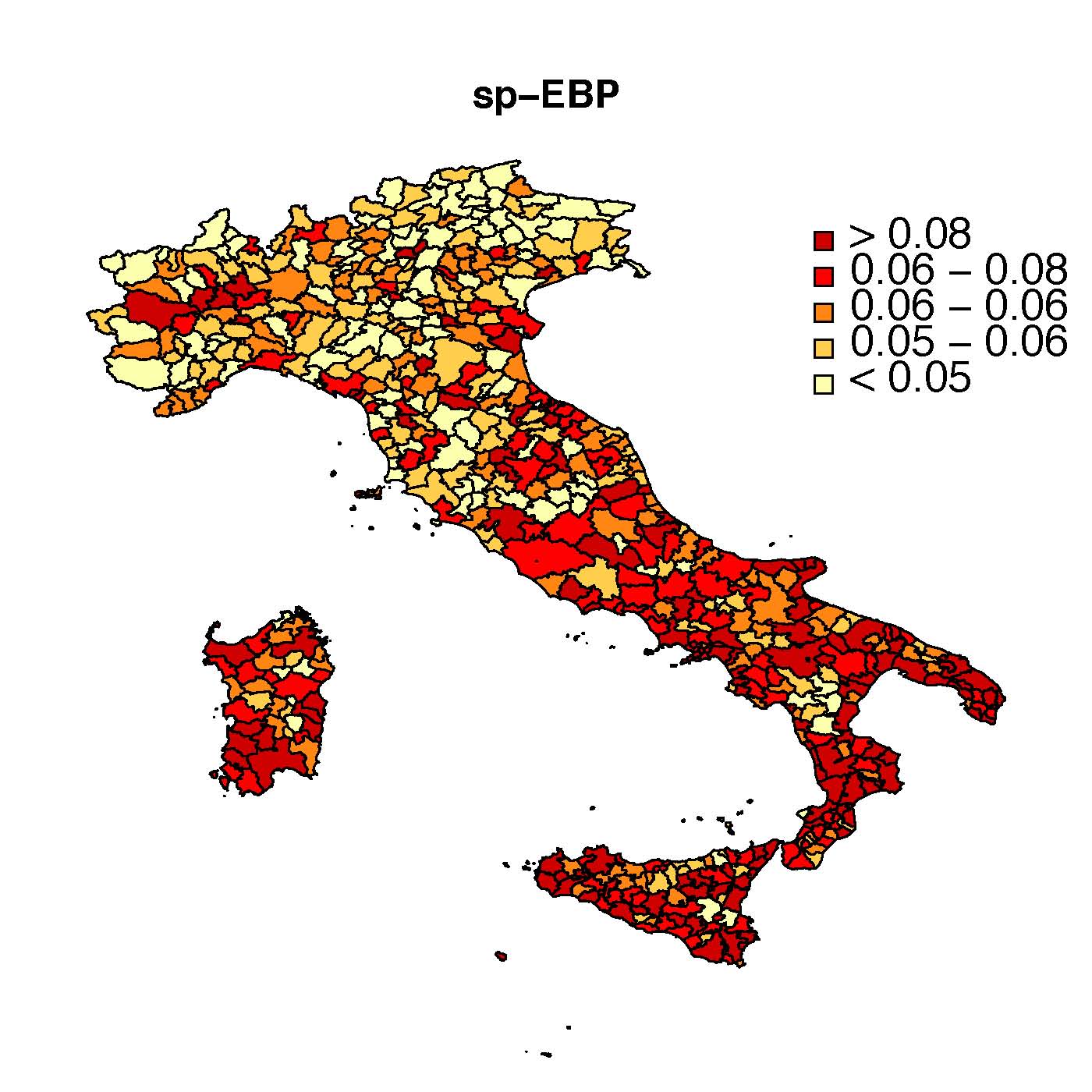}\\}{\includegraphics[scale=0.3]{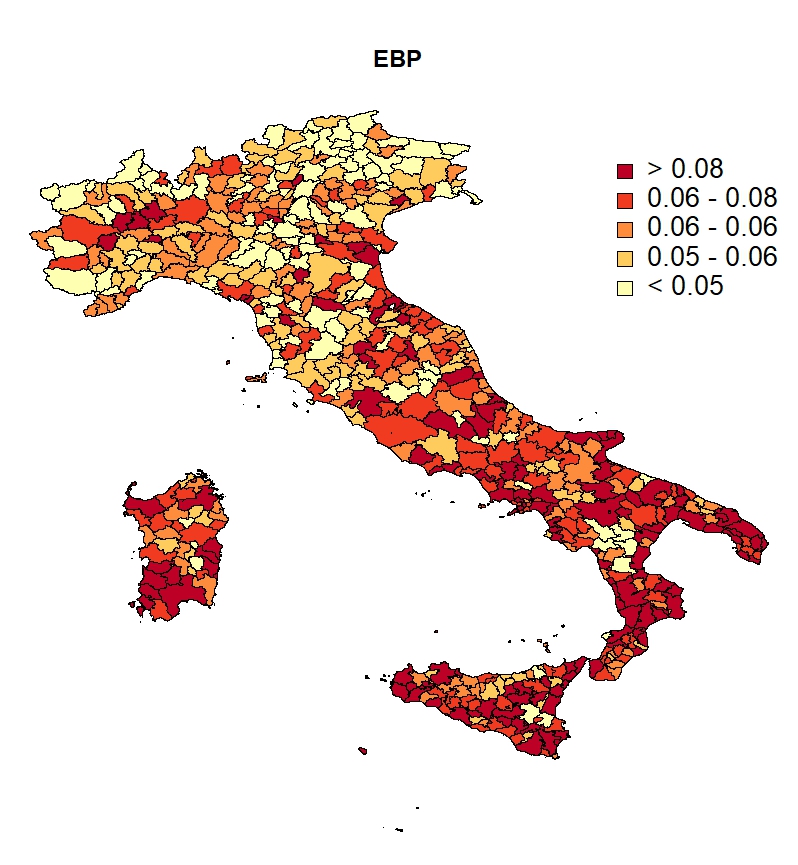}}
{\includegraphics[scale=0.3]{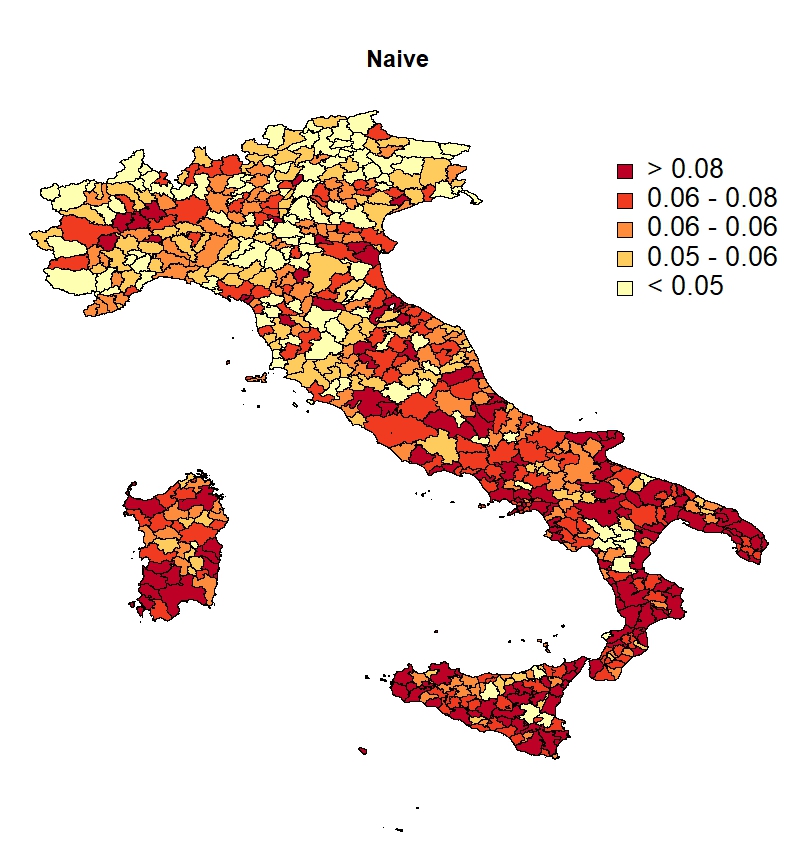}}
\end{figure}

To assess the quality of predictions, we used a set of diagnostic tools based on the requirement that model-based small area estimates should be coherent  with, in the sense of being close to, the corresponding unbiased direct estimates, albeit more precise.  Figure \ref{fig2:app} shows the estimates derived from the sp-EBP approach versus  the direct, the EBP, and the Naive estimates, respectively. From this figure (first panel), we may observe that our approach leads to predictions which are close to those provided by a direct approach, with a correlation coefficient equal to $0.881$. From the remaining panels in Figure \ref{fig2:app}, it is evident that model-based estimates for unemployment incidence are all very close to each other, with correlation coefficients equal to $0.978$ (sp-EBP vs. EBP) and to  $0.977$ (sp-EBP vs. Naive).

\begin{figure}
\caption{sp-EBP estimates of small area proportions versus the corresponding direct   (left), EBP (centre), and Naive estimates (right). {Dots' size is proportional to the sample size}. \label{fig2:app}}
\vspace{2mm}
\centering
\includegraphics[scale=0.25]{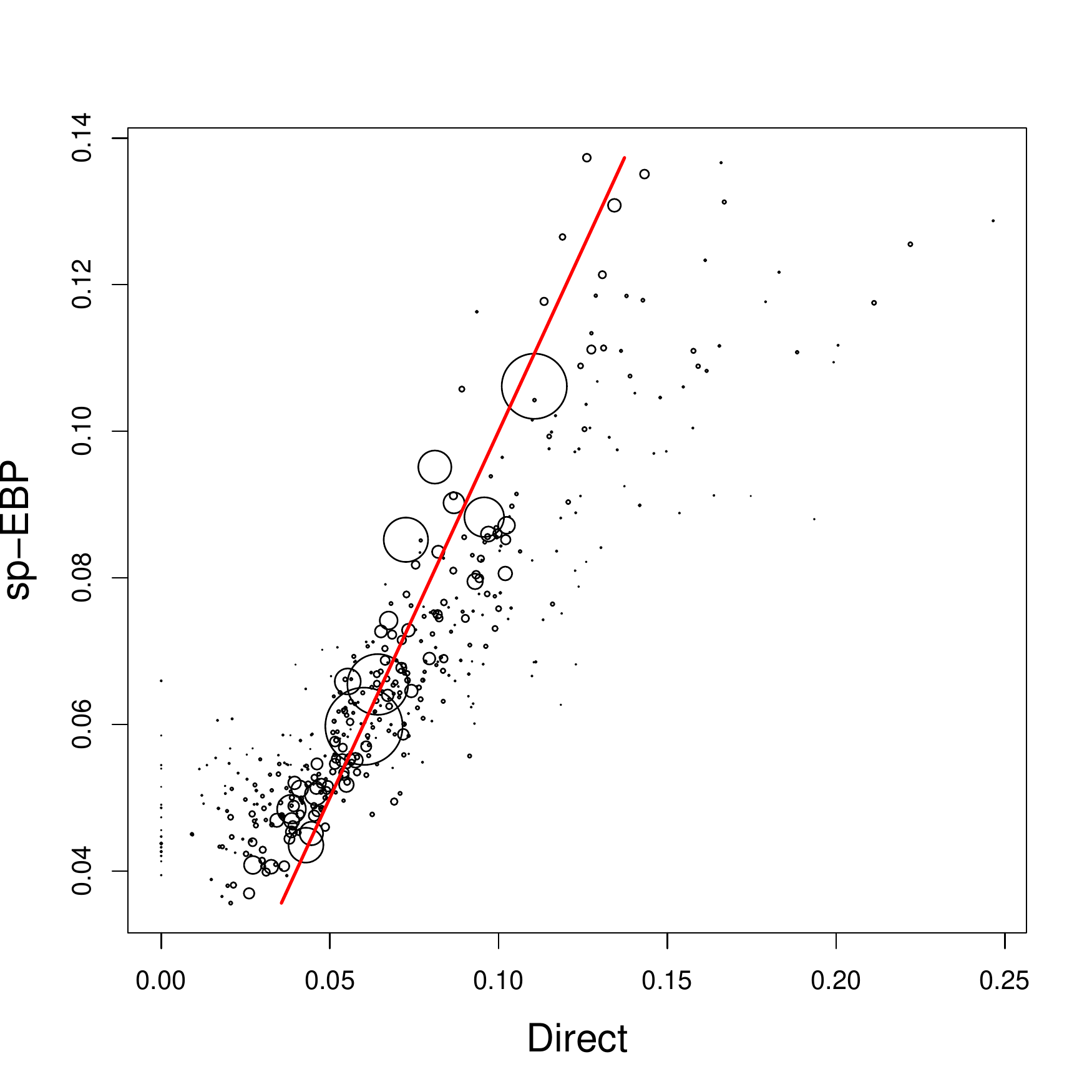}
\includegraphics[scale=0.25]{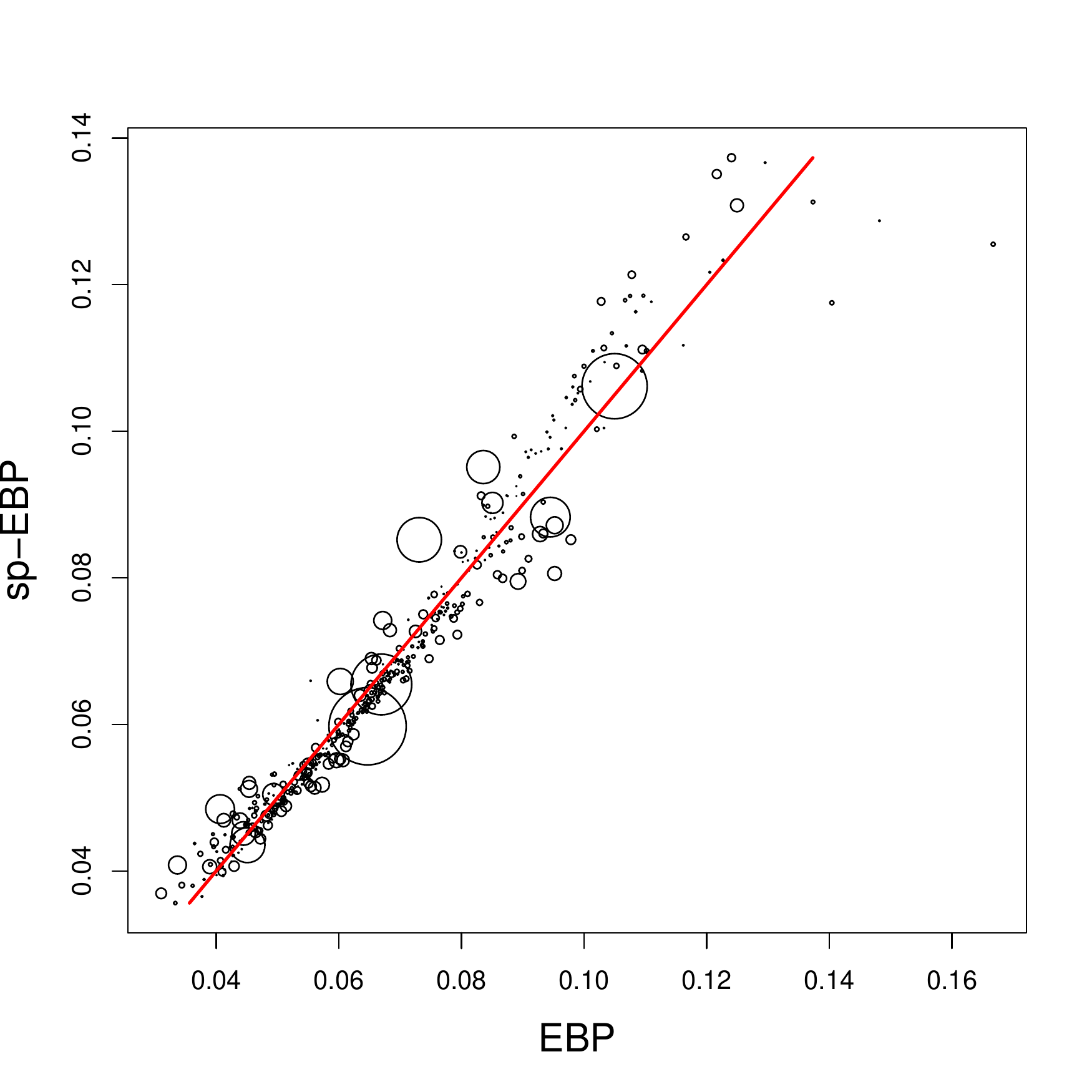}
\includegraphics[scale=0.25]{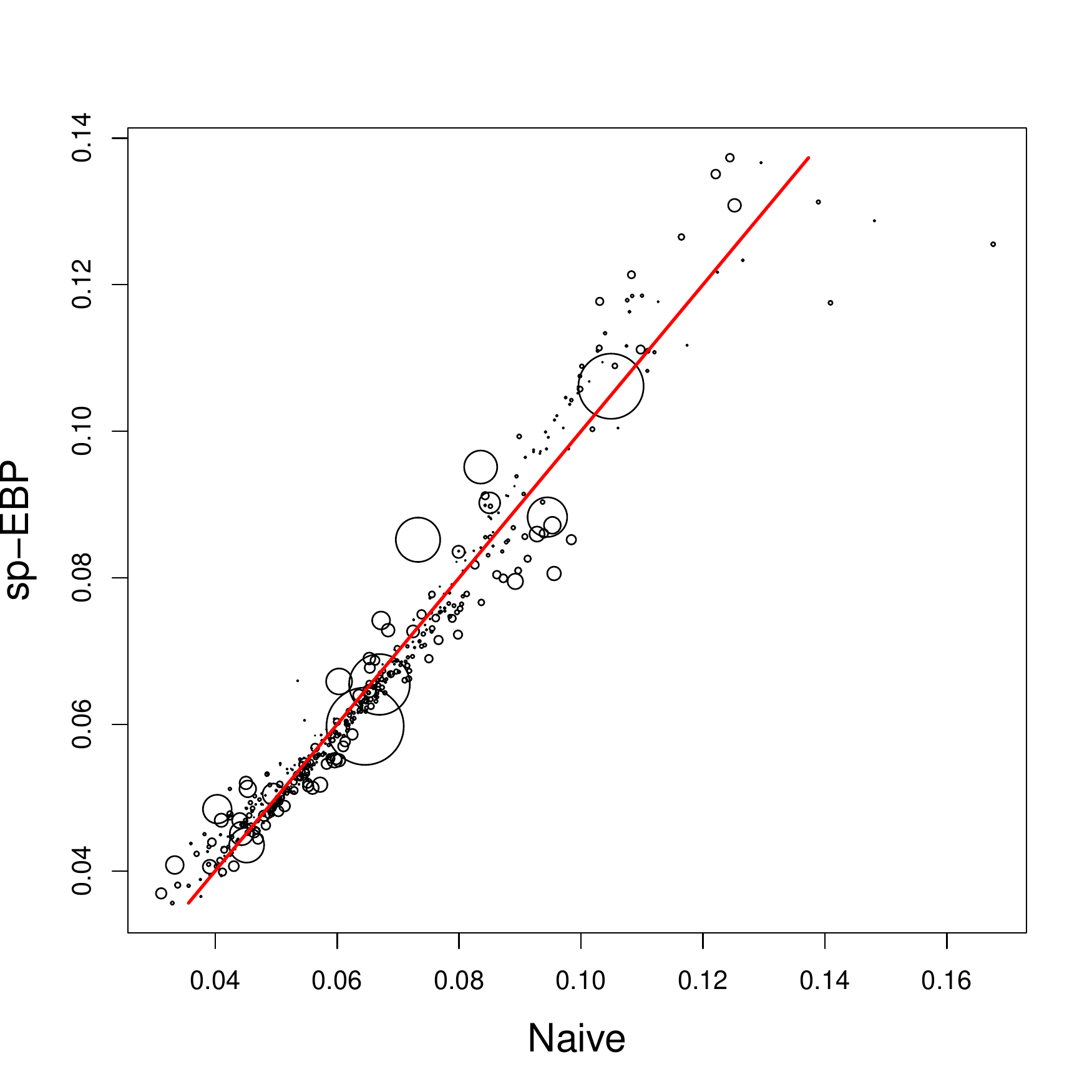}
\end{figure}

Coherence of direct and sp-EBP estimates can be also evaluated by computing a goodness-of-fit diagnostic \citep{BCHH:2001}, which is obtained from the following Wald-test statistic:
\begin{equation}
\label{eq:Wald}
W=\sum_{i=1}^{m}\frac{(\hat{p}_i^{\ml{Direct}}-\hat p_i^{\ml{sp-EBP}})^2}{\widehat{\m{Var}}(\hat{p}_i^{\ml{Direct}})+\widehat{\m{MSE}}^*(\hat p_i^{\ml{sp-EBP}})},
\end{equation}
where the estimated MSE of the sp-EBP is calculated by using formulas in Section \ref{sec:binary}. Considering the results of the simulation experiments with $m=500$, we decided to consider the bias-corrected MSE reported in equation \eqref{eq:bias_mseNP}. The above test is based on the idea that, should model-based estimates be close to the ``true'' small area parameters of interest, the unbiased direct estimates could be considered as random variables with expected value equal to the value of the corresponding model-based estimates. Here, $W=360.56$ and such a value needs to be compared to the $95$-th percentile of a $\chi^2$ distribution with $452$ d.f., $\chi^2_{452,0.95}=502.56$. In this respect, we may conclude that model-based estimates are not significantly different from direct estimates. 

To assess the potential gain in precision we obtain by using the proposed sp-EBP approach in place of the direct one, we compare in Figure \ref{fig_CV} the  empirical cumulative density functions (ecdfs) of the vcoefficients of variation (CV) of both estimators. The first panel uses CV's from all areas, while the second (third) one focuses on small areas with sample size smaller (higher) than $100$. As it is clear, by looking at the first panel, the ecdf corresponding to sp-EBP almost always dominates the one for the direct estimates, highlighting that CV values for the former approach are lower than those estimated with the latter. Only for very small CV values, the ecdfs show an inverse relation: CV's for direct predictions are smaller than those for sp-EBP. This is more evident in the second panel and is related to the presence of some areas with a small sample size for which $\hat p_i^{\ml{Direct}}$ is zero or is very close to zero, and so is $\widehat{\m{RMSE}}(\hat p_i^{\ml{Direct}})$. However, also in this case, about the $60\%$ of the Italian LLMAs, CVs associated to the direct estimator are above the standard $33\%$ threshold which is typically considered for reliability in the SAE context. Such a percentage reduces to about $20\%$ when considering the proposed sp-EBP approach; in addition, less than 5\% of the estimates have CV\% larger than 40\%.  When we move to higher CVs, the sp-EBP approach always provides smaller CV values when compared to the direct approach and such CVs are always smaller than $40\%$. When focusing on the third panel in Figure \ref{fig_CV}, it is evident that, as the sample size gets larger, direct and model-based estimates tend to have quite similar CV values, although those associated with model-based estimates are still consistently smaller. 

\begin{figure}
\centering
\caption{CV's empirical cumulative density functions for the sp-EBP and the direct estimator. }
\label{fig_CV}
\includegraphics[scale=0.40]{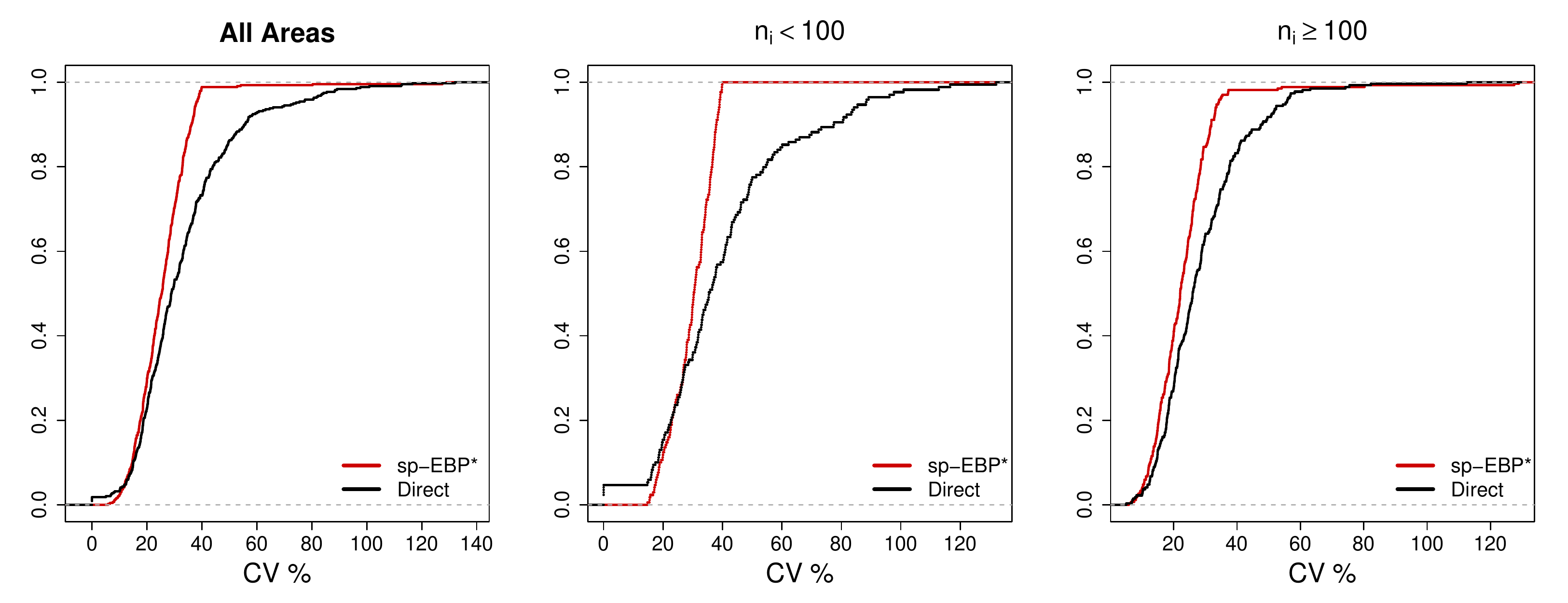} 
\end{figure}

\section{Conclusions} \label{sec:concl}

The paper described some tools to derive Best Predictions for responses with distribution in the Exponential Family in the presence of clustered data. In particular, we proposed a semi-parametric version of the EBP and the corresponding second-order, bias-corrected, MSE approximation using a NPML approach and leaving the distribution of the random effects unspecified. 
Motivated by a real application to data on unemployment incidence in LLMAs in Italy, we focused on a binary response modeled via a mixed logistic model with random intercepts, which represents a relevant case in the SAE framework.

Simulation experiments showed that the proposed estimator performs equally or better than the competitors. In particular, when moving far from the assumption of Gaussian distributed random effects, the proposed semi-parametric approach performed better than the corresponding parametric versions. Also, when compared to the parametric EBP, simulation results highlighted better performance of the proposed approach in terms of computational load required to get predictions and the corresponding MSE. 
The simulation study, where different sample sizes were considered, showed that the semi-parametric approach is always reliable, especially for large $m$.  Such a gain comes from the discrete nature of the mixing distribution estimate which substantially simplifies the calculations to get the EBPs and the corresponding MSEs. 

We illustrated the benefits of our proposal discussing the estimation of unemployment incidence for Italian LLMAs in Italy in $2012$.
In this context, direct estimates cannot be published for most of the LLMAs  given the unacceptable large value of the coefficient of variation for those areas with a small sample size. In this respect, model-based approaches represent a necessary strategy. Since the sample size and the number of small areas are particularly large in this application, the implementation of the EBP turns to be particularly cumbersome, and the evaluation of its precision prohibitive. This application indicated that the proposed methodology leads to estimates which are coherent with, but more efficient than, the direct estimates, still being comparable with alternative model-based estimates.

Although the approach we propose is presented for responses with density in the Exponential Family, we did not explore the behavior of the small area sp-EBPs for counts or multinomial responses. However, a possible extension to multi-category outcomes is quite straightforward. Also, we notice that suitable extensions of the proposed approach to allow for spatial correlation could be envisioned by properly modeling, for each small area, the prior mixture probabilities as a function of neighborhood components membership. {Last, developing design-consistent small area estimators under the proposed methodology represent a topic of interest, especially for those researchers working in survey sampling from a design-based or a model-assisted perspective. More specifically, we could adopt a model-assisted approach, thereby the model is used only to motivate the predictors, but their properties are evaluated only with respect to the randomization distribution induced by the sampling design.}

\section*{Acknowledgements}
The work by Marino, Ranalli and Salvati has been developed under the support of the project PRIN-SURWEY (grant 2012F42NS8, Italy). The work of Alf\'{o} has been supported by the project ``Mixture and latent variable models for causal inference and analysis of socio-economic data'' (grant RBFR12SHVV, FIRB - Futuro in Ricerca). The work of Salvati was carried out with the support of the project InGRID 2 (Grant Agreement No 730998, EU).

\end{document}

\maketitle

\section{EM algorithm for parameter estimation }\label{sec:appA:EM_algorithm}
In this section, we provide computational details of the the EM algorithm required to get model parameter estimates. 
Let us start from the observed data likelihood in equation (6) of the manuscript:
\begin{equation*} 
L (\bs \Phi ) =  \prod_{i = 1}^m \sum_{g = 1}^G f_{y\mid \alpha} (\b y_{i} \mid  \b \xi_g; \b X_i) \pi_g,
\end{equation*} 
where $f_{y\mid \alpha} (\b y_{i} \mid  \b \xi_g; \b X_i) = \prod_{j \in s_i} f_{y\mid \alpha} ( y_{ij} \mid  \b \alpha_i = \b \xi_g; \b x_{ij})$. In the following, we will denote this quantity as $f_{ig}$ to simplify notation. 

As we highlighted before, even if a direct maximization of the above expression is always affordable, an indirect approach, based on the use of an EM algorithm, is frequently adopted. 
For this purpose, let us define the binary indicator variable $z_{ig}$ which is equal to one if area $i$ belongs to the $g$-th component of the finite mixture. The EM algorithm starts from the definition of the complete-data log-likelihood:
\begin{align}
\ell_c(\bs \Phi) &  
= \sum_{i = 1}^m \sum_{g = 1}^{G} z_{ig}
\left[
\log f_{ig}+ \log (\pi_{g})
\right],
\label{eq:compLk}
\end{align} 
In the E-step of the algorithm, we compute the expected value of the complete data log-likelihood in equation  \eqref{eq:compLk}, conditional on the observed data and the current parameter estimates, say $\hat{\b \Phi}^{(t)}$. Due to linearity in the indicator variables $z_{ig}$, at the generic $(t+1)$-th iteration of the algorithm, we need to derive the following quantities:
\begin{equation}
\tau _{ig}^{(t+1)} = E_{\hal \Phi^{(t)}}\left[z_{ig} \mid \b y_i,\ha\Phi^{(t)} \right] = \Pr\left(z_{ig} =1 \mid \b y_i,\ha\Phi^{(t)}\right) =
\frac{\pi_g^{(t)} f_{ig}^{(t)}}
{\sum_{l =1}^G \pi_l^{(t)} f_{il}^{(t)}},
\label{eq:tau_ig}
\end{equation}
where $f_{ig}^{(t)}$ denotes the joint conditional distribution of the $i$-th small area in the $g$-th component of the finite mixture, under the current estimate of model parameters $\b \Phi^{(t)}$. 
The (conditional) expected value of the complete data log-likelihood is given by
\begin{equation}\label{expected_compLk}
Q\left(\bs \Phi \mid \ha \Phi^{(t)}\right) =  \sum_{i = 1}^m \sum_{g = 1}^{G}  \tau _{ig}^{(t+1)}
\left[
\log (f_{ig}) + \log (\pi_{g})
\right].
\end{equation}

In the M-step, parameter estimates are obtained by maximizing \eqref{expected_compLk} with respect to model parameters $\b \Phi$. 
A closed form expression is available for the mixture component probabilities: 
\begin{equation*}
\pi_g^{(t+1)} = \sum_{i = 1}^m \frac{ \tau_{ig}^{(t+1)}}{m}.
\end{equation*}
For the parameters in the regression model, updates depend on the nature of the conditional distribution for $Y_{ij}$. Generally, the problem reduces to a weighted ML estimation problem for generalized linear models, which can be solved by using standard Newton-type recursions.

The E- and the M-step of the algorithm are iterated until convergence, which can be specified in terms of log-likelihood or parameter values, using relative or absolute norms.
{We report in the following an example of the algorithm structure in a programmable form. }
\begin{algorithm}
{
\caption{Pseudo-code of the EM algorithm for parameter estimation }
\Begin{
\textbf{Initialize} $\b \Phi$ and $\tau_{ig}, i = 1, \dots, n, g = 1, G$
\Repeat{$\ell(\b \Phi^{(t)}) - \ell( \b \Phi^{(t-1)}) > \varepsilon$}{
update $\tau_{ig} \quad \quad\quad \quad $ {{Expectation step}}\,\\ 
update $\b \Phi  \quad \quad \quad \quad$ {{Maximization step}}\, 
}
}
}
\end{algorithm}
{
A crucial point in this framework is represented by the initialization of model parameters. To avoid local maxima and/or spurious solutions, a multi-start strategy is frequently adopted by based both on a deterministic and a random starting rule. For the former, fixed model parameters can be initialized at the corresponding estimates from a homogeneous generalized linear model, while random parameters can be set equal to the $G$ locations of a standard Gaussian quadrature approximation. On the other hand, random starting solutions can be obtained by randomly perturbing the deterministic ones. Overall, for a given $G$, the solution that at convergence of the algorithm corresponds to the highest log-likelihood value is retained as the optimal one.
}

\section{Computing the standard errors for parameter estimates}\label{sec:appB_SE}
A disadvantage of the EM algorithm is that it does not directly provide estimates for the covariance matrix of model parameter estimates. 
However, we may rely on the approach discussed by \cite{Louis1982} and, for practical purposes, on the formula described by \cite{Oakes1999}. Let 
$$
J (\hat{\bs \Phi}) = - \left[\frac{\partial^2 \ell(\bs \Phi)}{\partial \bs \Phi \partial \bs \Phi^\prime}\right]\Bigg\rvert_{\bs \Phi = \hat{\bs \Phi}}$$
denote the observed information matrix. According to \cite{Oakes1999}, we may write 
\begin{equation}
{J}(\hat{\bs\Phi})=-\left\{\frac{\partial^2 Q({\bs  \Phi} 
 \mid \hat{\bs
 \Phi})}{\partial {\bs \Phi} \partial {\bs  \Phi}^{\prime}}\right|_{{\bs  \Phi}=\hat{\bs \Phi}} + \left. \left. \frac{\partial^2 Q({\bs
 \Phi} \mid \hat{\bs
 \Phi})}{\partial {\bs \Phi} \partial \hat{\bs
 \Phi}^{\prime}}\right|_{{\bs \Phi}=\hat{\bs \Phi}
}\right\}, \label{eq:oakes}
\end{equation}
where the first term denotes the expected  complete-data Hessian matrix conditional on the observed data and the parameter estimates.
To derive such a quantity, we proceed as follows. Let us denote by $h[\cdot]$ the inverse link function. As a first step, we need to compute the score functions associated with each element of $\bs \Phi$, based on the sample data only; these are given by 
\begin{align}
\label{eq:scorefncs}
&S(\b \alpha_g) = \frac{\partial Q(\bs \Phi \mid \hat{\bs \Phi})}{\partial {\b \alpha_g}} =   \sum_{i=1}^m \tau_{ig} S_{ig}( \b \alpha_g) = \sum_{i=1}^m \tau_{ig} \sum_{j\in s_i} \left\{y_{ij} -h[\eta_{ijg}]\right\} \b w_{ij}, \quad g = 1, \dots, G, \nonumber \\[0.2cm]
&S(\bs \beta) = \frac{\partial Q(\bs \Phi \mid \hat{\bs \Phi})}{\partial {\bs \beta}}=   \sum_{i=1}^m \sum_{g = 1}^G  \tau_{ig} S_{ig}(\bs \beta) = \sum_{i=1}^m \sum_{g = 1}^G \tau_{ig} \sum_{j\in s_i} \left\{y_{ij} -h[\eta_{ijg}]\right\} \b x_{ij}, \nonumber\\[0.2cm]
&S(\pi_{g}) = \frac{\partial Q(\bs \Phi \mid \hat{\bs \Phi})}{\partial {\pi_g}} = 
\sum_{i=1}^m \left[\frac{ \tau_{ig}}{\pi_{g}}-\frac{\tau_{iG}}{\pi_{G}}\right],
\quad g = 1, \dots, G-1,
\end{align}
where $\tau_{ig}$ represents the posterior probability for the $i$-th small area to belong to the $g$-th component of the finite mixture.
Given the score functions above, the expectation of the complete data Hessian matrix can be obtained from the following equations: 
\begin{align*}
&H( \b \alpha_g,  \b \alpha_g) = \frac{\partial S(\b \alpha_g)}{\partial \b \alpha_g^\prime} = -\sum_{i=1}^m \tau_{ig} \sum_{j\in s_i} \left\{\frac{\partial h[\eta_{ijg}]}{\partial \eta_{ijg}}  \right\}\b w_{ij} \b w_{ij}^\prime, \quad g = 1, \dots, G, \\[0.1cm]
&H(\bs \beta,  \b \alpha_g) = \frac{\partial S(\b \beta)}{\partial \b \alpha_g^\prime} = -\sum_{i=1}^m \tau_{ig} \sum_{j\in s_i} \left\{ \frac{\partial h[\eta_{ijg}]}{\partial \eta_{ijg}}  \right\}\b x_{ij}\b w_{ij}^\prime,\quad g = 1, \dots, G, \\[0.1cm]
&H(\bs \beta, \bs \beta) =  \frac{\partial S(\b \beta)}{\partial \b \beta^\prime} = -\sum_{i=1}^m \sum_{g=1}^G \tau_{ig} \sum_{j\in s_i} \left\{ \frac{\partial h(\eta_{ijg})}{\partial \eta_{ijg}}  \right\}\b x_{ij} \b x_{ij}^\prime,  \\[0.1cm]
&H(\pi_g, \pi_l) =  \frac{\partial S(\pi_g)}{\partial \pi_g} = -\sum_{i=1}^m \left[ \frac{\tau_{ig}}{\pi_g^2} \mathds{1} 1(g = l) + 
\frac{\tau_{iG}}{\pi_G^2}\right],  \quad g = 1, \dots, G-1.
\end{align*}
The remaining (non-redundant) elements of the expected complete data Hessian matrix,  that is $H(\b \alpha_g, \bs \alpha_k)$,  $H(\b \beta, \pi_g)$, and $H( \b \alpha_g, \pi_g)$, are all null due to parameter distinctiveness.

The second component involved in expression \eqref{eq:oakes} is the first derivative of the observed data score with respect to model parameter estimates (at convergence). 
For simplicity, we compute this quantity by numerical differentiation. 

According to \cite{FriedlKauermann2000}, the covariance matrix of parameter estimates may be based on the standard sandwich formula \citep{Huber:1967, white:1980}: 
\begin{equation}\label{eq:var}
 {V} (\hat{\bs  \Phi}  )={J}(\hat{\bs  \Phi} )^{-1} {V^\star}(\hat{\bs  \Phi} ){J}(\hat{\bs  \Phi})^{-1},
\end{equation}
where $ {V^\star}(\hat{\bs  \Phi} )= \sum_{i=1}^{m} S_{i}(\hat{\bs \Phi})S_{i}(\hat{\bs \Phi})^{\prime}$ denotes the estimate of the covariance matrix of the score $S(\bs \Phi)$ and $S_{i}(\hat{\bs \Phi})$ is the individual score vector. Such an approach helps stabilize the estimate of the observed information and ensures robustness to potential model misspecification.

\section{Analytic computation of the bias correction term for the MSE estimation}\label{sec:corr:term}
We provide computational details for the bias correction term required to get a second-order, bias-corrected, MSE estimator of the proposed np-EBP. 
Let $\b \Phi_0$ denote the ``true'' vector of model parameters and let us consider a second-order Taylor expansion of $d^{\ml{np}}({\b \Phi})$ around $\b \Phi_0$ evaluated at $\hat{\b \Phi}$:
\[
d^{\ml{np}}(\hat{\b \Phi}) = d^{\ml{np}}(\b \Phi_0) + \left(\frac{\partial d^{\ml{np}}}{\partial \bs \Phi} \right)^\prime\Bigg\rvert_{\b \Phi_0} (\hat{\bs \Phi} - \bs \Phi_0) + \frac{1}{2} (\hat{\bs \Phi} - \bs \Phi_0)^\prime \left(\frac{\partial^2 d^{\ml{np}}}{\partial \bs \Phi \bs \Phi^\prime} \right)(\hat{\bs \Phi} - \bs \Phi_0) + o_p(m^{-1}),
\]
where $ d^{\ml{np}}$ is a short hand for $ d^{\ml{np}}(\b \Phi)$. From the expression above, we get 
$$
E[d^{\ml{np}}(\hat{\b \Phi})] = d^{\ml{np}}({\b \Phi}) 
+ \frac{1}{m} b^{\ml{np}}(\b \Phi) +  o_p(m^{-1}),
$$
where $b^{\ml{np}}(\b \Phi)$ denotes a bias correction term which is given by:
\begin{align}\label{eq:biasNP_App}
b^{\ml{np}}(\bs \Phi) & = \left(\frac{\partial d^{\ml{np}}}{\partial \bs \Phi} \right)^\prime\Bigg\rvert_{\b \Phi_0} m E(\hat{\bs \Phi} - \bs \Phi_0) + \frac{m}{2}E\left[(\hat{\bs \Phi} - \bs \Phi_0)^\prime \left(\frac{\partial^2 d^{\ml{np}}}{\partial \bs \Phi \bs \Phi^\prime} \right)\Bigg\rvert_{\b \Phi_0}(\hat{\bs \Phi} - \bs \Phi_0)\right] \nonumber\\
& = b_1^{\ml{np}}(\hat{\b \Phi}) + b_2^{\ml{np}}(\hat{\b \Phi}).
\end{align}
To derive $b_2^{\ml{np}}(\hat{\b \Phi})$, we proceed as follows:
\begin{align*}
b_2^{\ml{np}}(\hat{\b \Phi}) &= \frac{m}{2}E\left[ (\hat{\bs \Phi} - \bs \Phi_0)^\prime \left(\frac{\partial^2 d^{\ml{np}}}{\partial \bs \Phi \partial \bs \Phi^\prime} \right)\Bigg\rvert_{\b \Phi_0} (\hat{\bs \Phi} - \bs \Phi_0)\right]  
\\
& = \frac{{m}}{2}E\left\{tr \left[\left(\frac{\partial^2 d^{\ml{np}}}{\partial \bs \Phi \partial\bs \Phi^\prime} \right)\Bigg\rvert_{\b \Phi_0} (\hat{\bs \Phi} - \bs \Phi_0)(\hat{\bs \Phi} - \bs \Phi_o)^\prime \right] \right\} 
\\
& = \frac{{m}}{2}tr \left\{E\left[\left(\frac{\partial^2 d^{\ml{np}}}{\partial \bs \Phi \partial \bs \Phi^\prime} \right)\Bigg\rvert_{\b \Phi_0} (\hat{\bs \Phi} - \bs \Phi_0) (\hat{\bs \Phi} - \bs \Phi_0)^\prime \right] \right\} 
\\
& = \frac{{m}}{2}tr \left\{\left(\frac{\partial^2 d^{\ml{np}}}{\partial \bs \Phi \partial \bs \Phi^\prime} \right) \Bigg\rvert_{\b \Phi_0}
E\left[(\hat{\bs \Phi} - \bs \Phi_0)(\hat{\bs \Phi} - \bs \Phi_0)^\prime\right] \right\} 
\\
& = \frac{{m}}{2}tr \left\{\left(\frac{\partial^2 d^{\ml{np}}}{\partial \bs \Phi \partial \bs \Phi^\prime} \right)\Bigg\rvert_{\b \Phi_0} V(\hat{\b \Phi}) \right\} .
\end{align*}
The first term in the right hand side of expression \eqref{eq:biasNP_App} requires a more complex computation. 
Let us consider a first-order Taylor expansion of the score function $S(\b \Phi )= {\partial \log L(\b \Phi})/{\partial \b \Phi}$ around $\b \Phi_0$ evaluated at $\hat{\b \Phi}$: 
\begin{align*}
0 = S(\hat{\b \Phi}) & = S({\b \Phi_0}) + \left[\frac{\partial S(\b \Phi)}{\partial \b \Phi}\right]\Bigg\rvert_{\b \Phi_0} (\hat{\b \Phi} - {\b \Phi_0}) + o_p(\lvert \lvert \hat{\b \Phi} - {\b \Phi_0} \rvert \rvert).
\end{align*}
Based on the above expression, we have
\begin{align}\label{eq:firstOrderScore}
\hat{\b \Phi} - {\b \Phi_0} &= 
J(\b \Phi_0)^{-1} S(\b \Phi_0) +  o_p(\lvert \lvert \hat{\b \Phi} - {\b \Phi_0} \rvert \rvert), 
\end{align}
where $J(\b \Phi_0) = -[{\partial S(\b \Phi)}/{\partial \b \Phi}]_{\b \Phi_0}$ denotes the Fisher information matrix.
Then, let us notice that 
\[
J(\b \Phi_0) = I_e(\b \Phi_0) + o_p(1),
\]
where $I_e(\b \Phi_0) $ corresponds to the expected information matrix, so that
\begin{equation}\label{eq:firstOrderScore2}
\hat{\b \Phi} - {\b \Phi_0} = 
I_e(\b \Phi_0)^{-1} S(\b \Phi_0) +  o_p(\lvert \lvert \hat{\b \Phi} - {\b \Phi_0} \rvert \rvert), 
\end{equation}

Let us now consider a second-order Taylor expansion of the $k$-th element of the score function, $S^k(\b \Phi)$, around $\b \Phi_0$ evaluated at $\hat{\b \Phi}$, with $k = 1, \dots, K,$ and $K$ being the number of free model parameters: 
\begin{align*}
0 = S^k(\hat{\b \Phi}) &=
S^k({\b \Phi_0}) + \left[\frac{\partial S^k(\b \Phi)}{\partial \b \Phi}\right]\Bigg\rvert_{\b \Phi_0} (\hat{\b \Phi} - {\b \Phi_0})\nonumber 	\\
& +\frac{1}{2} \left[(\hat{\b \Phi} - {\b \Phi_0})^\prime
\left[\frac{\partial^2 S^k(\b \Phi)}{\partial \b \Phi \partial \b \Phi^\prime}\right]\Bigg\rvert_{\b \Phi_0}(\hat{\b \Phi} - {\b \Phi_0}) \right] + o_p(\lvert \lvert \hat{\b \Phi} - {\b \Phi} \rvert \rvert^2),
\end{align*}
which, according to expression \eqref{eq:firstOrderScore2}, may be written as
\begin{align*}
0 = S^k(\hat{\b \Phi}) & = S^k({\b \Phi_0}) + \left[\frac{\partial S^k(\b \Phi_0)}{\partial \b \Phi}\right]\Bigg\rvert_{\b \Phi_0} (\hat{\b \Phi} - {\b \Phi_0}) \\
&+ 
\frac{1}{2} \left[S(\b \Phi_0)^\prime I_e(\b \Phi_0)^{-1}   
\left[\frac{\partial^2 S^k(\b \Phi_0)}{\partial \b \Phi \partial \b \Phi^\prime}\right]\Bigg\rvert_{\b \Phi_0}
I_e(\b \Phi_0)^{-1} S(\b \Phi_0)  \right] + o_p(\lvert \lvert \hat{\b \Phi} - {\b \Phi_0} \rvert \rvert^2).
\end{align*}
The corresponding second order multivariate Taylor expansion of $S(\b \Phi)$ around $\b \Phi_0$ evaluated at $\hat{\b \Phi}$ may be obtained by stacking those for $S^k(\b \Phi)$, with for $k = 1, \dots, K$, and is given by
\begin{align*}
0 = S(\hat{\b \Phi}) &= S({\b \Phi_0}) + \left[\frac{\partial S(\b \Phi)}{\partial \b \Phi}\right]\Bigg\rvert_{\b \Phi_0} (\hat{\b \Phi} - {\b \Phi_0}) \\
& + \frac{1}{2} \left[S(\b \Phi_0)^\prime I_e(\b \Phi_0)^{-1} 
\left[\frac{\partial^2 S^k(\b \Phi)}{\partial \b \Phi \partial \b \Phi^\prime}\right]\Bigg\rvert_{\b \Phi_0}  I_e(\b \Phi_0)^{-1} S(\b \Phi_0) \right]_{1\leq k \leq K} + o_p(\lvert \lvert \hat{\b \Phi} - {\b \Phi_0} \rvert \rvert^2).
\end{align*}
Solving for $(\hat{\b \Phi} - {\b \Phi_0})$, we obtain
\[
\hat{\b \Phi} - {\b \Phi_0} = 
I_e(\b \Phi_0)^{-1} \left\{
S(\b \Phi_0) + \frac{1}{2} \left[S(\b \Phi_0)^\prime I_e(\b \Phi_0)^{-1}   
\left[\frac{\partial^2 S^k(\b \Phi)}{\partial \b \Phi \partial \b \Phi^\prime}\right]\Bigg\rvert_{\b \Phi_0} I_e(\b \Phi_0)^{-1} S(\b \Phi_0)  \right]_{1\leq k \leq K} \right\}+ o_p(\lvert \lvert \hat{\b \Phi} - {\b \Phi_0} \rvert \rvert^2).
\]

Going back to the computation of the bias correction term, we may substitute the above expression into $b_1^{\ml{np}}(\hat{\b \Phi})$ and obtain the following approximation:
\begin{align*}
b_1^{\ml{np}}(\hat{\b \Phi}) &= \left(\frac{\partial d^{\ml{np}}}{\partial \bs \Phi} \right)^\prime\Bigg \rvert_{\b \Phi_0} m E(\hat{\bs \Phi} - \bs \Phi) 
\\
&= \left(\frac{\partial d^{\ml{np}}}{\partial \bs \Phi} \right)^\prime\Bigg \rvert_{\b \Phi_0}  m \Bigg \{ I_e({\bs \Phi_0})^{-1} E[S(\b \Phi_0)]\\
& \quad  + \frac{1}{2} I_e({\bs \Phi_0})^{-1} E\left[S(\b \Phi_0)^\prime I_e(\b \Phi_0)^{-1}   
\left[\frac{\partial^2 S^k(\b \Phi)}{\partial \b \Phi \partial \b \Phi^\prime}\right]\Bigg\rvert_{\b \Phi_0} I_e(\b \Phi_0)^{-1} S(\b \Phi_0)  \right]_{1\leq k \leq K}
\Bigg\}
\end{align*}
Clearly, due to $E[S(\b \Phi_0)] = 0$, the first term in the curly brackets vanishes. The remaining terms are computed as follows
\begin{align*}
b_1^{\ml{np}}(\hat{\b \Phi}) &= \left(\frac{\partial d^{\ml{np}}}{\partial \bs \Phi} \right)^\prime \Bigg \rvert_{\b \Phi_0}  \frac{m}{2} I_e({\bs \Phi_0})^{-1} E\left\{ tr\left[S(\b \Phi_0)^\prime I_e(\b \Phi_0)^{-1}   
\left[\frac{\partial^2 S^k(\b \Phi)}{\partial \b \Phi \partial \b \Phi^\prime}\right]\Bigg\rvert_{\b \Phi_0} I_e(\b \Phi_0)^{-1} S(\b \Phi_0)  \right]_{1\leq k \leq K}
\right\}
\\
&= \left(\frac{\partial d^{\ml{np}}}{\partial \bs \Phi} \right)^\prime\Bigg \rvert_{\b \Phi_0}  \frac{m}{2} I_e({\bs \Phi_0})^{-1} E\left\{ tr\left[
I_e(\b \Phi_0)^{-1}   
\left[\frac{\partial^2 S^k(\b \Phi)}{\partial \b \Phi \partial \b \Phi^\prime}\right]\Bigg\rvert_{\b \Phi_0} I_e(\b \Phi)^{-1} S(\b \Phi_0)
S(\b \Phi_0)^\prime   \right]_{1\leq k \leq K}
\right\}
\\
&= \left(\frac{\partial d^{\ml{np}}}{\partial \bs \Phi} \right)^\prime\Bigg \rvert_{\b \Phi_0}  \frac{m}{2} I_e({\bs \Phi_0})^{-1} E\left\{ tr\left[
I_e(\b \Phi_0)^{-1}   
\left[\frac{\partial^2 S^k(\b \Phi)}{\partial \b \Phi \partial \b \Phi^\prime}\right]\Bigg\rvert_{\b \Phi_0} I_e(\b \Phi_0)^{-1} I_e(\b \Phi_0)  \right]_{1\leq k \leq K}
\right\}
\\
& -\left(\frac{\partial d^{\ml{np}}}{\partial \bs \Phi} \right)^\prime\Bigg \rvert_{\b \Phi_0}  \frac{m}{2} I_e({\bs \Phi_0})^{-1} tr\left\{ I_e(\b \Phi_0)^{-1}   
\left[
\frac{  \partial I_{e}^{k}(\b \Phi_0)}{ \partial \b \Phi^\prime}\right]_{1\leq k \leq K}
\right\},
\end{align*}
where $I_{e}^{k}(\b \Phi_0)$ denotes the $k$-th row of the expected information matrix and $S(\b \Phi_0) S(\b \Phi_0)^\prime $ is approximated by $I_e(\b \Phi_0)$.

\section{Model derivatives for Bernoulli responses}\label{sec:appB:deriv}
In this section, we provide explicit formulas for computing model derivatives in the case of binary data. As before, let $\tilde p_{i(h)}$ denote the np-BP of $p_i$, conditional on $y_{i\cdot} = h$, with $y_{i\cdot} = \sum_{j \in s_i} y_{ij}$ and let $\tau_{ig(h)}$ be the posterior probability for the $i$-th small area to belong to the $g$-th component of the finite mixture,  conditional on $y_{i\cdot} = h$. This latter quantity is defined as
\[
\tau_{ig(h)} = \frac{\exp \left[ \alpha_g h - \sum_{j\in s_i} \log \left(1 + e^{\eta_{ijg}}\right)
\right]\pi_g}
{\sum_{l = 1}^G \exp \left[ \alpha_l h - \sum_{j\in s_i}\log \left(1 + e^{\eta_{ijl}})\right)
\right]\pi_l} .
\]
The first derivative vector of $\tilde p_{i(h)}$ with respect to model parameters has elements: 
\begin{align*}
\footnotesize
\frac{\partial \tilde p_{i(h)}^{\ml{np-BP}}}{\partial \alpha_g} &= 
 \frac{1}{N_i}  \sum_{j = 1}^{N_i}  \left\{ \frac{\partial} {\partial \alpha_g} \left(\sum_{l=1}^G p_{ijl}\,  \tau_{il(h)} \right) \right\} \\
 & =  \frac{1}{N_i} \sum_{j = 1}^{N_i}  \left\{\left[ \tau_{ig(h)} \frac{\partial p_{ijg}}{\partial \alpha_g} \right] + \left[ \sum_{l=1}^G p_{ijl} \frac{\partial \tau_{il(h)}}{\partial \alpha_g} \right]\right\}  \\[2mm]
 &= \frac{1}{N_i}  \sum_{j = 1}^{N_i} \left\{
\tau_{ig(h)}  p_{ijg} (1-p_{ijg}) + \right.  \\[2mm]
&\left. + \sum_{l=1}^G p_{ijl} \left[ 
\frac{ \pi_l  f_{il(h)}S_{il(h)}(\alpha_l)
}{\sum_{k=1}^G \pi_k f_{ik(h)}} \mathds 1(l = g) - \frac{ 
\pi_l f_{il(h)} \pi_g f_{ig(h)}S_{ig(h)}(\alpha_g)
}
{\left[\sum_{k=1}^G \pi_k f_{ik(h)}\right]^2} \right]\right\}\\[2mm]
&=\frac{1}{N_i} \sum_{j = 1}^{N_i} \left\{ \tau_{ig(h)} p_{ijg} (1-p_{ijg}) 
+
\sum_{l=1}^G p_{ijl}\,  \tau_{il(h)} S_{il(h)}(\alpha_l) \mathds 1(l = g)
- \sum_{l=1}^G p_{ijl}
\tau_{il(h)} \tau_{ig(h)}S_{ig(h)}(\alpha_g),
 \right\}.
\end{align*}
with $f_{ig(h)}$ and  $S_{ig(h)} (\alpha_g) = h - \sum_{r \in s_i} p_{irg}$ denoting the joint conditional distribution and the score function for the $i$-th small area, respectively, both conditional on the $g$-th component of the finite mixture and $y_{i\cdot} = h$.

For the fixed parameters $\bs \beta$, model derivatives are obtained as follows:
\begin{align*}
\footnotesize
  \frac{\partial \tilde p_{i(h)}^{\ml{np-BP}}}{\partial \bs \beta}  &= \frac{1}{N_i} \sum_{j=1}^{N_i} \left\{
 \frac{\partial }
{\partial \bs \beta} \left(\sum_{l=1}^G p_{ijl} \tau_{il(h)} \right) \right\}\\[2mm]
& = \frac{1}{N_i} \sum_{j=1}^{N_i} \left\{
 \left[\sum_{l = 1}^G \tau_{il(h)} \frac{\partial p_{ijl}}{\partial \bs\beta} \right] + \left[ \sum_{l=1}^G p_{ijl} \frac{\partial \tau_{il(h)}}{\partial \bs \beta}
\right] \right\}\\[2mm]
&= \frac{1}{N_i} \sum_{j=1}^{N_i} \left\{
\left[\sum_{l=1}^G \tau_{il(h)}   p_{ijl} (1-p_{ijl}) \b x_{ij} \right] + \right.\\[2mm]
&\left.  +  \sum_{l=1}^G p_{ijl} \left[ 
\frac{\pi_l f_{il(h)}S_{il(h)}(\b \beta)}{\sum_{k=1}^G \pi_k f_{ik(h)}} - \frac{
\pi_l f_{il(h)} \sum_{u = 1}^G \pi_u f_{iu(h)} S_{iu(h)}(\b \beta) }{\left[\sum_{k=1}^G \pi_k f_{ik(h)}\right]^2} \right]\right\}\\[2mm] 
&=\frac{1}{N_i} \sum_{j =1}^{N_i} \left\{\sum_{l=1}^G \tau_{il(h)}  p_{ijl} (1-p_{ijl}) \b x_{ij}^\prime 
+
\sum_{l=1}^G p_{ijl} \tau_{il(h)} S_{il(h)}(\b \beta)  +\right.\\[2mm]
& \left.- \sum_{l=1}^G p_{ijl} 
\tau_{il(h)} \sum_{k=1}^G \tau_{ik(h)}  S_{ik(h)}(\b \beta)
 \right\},
\end{align*}
with $S_{ig(h)}(\b \beta) = -\sum_{r \in s_i} p_{irg} \b x_{ir}$.
Last, for the component probabilities $\pi_g$, we have: 
\begin{align*}
\frac{\partial \tilde p_{i(h)}^{\ml{np-BP}}}{\partial \pi_g} &=
\frac{1}{N_i} \sum_{j=1}^{N_i} \left\{ \sum_{l=1}^G p_{ijl}\left[\frac{f_{il(h)} }
{\sum_{k=1}^G \pi_k f_{ik(h)}} \mathds 1(g = l)-
\frac{ f_{il(h)} \pi_l f_{ig(h)}}{(\sum_{k=1}^G \pi_k f_{ik(h)})^2}\right]\right\}\\
& = \frac{1}{N_i} \sum_{j=1}^{N_i} \left\{ \sum_{l=1}^G p_{ijl}\left[\frac{\tau_{il(h)}}{\pi_l} \mathds 1(g = l)- \frac{\tau_{il(h)}\tau_{ig(h)}}{\pi_g }\right] \right\}.
 \end{align*}

{
\section{Simulation results: Mean Absolute Error of the estimators}
In this section, we report the distribution of the Mean Absolute Bias (MAE) across small areas for the three estimators under comparison in the simulation study. For each area, the MAE index is computed as follows:
$$
\mbox{MAE}_i=T^{-1}\sum_{t=1}^{T}|\hat{p}_{it}^{\ml{Model}}-p_{it}|, \quad i = 1, \dots, m.
$$
}

\begin{figure}[h]
\caption{Distribution of the MAE over areas for the np-EBP, the the EBP, and the Naive approach, for $m=100$ (left panel), $m = 200$ (central panel), and $m = 500$ (right panel), under Scenario $G$ (upper panel) and Scenario M (lower panel). \label{fig:predictionBias}}
\vspace{2mm}
\centering

\frame{\includegraphics[scale=0.13]{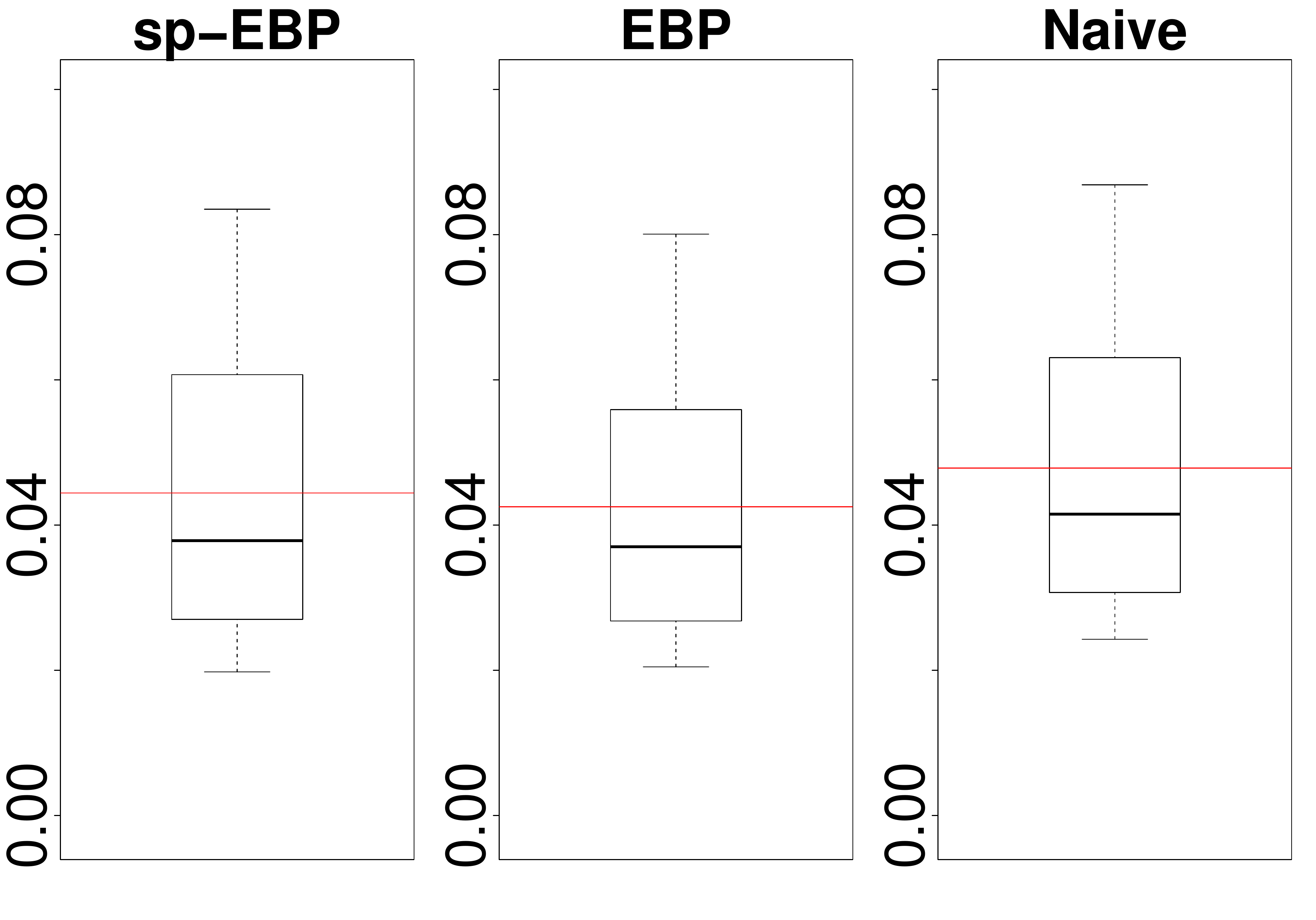}}
\frame{\includegraphics[scale=0.13]{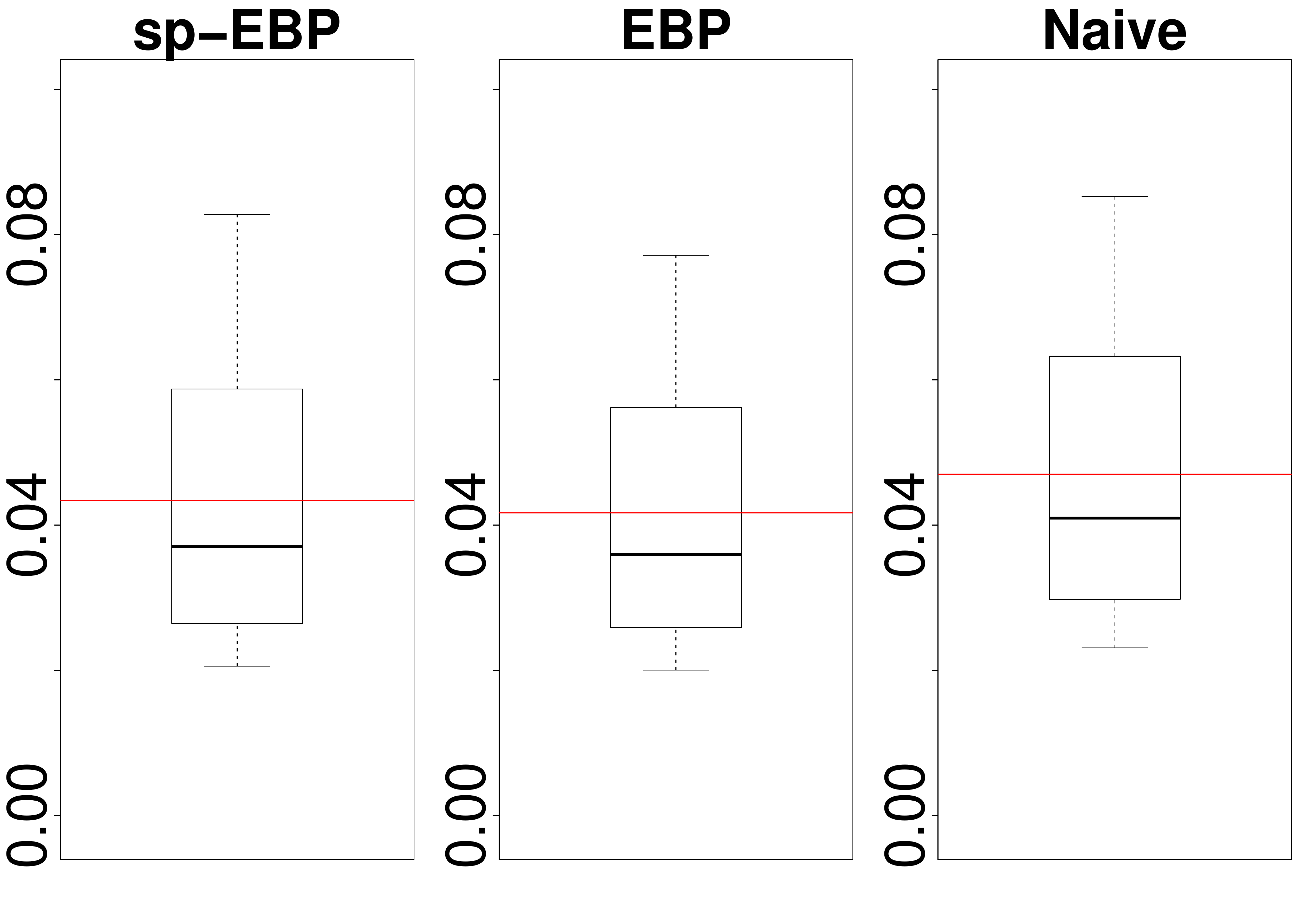}}
\frame{\includegraphics[scale=0.13]{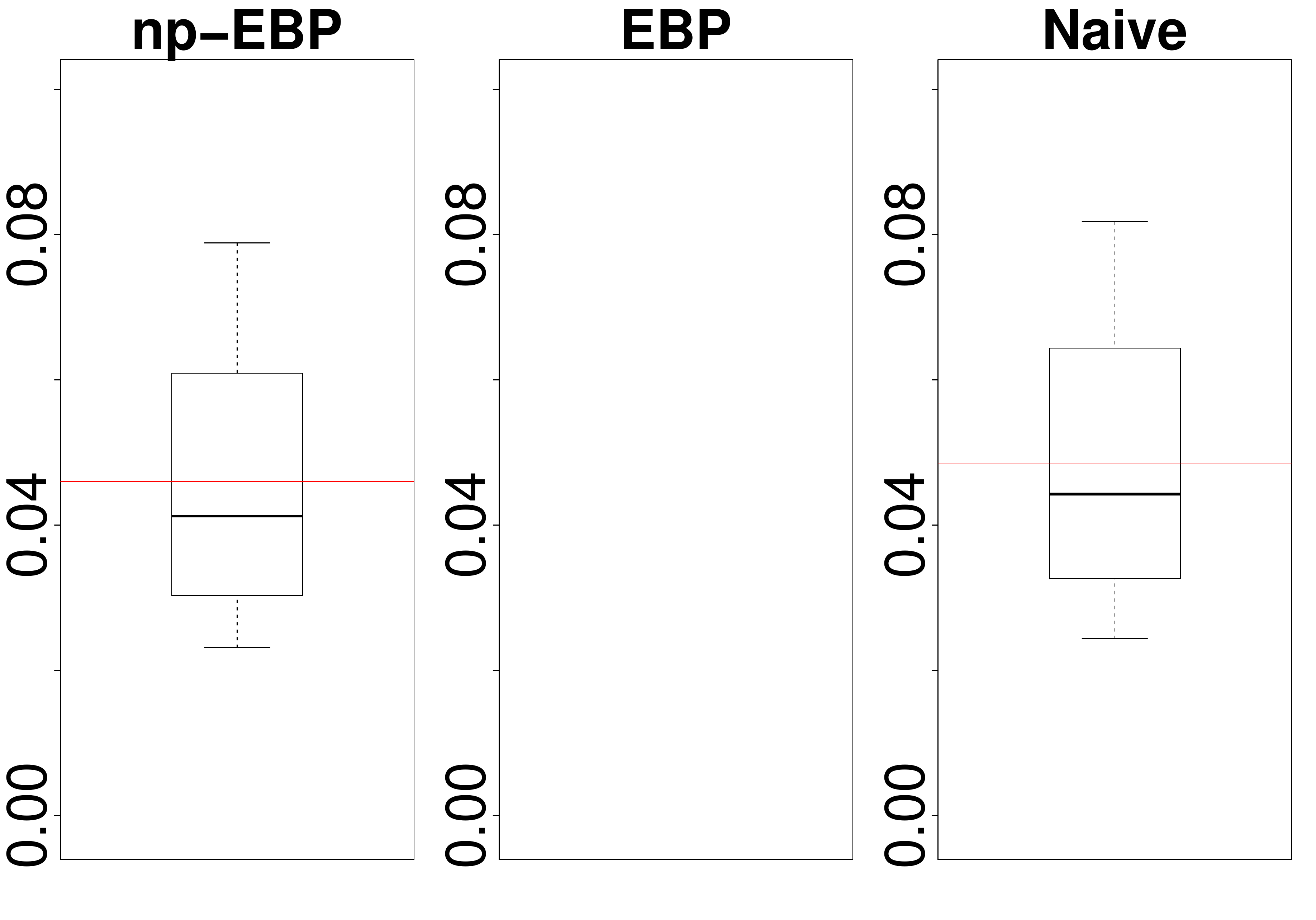}}
\\[5mm]

\frame{\includegraphics[scale=0.13]{bias_pred_m100.pdf}}
\frame{\includegraphics[scale=0.13]{bias_pred_m200.pdf}}
\frame{\includegraphics[scale=0.13]{bias_pred_m500.pdf}}

\end{figure}

\newpage
\bibliographystyle{apalike}